\documentclass[11pt]{article}

\usepackage[usenames,dvipsnames]{pstricks}
\usepackage{epsfig}
\usepackage{pst-grad} 
\usepackage[space]{grffile} 
\usepackage{etoolbox} 
\makeatletter 
\patchcmd\Gread@eps{\@inputcheck#1 }{\@inputcheck"#1"\relax}{}{}
\makeatother

\usepackage{graphicx}
\usepackage{amsmath}
\usepackage{psfrag}
\usepackage{color,soul}
\usepackage{natbib}
\usepackage{siunitx}
\usepackage{float}
\usepackage{overpic}
\usepackage{tikz}
\usepackage{multirow}
\usepackage{bm}
\usepackage{textgreek}
\usepackage{url}
\usepackage{amssymb}
\usepackage{wasysym}
\usepackage{xfrac}
\usepackage{setspace}
\usepackage{gensymb}
\usepackage{mathtools}
\usepackage{arydshln}

\def\NoNumber#1{{\def\alglinenumber##1{}\State #1}\addtocounter{ALG@line}{-1}}
\usepackage{booktabs} 
\usepackage{algorithm}
\usepackage[noend]{algpseudocode}

\def \eg{\emph{e.g.}, }
\def \ie{\emph{i.e.}, }

\usepackage{hyperref}
\definecolor{darkblue}{rgb}{0,0,0.80}
\hypersetup{
    colorlinks=true,
    linkcolor={darkblue},
    citecolor={darkblue},
    filecolor=black,
    urlcolor={darkblue},
    plainpages=false,
}

\usepackage[noblocks]{authblk}
\usepackage[margin=1in]{geometry}
\usepackage[title]{appendix}

\newcommand\affiliation[1]{\gdef\@affiliation{\let\aff\aff@inst#1}}
\gdef\@affiliation{}
\def\email#1{Email address for correspondence: #1}
\def\aff#1{\ignorespaces\textsuperscript{#1}}
\def\corresp#1{\unskip\thanks{#1}}
\numberwithin{equation}{section}

\renewenvironment{abstract}
{\begin{quote}
\noindent \rule{\linewidth}{.5pt}\par{\bfseries \abstractname.}}
{\medskip\noindent \rule{\linewidth}{.5pt}
\end{quote}
}

  \DeclareTextFontCommand\textsfi{\usefont{OT1}{cmss}{m}{sl}}
  \DeclareMathAlphabet\mathsfi            {OT1}{cmss}{m}{sl}
  \DeclareTextFontCommand\textsfb{\usefont{OT1}{cmss}{bx}{n}}
  \DeclareMathAlphabet\mathsfb            {OT1}{cmss}{bx}{n}
  \DeclareTextFontCommand\textsfbi{\usefont{OT1}{cmss}{m}{sl}}
  \DeclareMathAlphabet\mathsfbi            {OT1}{cmss}{m}{sl}
  \IfFileExists{t1phv.fd}
  {\DeclareTextFontCommand\textsfbi{\usefont{T1}{phv}{b}{it}}
  \DeclareMathAlphabet\mathsfbi            {T1}{phv}{b}{it}
  }{}
  \IfFileExists{ot1phv.fd}
  {\DeclareTextFontCommand\textsfbi{\usefont{OT1}{phv}{b}{it}}
  \DeclareMathAlphabet\mathsfbi            {OT1}{phv}{b}{it}
  }{}
  
\DeclareSymbolFont{matha}{OML}{txmi}{m}{it}

\newcommand{\edit}[1]{\color{black}#1 \color{black}}

\title{\bf Scalable resolvent analysis for three-dimensional flows}

\author[1]{\bf Ali Farghadan}
\author[2]{\bf Eduardo Martini}
\author[1]{\bf Aaron Towne\corresp{\email{towne@umich.edu}}}

\affil[1]{\normalsize Department of Mechanical Engineering, University of Michigan, Ann Arbor, MI, USA }
\affil[2]{\normalsize Institut Pprime, CNRS, Universit\'e de Poitiers ISAE-ENSMA, Poitiers, France	}

\date{}
\begin{document}
\maketitle

\begin{abstract}

Resolvent analysis is a powerful tool for studying coherent structures in turbulent flows. However, its application beyond canonical flows with symmetries that can be used to simplify the problem to inherently three-dimensional flows and other large systems has been hindered by the computational cost of computing resolvent modes. In particular, the CPU and memory requirements of state-of-the-art algorithms scale poorly with the problem dimension, \ie the number of discrete degrees of freedom. In this paper, we present RSVD-$\Delta t$, a novel approach that overcomes these limitations by combining randomized singular value decomposition with an optimized time-stepping method for computing the action of the resolvent operator. Critically, the CPU cost and memory requirements of the algorithm scale linearly with the problem dimension\edit{.} We develop additional strategies to minimize these costs and control errors. We validate the algorithm using a Ginzburg-Landau test problem and demonstrate \edit{RSVD-$\Delta t$'s} low cost and improved scaling using a three-dimensional discretization of a turbulent jet. Lastly, we use it to study the impact of low-speed streaks on the development of Kelvin-Helmholtz wavepackets in the jet via secondary stability analysis, a problem that would have been intractable using previous algorithms.

\end{abstract}


\section{Introduction} \label{sec:intro}

\edit{Turbulent flows, though chaotic, often exhibit recurring patterns that are essential to their dynamics. For instance, near-wall streaks affect both momentum transfer and energy dissipation \citep{Zhouetal22}, while large-scale motions contribute to turbulence by shaping the energy cascade \citep{HutchinsMarusic07}, and hairpin vortices play a key role in energy and momentum transport within turbulent boundary layers \citep{Adrian07}. In free-shear flows, coherent structures arising from the Kelvin-Helmholtz instability, Orr mechanism, and lift-up mechanism have been observed in free jets, where they are integral to energy transfer and noise generation \citep{JordanColonius13, Pickeringetal20}. Coherent structures are also critical to the transition from laminar to turbulent flow \citep{SchmidHenningson01} and to sustaining turbulence \citep{Mckeon17}.} Popular data-driven methods include proper orthogonal decomposition (POD) \citep{Sirovich87_1}, dynamic mode decomposition (DMD)  \citep{Schmid10}, and spectral proper orthogonal decomposition (SPOD) \citep{Lumley67, Towneetal18}. In particular, SPOD identifies energy-ranked, single-frequency structures that evolve coherently in space and time.

\par Resolvent (or input-output) analysis originates from classical control theory \citep{DunfordSchwartz58, Kato13} and has become arguably the most important operator-theoretic modal decomposition technique in fluid mechanics \citep{McKeonSharma10, Tairaetal17, Jovanovic21}. Resolvent analysis has been applied to a wide variety of flows, including canonical wall-bounded flows \citep{DawsonMcKeon19, Morraetal19}, turbulent jets \citep{Jeunetal16, Schmidtetal18, Lesshafftetal19, Pickeringetal20}, and airfoils \citep{ThomareisPapadakis18, Yehetal20}. It has been used for diverse tasks including design optimization \citep{ChavarinLuhar20, Ranetal21}, receptivity analysis \citep{Kamaletal23, CookNichols23}, and flow control \citep{YehTaira19, Towneetal20, Martinietal20, Martinietal22}. Singular value decomposition (SVD) of the resolvent operator is at the heart of input-output-based studies. The left singular vectors of the resolvent operator, known as the response modes, are often related to the coherent motions in the flow \citep{McKeonSharma10, Towneetal18}. Specifically, the resolvent modes associated with the largest singular values \edit{can} provide an approximation of the leading SPOD modes \citep{Towneetal18} and, in some cases, capture the majority of the power spectral density (PSD) of the flow \citep{Symonetal19}. The right singular vectors, also known as the forcing modes, describe the optimal inputs that lead to the most amplified responses, characterized by the largest singular values, and offer information about the mechanisms driving these responses. \edit{The singular values, referred to as the gains, quantify the amplification from unit-norm forcing to the corresponding response.}

\par Resolvent analysis can be computationally demanding. Two steps constitute most of the cost:  $(i)$ forming the resolvent operator, which involves computing an inverse, and $(ii)$ performing the SVD. Both steps nominally scale like $O(N^3)$, where $N$ is the state dimension. State-of-the-art methods, described below, improve on this scaling, but the computational cost remains a strong function of the state dimension $N$. The state dimension, in turn, depends acutely on the number of spatial dimensions that must be numerically discretized. While the linearized Navier–Stokes equations are nominally three-dimensional, they can be simplified by expanding the flow variables into Fourier modes in homogenous dimensions, \ie those in which the base flow about which the equations are linearized does not vary. This markedly reduces the size of the discretized operators that must be manipulated, decreasing the computational cost. Accordingly, inherently three-dimensional flows that do not contain homogeneous directions or other simplifying symmetries are particularly challenging.

\par Recent advancements aim to overcome these two computational bottlenecks. The second bottleneck can be alleviated by using efficient algorithms to compute only the SVD modes with the largest singular values, which are typically of primary interest, rather than the complete decomposition. Standard methods like power iteration and various Arnoldi methods have been frequently applied for this purpose. More recently, randomized singular value decomposition (RSVD) \citep{Halkoetal11} has been shown to further reduce the cost of resolvent analysis of one- \citep{Moarrefetal13} and two-dimensional \citep{Ribeiroetal20} problems.

\par Regarding the first bottleneck, forming the resolvent operator by computing an inverse is feasible only for small systems, \eg one-dimensional ones. Fortunately, the aforementioned SVD algorithms do not require direct access to the resolvent operator, but rather its action on a specified forcing vector, \ie the result of applying the resolvent operator to that vector. Accordingly, we can recast the first bottleneck in terms of the computational cost of computing the action of the resolvent operator on a vector. The standard approach for doing so is to solve a linear system whose solution yields the action of the resolvent operator on the right-hand-side vector via LU decomposition of the inverse of the resolvent operator (which can be directly formed in terms of the linearized Navier-Stokes operator; see $\S$\ref{sec:RSVD} for details). The computational cost of this approach typically scales like $O(N^{1.5})$ or $O(N^2)$ for two- and three-dimensional problems, respectively, which is tolerable for most two-dimensional problems but quickly becomes intractable for three-dimensional problems. Numerous authors have used this LU-based approach along with Arnoldi methods \citep{SippMarquet13, Jeunetal16, Schmidtetal18, Karbanetal20}. \citet{Brynjelletal17} used LU decomposition along with a power iteration with a Laplace preconditioner to increase the convergence rate of the resolvent modes. \citet{Ribeiroetal20} used LU decomposition along with RSVD, which we call ``RSVD-LU'' in this study, and demonstrated significant CPU savings compared to using an Arnoldi iteration. However, the poor cost scaling of the LU decomposition with problem size $N$\edit{, typically between $O(N^{1.5})$ to $O(N^2)$,} remains a limiting factor, impeding the investigation of three-dimensional flows and other large systems. Recently, \citet{Houtmanetal23} used an iterative solver as an alternative to LU decomposition to handle large systems. While this approach is attractive due to its potential to achieve $O(N)$ scaling, the absolute cost of iterative solvers depends heavily on the availability of an effective preconditioner, which is typically problem dependent, especially for the ill-conditioned matrices obtained from the linearized Navier-Stokes equations.

\par Resolvent modes can be computed at a reduced cost for slowly varying flows, \ie flows whose mean changes gradually in some spatial direction, by using spatial marching methods to approximate the action of the resolvent operator. Spatial marching methods approximately evolve perturbations in the slowly varying direction. The best-known spatial marching method is the parabolized stability equations (PSE), but the inherent ill-posedness of PSE \citep{LiMalik96} requires deleterious regularization that makes it ill-suited to compute resolvent modes in most cases \citep{Towneetal19}. One exception is very low frequencies, where PSE has been used to compute resolvent modes corresponding to boundary-layer streaks \citep{Sasakietal22}. The one-way Navier–Stokes (OWNS) equations \citep{TowneColonius15} overcome many of the limitations of PSE; they are formally well-posed and capture the complete downstream response of the flow. The original formulation did not include a right-hand-side forcing on the linearized equations, which is fundamental to resolvent analysis. This was addressed by a second OWNS variant formulated in terms of a projection operator that splits both the solution and forcing into upstream- and downstream-traveling components \citep{Towneetal22}. This method has been combined with a power-iteration approach to accurately and efficiently approximate resolvent modes for a range of slowly varying flows ranging from incompressible boundary layers to supersonic jets to hypersonic boundary layers. Recently, the cost of this approach was further reduced by a new recursive OWNS formulation \citep{ZhuTowne23}. The fundamental limitation of OWNS-based approaches is their restriction to (mostly) canonical flows that contain a slowly varying direction.

\par Several data-driven methods for computing resolvent modes have been proposed, which avoid working directly with the resolvent operator at all. \citet{Towneetal15} and \citet{Towne16} introduced empirical resolvent decomposition (ERD). Starting with data in the form of a set of forcing and response pairs, ERD solves an optimization problem to identify modes within the span of the data that maximizes the gain. Another recent approach uses dynamic mode decomposition (DMD) \citep{Schmid10} to estimate the resolvent modes from data \citep{Herrmannetal21}. This approach benefits from the advancements in DMD \citep{Schmid22} and is robust, but to accurately approximate the resolvent modes, many random initial conditions may need to be simulated.

\citet{Bartheletal22} recently proposed a reformulation of resolvent analysis called variational resolvent analysis (VRA). Using the same mathematics that underly ERD, VRA computes resolvent modes by solving a Rayleigh quotient, avoiding the inverse that appears in the definition of the resolvent operator. To make the method computationally advantageous, the response modes are constrained to lie within the span of some other reduced-order basis. \citet{Bartheletal22} obtain this basis from a series of locally parallel resolvent analyses; if the basis is taken from data, VRA becomes ERD. VRA showed speed-up compared to standard approaches for a canonical boundary layer, but it remains to be investigated for more complex scenarios where an effective basis is not evident.

\par Time-stepping methods offer an alternative approach to overcome the first bottleneck (these methods are sometimes referred to as ``matrix-free'' approaches, as forming the LNS operator is not necessary). The central idea is to obtain the action of the resolvent operator on a vector by solving the linearized equations in the time domain. A pioneering study by \citet{Monokrousosetal10} used time stepping along with power iteration to compute resolvent modes for a flat-plate boundary-layer flow. Modes at a particular frequency of interest were computed by forcing the linearized equations exclusively at that frequency and time stepping until a steady-state solution is obtained. \citet{Gomezetal16} proposed an iterative procedure for updating the initial conditions to reduce the time required to reach the steady-state solution. This resulted in an 80\% reduction of CPU time for a test problem, but only the leading mode at each frequency was obtained. \citet{Martinietal21} introduced two additional variations of time-stepping approaches for computing resolvent modes with improved efficiency. The first, referred to as the transient response method, evaluates the transitional response of the LNS to temporally compact forcing. The second variation, known as the steady-state response method, computes the steady-state solution of the LNS when it is forced with a set of harmonic frequencies. Both methods allow all frequencies of interest to be simultaneously computed by isolating each frequency in the flow response using a discrete Fourier transform. Additionally, the steady-state method can be easily paired with more advanced SVD algorithms (\eg Arnoldi, rather than power iteration) to obtain multiple resolvent modes at each frequency.  

\par Time-stepping methods for computing resolvent modes are potentially powerful because they obtain the action of the resolvent operator without the need for inverses or LU decomposition. Indeed, we will show that time time-stepping methods can achieve linear cost scaling with the problem dimension $N$. However, achieving this potential and overall low CPU and memory costs requires careful consideration of numerous factors \edit{including simultaneous time integration for all frequencies of interest, the use of explicit solvers when implicit solvers require costly LU decomposition or preconditioning, efficient removal of undesired transient responses—especially if they decay slowly—and leveraging streaming calculations.}

\par In this paper, we present a novel approach, abbreviated as ``RSVD-$\Delta t$'', that combines the benefits of RSVD with the advantages of time stepping. In short, the method eliminates the bottleneck in the RSVD-LU approach created by the LU decomposition by obtaining the action of the resolvent operator via an optimized time-stepping approach. All frequencies of interest as computed simultaneously using a steady-state response approach as in \citet{Martinietal21}. Additionally, we develop a novel technique to remove the undesired transient component of the response, shortening the temporal interval over which the equations are integrated and reducing the CPU cost by an order of magnitude in most cases. To minimize memory usage, we utilize streaming calculations for transferring data between the Fourier and time domains. The RSVD-$\Delta t$ algorithm is shown to exhibit linear scalability both in terms of computational complexity and memory requirements and can be efficiently parallelized. Overall, these capabilities allow us to compute resolvent modes for three-dimensional flows and other large systems that were previously out of reach. \edit{An open-source, parallelized implementation of our algorithm is available on GitHub (\url{https://github.com/AliFarghadan/RSVD-Delta-t}).}

\par In the remainder of the paper, we provide a brief review of the formulation and computation of resolvent analysis in $\S$\ref{sec:resolvent}, discuss the RSVD-LU algorithm in $\S$\ref{sec:RSVD}, explain the time-stepping method in $\S$\ref{sec:TS}, and introduce our RSVD-$\Delta t$ algorithm in $\S$\ref{sec:RSVDt}. An overview of the computational complexity of all approaches is given in $\S$\ref{sec:comp_theory}, the sources of errors of our algorithm are detailed in $\S$\ref{sec:ErrSource}, and approaches to optimize the algorithm are developed in $\S$\ref{sec:Optimizing}. Two test cases are defined in $\S$\ref{sec:test_cases} to validate, examine and compare the accuracy and performance of RSVD-$\Delta t$ against other approaches. In $\S$\ref{sec:application}, we use RSVD-$\Delta t$ to study the impact of streaks on the Kelvin-Helmholtz wavepackets in a jet. Concluding remarks are made in $\S$\ref{sec:con}. 

\section{Resolvent analysis} \label{sec:resolvent}

\subsection{Formulation}

\edit{Our starting point is the compressible Navier-Stokes equations, written as
\begin{equation}
\frac{\partial \bm q}{\partial t} = \boldsymbol{\mathcal N}(\bm q),
\label{eqn:NS}
\end{equation}
where the nonlinear Navier-Stokes operator $\boldsymbol{\mathcal N}$ acts on the state vector ${\bm q} \in \mathbb{C}^{N}$, which describes the flow discretized in all inhomogeneous directions. A standard Reynolds decomposition
\begin{equation}
\bm q(\bm x, t) = \bar{\bm q}(\bm x) + \bm q'(\bm x, t)
\label{eqn:RD}
\end{equation}
partitions the flow state into the time-averaged mean ${\bar{\bm q}}$ and the fluctuation ${\bm q'}$. Substituting \eqref{eqn:RD} into \eqref{eqn:NS} and applying a Taylor expansion of $\boldsymbol{\mathcal N}$ around ${\bar{\bm q}}$,
\begin{equation}
\bm A(\bar{\bm q}) = \left. \frac{\partial \boldsymbol{\mathcal N}}{\partial \bm q} \right|_{\bm q = \bar{\bm q}} \in \mathbb{C}^{N \times N},
\end{equation}
leads to
\begin{equation}
\begin{gathered}
\frac{\partial \bm q'}{\partial t} = \bm A(\bar{\bm q})\bm q' + \bm B\bm f'(\bar{\bm q}, \bm q'),
\\ 
\centering
\bm y' = \bm C \bm q',
\end{gathered}
\label{eqn:linsys_fil}
\end{equation}
where $\bm A$ is the linearized Navier-Stokes (LNS) operator, $\bm B \in \mathbb{C}^{N \times N_f}$ is an input matrix that can be used to restrict the forcing $\bm f' \in \mathbb{C}^{N_f}$, and $\bm C \in \mathbb{C}^{N_y \times N}$ is an output matrix that extracts the output of interest $\bm y' \in \mathbb{C}^{N_y}$ from the state. Here, the forcing term $\bm f'(\bar{\bm q}, \bm q')$ encapsulates all nonlinear terms from the Taylor expansion of $\boldsymbol{\mathcal{N}}$ around $\bar{\bm q}$, excluding the linear term represented by $\bm A(\bar{\bm q}) \bm q'$. It can also represent exogenous forcing. The matrices $\bm B$ and $\bm C$ serve as \emph{masks} in equation \eqref{eqn:linsys_fil}, providing generality and flexibility.

\par Resolvent analysis is most natural when $\bm A$ is stable, \ie all of its eigenvalues lie in the left-half plane. If $\bm A$ is unstable, discounting can be used to obtain a stable system \citep{Jovanovic04, YehTaira19}. For more details on this procedure, see \citet{Rolandietal24}. We assume that, if necessary, discounting has already been performed so that $\bm A$ is strictly stable.}

\par Resolvent analysis seeks the forcing that produces the largest steady-state response. Since the steady state is of interest, the solution can be obtained in the frequency domain. Taking the Fourier transform 
\begin{equation}
\mathcal F(\cdot) = \hat{(\cdot)}(\omega) = \int_{-\infty}^{+\infty} (\cdot)e^{-\text{i}\omega t} \,dt
\label{eqn:FT}
\end{equation}
of \eqref{eqn:linsys_fil} and solving for the output yields 
\begin{equation}
\hat{\bm y}(\omega) = \bm R(\omega)\hat{\bm f}(\omega),
\label{eqn:resolvent_eqn}
\end{equation}
where $\omega$ is the frequency and $\hat{(\cdot)}$ denotes the frequency counterpart of the time domain vector. The resolvent operator
\begin{equation}
\bm R = \bm C(\text{i}\omega \bm I - \bm A)^{-1}\bm B
\label{eqn:resolvent}
\end{equation}
maps the input forcing to the output response (here, $\text{i} = \sqrt{-1}$ and $\bm I$ is the identity matrix.)

\par The optimization problem for the most amplified forcing is formally defined as maximizing
\begin{equation}
\sigma = \frac{||\hat{\bm y}||_q}{||\hat{\bm f}||_f} = \frac{||\bm R\hat{\bm f}||_q}{||\hat{\bm f}||_f},
\label{eqn:optimization}
\end{equation}
where $||\boldsymbol{x}||^2_f = \; \langle \boldsymbol{x}, \boldsymbol{x} \rangle_f \;= \boldsymbol{x}^*\bm W_f\boldsymbol{x}$ computes the $f$-norm of any vector $\boldsymbol{x}$ and $(\cdot)^*$ denotes the conjugate transpose. $\bm W_f$ is a weight matrix that accounts for numerical quadrature and allows us to define arbitrary norms. Note that input and output norms can be different, \ie $||\cdot||_q = ||\cdot||_f$ is not required. For notational brevity, we assume identity matrices for the weight, input, and output matrices in what follows. The minor adjustments to our algorithm to accommodate non-identity weight, input, and output matrices are outlined in Appendix \ref{appA}.

\par Solving the Rayleigh quotient \eqref{eqn:optimization} is equivalent to computing the SVD of the resolvent operator \citep{Stewart93}
\begin{equation}
\bm R = \bm U \boldsymbol{\varSigma} \bm V^*,
\label{eqn:svd}
\end{equation}
where $\boldsymbol{\varSigma}$ contains the singular values (a.k.a. \emph{gains}), and $\bm V$ and $\bm U$ are right and left singular vectors corresponding to input and output vectors (a.k.a. forcing and response \emph{modes}), respectively.

\subsection{Computation} \label{sec:comp}

Computing resolvent modes by following the definitions from the previous $\S$ involves two computationally intensive steps: $(i)$ forming the resolvent operator by computing the inverse in \eqref{eqn:resolvent} and $(ii)$ computing the full singular value decomposition in \eqref{eqn:svd}. Both of these steps nominally require $O(N^3)$ operations. This is workable for one-dimensional problems, \eg a channel flow \citep{Moarrefetal13}, but quickly becomes intractable for two- and three-dimensional problems. 

\par Instead, most applications of resolvent analysis to two-dimensional problems have adopted an alternative approach that leverages LU decomposition and iterative eigenvalue solvers \citep{SippMarquet13, Jeunetal16, Schmidtetal18, ThomareisPapadakis18, Karbanetal20}. This approach utilizes a mathematical equivalence to compute the resolvent modes faster than the natural approach. The right singular vectors of the resolvent operator \edit{are defined as} the eigenfunctions of $\bm R^*\bm R$, \ie $\bm R^*\bm R = \bm V \boldsymbol{\varSigma}^2 \bm V^*$. By computing the leading eigenmodes of $\bm R^*\bm R$, both right singular vectors and square of singular values of the resolvent operator are obtained. Recovering the left singular vectors is done via $\bm U = \bm R \bm V\boldsymbol{\varSigma}^{-1}$. The leading eigenvalues and eigenvectors can be efficiently computed via Arnoldi iteration \citep{Arnoldi51}. The cost of the Arnoldi method relies on the desired number of modes and the convergence threshold. \edit{The Arnoldi algorithm requires the repeated computation of $\bm{R}^*\bm{R} \bm{v}$ for a given vector $\bm{v}$.} Computing the LU decomposition of $(\text{i}\omega \bm{I} - \bm{A})$ circumvents computing $\bm R$ directly. \edit{This is because solving $(\text{i}\omega \bm{I} - \bm{A}) \bm{v} = \bm{b}$ using LU decomposition is equivalent to finding $\bm{v} = (\text{i}\omega \bm{I} - \bm{A})^{-1} \bm{b} = \bm R \bm b$ without explicit inversion.} This is a common practice to speed up the process of constructing the orthonormal basis of the Krylov subspace \citep{Theofilis11}. However, the $O(N^2)$ scaling \edit{of the LU decomposition} remains burdensome for three-dimensional systems.

\par The main objective of this paper is to enable resolvent analysis for high-dimensional systems. Therefore, we discuss state-of-the-art approaches and introduce an improved algorithm specifically designed to tackle three-dimensional flows.

\section{Computing resolvent modes using RSVD} \label{sec:RSVD}

RSVD is a recent randomized linear algebra technique that provides a low-cost approximation of  the leading singular modes of a matrix \citep{Halkoetal11} by sampling its image and range. In the following two subsections, we introduce the RSVD algorithm and discuss its application to resolvent analysis.

\subsection{RSVD algorithm} \label{sec:RSVDLU}

\begin{algorithm} 
\caption{RSVD-LU}
\label{alg:alg_RSVD}
\begin{algorithmic}[1]

\State \textbf{Inputs:} $\bm R, k, q$

    \State $\boldsymbol{\varTheta} \leftarrow \;$randn$(N,k)$ 
    \label{alg:stp1}
    \Comment{Create random test matrices}
    
    \State $\bm Y  \leftarrow \bm R \boldsymbol{\varTheta}$ \label{alg:stp2}
    \Comment{Sample the range of $\bm R$}
    
    \If{$q > 0$}  \Comment{Optional power iteration} \label{alg:stp3_1}
    \State $\bm Y \leftarrow \;  $\texttt{PI}$\,(\bm R, \bm Y, q)$
    \Comment{Algorithm \ref{alg:PI}}
    \label{alg:stp3_2}
    
    \EndIf
    
    \State $\bm Q \leftarrow \;$qr$(\bm Y)$ 
    \label{alg:stp4}
    \Comment{Build the orthonormal subspace $\bm Q$}
    
    \State $\bm S  \leftarrow \bm Q^* \bm R$\;$ $  \label{alg:stp5}
    \Comment{Sample the image of $\bm R$}
    
    \State $(\tilde{\bm U} , \boldsymbol{\varSigma} , \bm V ) \leftarrow \;$svd$(\bm S)$  \label{alg:stp6}
    \Comment{Obtain $\boldsymbol{\varSigma} , \bm V$}
    
    \State $\bm U  \leftarrow \; \bm Q \tilde{\bm U} $
    \Comment{Recover $\bm U$} \label{alg:stp7}

\State \textbf{Outputs:} $\bm U , \boldsymbol{\varSigma} , \bm V $ 

\NoNumber{\footnotesize{\edit{Algorithm 1. Inputs: resolvent operator $\bm R$, number of modes $k$, and number of power iterations $q$. Outputs: $k$ response modes $\bm U$, $k$ forcing modes $\bm V$ and $k$ gains $\boldsymbol{\varSigma}$.}}}

\end{algorithmic}
\end{algorithm}


There exist several variations of the RSVD algorithm; here, we outline the algorithm 
from \citet{Halkoetal11}. The first step is to sample the range of $\bm R$ by forming its sketch (line \ref{alg:stp2})
\begin{equation}
\bm Y = \bm R \boldsymbol{\varTheta}, 
\label{eqn:range}
\end{equation}
where $\boldsymbol{\varTheta} \in \mathbb{C}^{N\times k}$ is a dense random test matrix (line \ref{alg:stp1}) with $k \ll N$ columns that determines the number of leading modes to be approximated. Increasing the number of test vectors slightly beyond the desired number of modes enhances the accuracy of the leading modes. A feature of high-dimensional random vectors is that they form an orthonormal set with high probability \citep{Vershynin18}, such that, on average, $\boldsymbol{\varTheta}$ projects uniformly onto all of the right singular vectors of $\bm R$. Therefore, the sketch preserves the leading left singular vectors of $\bm R$. An orthonormal basis $\bm Q \in \mathbb{C}^{N\times k}$ for the sketch is obtained via QR decomposition (line \ref{alg:stp4}), which is then used to sample the image of $\bm R$ (line \ref{alg:stp5}) as
\begin{equation}
\bm S = \bm Q^* \bm R.
\label{eqn:image}
\end{equation}
Computing the SVD of $\bm S \in \mathbb{C}^{k\times N}$ (line \ref{alg:stp6}), which is inexpensive due to its reduced dimension, provides an approximation of the $k$ leading right singular vectors $\bm V \in \mathbb{C}^{N\times k}$ and singular values $\boldsymbol{\varSigma} \in \mathbb{C}^{k\times k}$ of $\bm R$. Finally, the corresponding approximations of the left singular vectors of $\bm R$ can be recovered as $\bm U = \bm Q \tilde{\bm U} \in \mathbb{C}^{N\times k}$ (line \ref{alg:stp7}).

\par RSVD accurately estimates the leading modes for matrices with rapidly decaying singular values. For systems with slowly decaying singular values, performing $q$ optional power iterations (lines \ref{alg:stp3_1}-\ref{alg:stp3_2} and Algorithm \ref{alg:PI}) enhances the accuracy of the estimates. The rationale of power iteration is to increase the effective gap between singular values within the sketch by exponentiating them, since
\begin{equation}
(\bm R\bm R^*)^q\bm Y = (\bm U \boldsymbol{\varSigma} (\bm V^* \bm V)\boldsymbol{\varSigma} \bm U^*)^q \bm Y =  (\bm U \boldsymbol{\varSigma}^{2}\bm U^*)^q \bm Y = (\bm U \boldsymbol{\varSigma}^{2q}\bm U^*) \bm Y.
\label{eqn:powerit_rationale}
\end{equation}
Raising the singular values to a high power artificially accelerates the decay rate of the singular values of $\bm R$, improving the effectiveness of the RSVD algorithm. The QR factorizations improve numerical stability, as discussed by \citet{Halkoetal11}. 

\begin{algorithm} 
\caption{Power iteration}
\label{alg:PI}
\begin{algorithmic}[1]

\State \textbf{Inputs:} $\bm R, \bm Y, q$
    
 \For{$i = 1:q$} 
 
    \State $\bm Q \leftarrow \;$qr$(\bm Y)$ \label{alg:PI_stp1}
    \Comment{For stabilization purposes}
        
    \State $\bm Y  \leftarrow \; \bm R^* \bm Q$   \label{alg:PI_stp2}
    \Comment{Sample the image of $\bm R$}
        
    \State $\bm Q \leftarrow \;$qr$(\bm Y)$ \label{alg:PI_stp3}
    \Comment{For stabilization purposes}
        
    \State $\bm Y  \leftarrow \; \bm R \bm Q$  \label{alg:PI_stp4}
    \Comment{Sample the range of $\bm R$}
    
    \EndFor

\State \textbf{Output:} $\bm Y$

\NoNumber{\footnotesize{\edit{Algorithm 2. Inputs: resolvent operator $\bm R$, $k$ response modes from the first direct action $\bm Y$, and number of power iterations $q$. Outputs: $k$ response modes $\bm Y$.}}}

\end{algorithmic}
\end{algorithm}


\subsection{RSVD for resolvent analysis} \label{sec:RSVD4resolvent}

The algorithm outlined in the previous section assumes direct access to the matrix $\bm R$. In the context of resolvent analysis, $\bm R$ is defined in terms of an inverse, which should be avoided. \citet{Ribeiroetal20} addressed this challenge by adopting the approach developed by \citet{Jeunetal16} for computing resolvent modes using an Arnoldi algorithm.  

\par The idea is to replace multiplication of $\bm R$ or $\bm R^*$ by solving an equivalent linear system. For example, $\bm Y = \bm R \boldsymbol{\varTheta}$ (line \ref{alg:stp2} of Algorithm \ref{alg:alg_RSVD}) can be obtained by solving the linear system  
\begin{equation}
(\text{i}\omega \bm I - \bm A)\bm Y = \boldsymbol{\varTheta}
\label{eqn:sketch_mod2}
\end{equation}
since $\bm R^{-1} = (\text{i}\omega \bm I - \bm A)$. Similarly, $\bm S = \bm Q^* \bm R$ (line \ref{alg:stp5} of Algorithm \ref{alg:alg_RSVD}) can be replaced with solving 
\begin{equation}
(\text{i}\omega \bm I - \bm A)^*\bm S^* = \bm Q.
\label{eqn:S_SVD_mod}
\end{equation}
The same concept can be used to replace multiplication by $\bm R$ and $\bm R^*$ in Algorithm \ref{alg:PI}.

\par Typically, the linear systems are solved by computing an LU decomposition
\begin{equation}
(\text{i}\omega \bm I - \bm A) = \bm L\bm P, 
\label{eqn:LU_dir}
\end{equation}
where $\bm L$ and $\bm P$ are the lower and upper triangular matrices (we use $\bm P$ to denote the upper triangular matrix instead of $\bm U$ to avoid confusion with the left singular vectors). The same LU decomposition can be used also for $(\text{i}\omega \bm I - \bm A)^*$ since 
\begin{equation}
(\text{i}\omega \bm I - \bm A)^* = (\bm L\bm P)^* = \bm P^*\bm L^*.
\label{eqn:LU_adj}
\end{equation}
Solving these linear systems is indeed significantly less computationally demanding than computing the inverse of $(\text{i}\omega \bm I - \bm A)$ to form $\bm R$ and performing subsequent matrix-matrix multiplication in the RSVD algorithm. The remaining steps of the algorithm incur negligible computational costs and are not altered. In the remainder of our paper, we will use the term ``RSVD-LU'' to refer to the modified version of RSVD that is compatible with resolvent analysis \citep{Ribeiroetal20}.

\section{Computing resolvent modes using time stepping} \label{sec:TS}

\edit{An alternative class of methods for computing resolvent modes utilizes time stepping. This idea was first proposed by \citet{Monokrousosetal10} and recently was improved upon by \citet{Martinietal21}, who introduced two methods: the transient response method and the steady-state response method. The latter was found to be better suited for complex algorithms, and we will employ and extend this method in the present paper. A key difference between this paper and the work of \citet{Martinietal21} is that while their time-stepping approaches are integrated into Arnoldi's method to compute the leading resolvent modes, we will incorporate it into the RSVD algorithm. RSVD has demonstrated superior speed compared to previous methods, including Arnoldi and standard SVD-based techniques \citep{Ribeiroetal20}. Additionally, we introduce new approaches to minimize the CPU and memory costs for any time-stepping method for computing resolvent modes.}

\subsection{The action of the resolvent operator via time stepping}

The central idea of the time-stepping approach is to obtain the action of the resolvent operator on a vector (or matrix) by solving the linear system that underlies the resolvent operator in the time domain. In this context, the action of a matrix $\bm R$ on a vector (or matrix) $\bm b$ is defined as follows; Given $\bm b$, our objective is to compute $\bm x = \bm R \bm b$, which is equivalent to solving the linear system $\bm R^{-1}\bm x = \bm b$ for $\bm x$. 

\par Starting with a harmonically forced ordinary differential equation (ODE)
\begin{equation}
\frac{d\bm q}{dt} = \bm A\bm q + \bm f,
\label{eqn:direct}
\end{equation}
where
\begin{equation}
\bm f(t) = \hat{\bm f} e^{\text{i} \omega t}
\end{equation}
is the harmonic forcing with frequency $\omega \in \mathbb{R}$ and $\hat{\bm f} \in \mathbb{C}^{N}$ is an arbitrary vector. The steady-state response of \eqref{eqn:direct} is 
\begin{equation}
\bm q(t) = \hat{\bm q}_s e^{\text{i} \omega t},
\end{equation}
where
\begin{equation}
\hat{\bm q}_s = (\text{i}\omega \bm I - \bm A)^{-1}\hat{\bm f} = \bm R \hat{\bm f}
\label{eqn:direct_FS}
\end{equation}
is the Fourier-domain solution. Therefore, the action of $\bm R$ can be obtained by computing the steady-state solution $\bm q(t)$ of \eqref{eqn:direct} and subsequently taking a Fourier transform to obtain $\hat{\bm q}_s$. Similarly, the action of $\bm R^*$ can be obtained by computing the steady-state response $\bm z(t)$ of the adjoint equation 
\begin{equation}
\begin{gathered}
-\frac{d\bm z}{dt} = \bm A^*\bm z + \bm f,
\\
\bm f = \hat{\bm f}e^{\text{i}\omega t},
\label{eqn:adjoint}
\end{gathered}
\end{equation}
backward in time and taking a Fourier transform to obtain
\begin{equation}
\hat{\bm z}_s = (-\text{i}\omega \bm I - \bm A^*)^{-1}\hat{\bm f} = \bm R^* \hat{\bm f}.
\label{eqn:adjoint_FS}
\end{equation}
\edit{We use a discrete adjoint, \ie we take the complex conjugate of the LNS operator, $\bm A^*$, as the adjoint operator. This is consistent with the common approach of computing resolvent modes from the SVD of $\bm R$ without explicitly defining an adjoint. We note, however, that using a continuous formulation of the adjoint equations or a tailored set of differentiation operators and boundary conditions such as the summation-by-parts, simultaneous approximation term (SBP-SAT) method \citep{Carpenteretal99} may be necessary in certain cases to prevent numerical artifacts \citep{BergNordstrom12, Vishnampetetal15}.}

\par The arbitrary harmonic forcing term $\hat{\bm f}$ can be a matrix instead of a vector by defining $\hat{\bm F} \in \mathbb{C}^{N \times k}$. In that case, each column of the solutions $\hat{\bm Q}$ and $\hat{\bm Z}$ corresponds to one specific column of the forcing matrix. 

\subsection{Direct and adjoint actions for a range of frequencies} \label{sec:rangeFreq}

\begin{figure}
\centering
\includegraphics[scale=0.6]{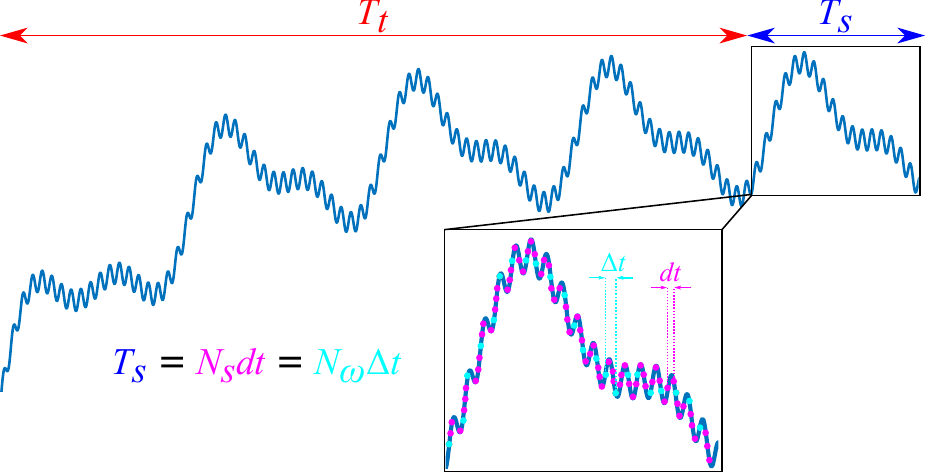}
\caption{Schematic of the response waveform. The solution contains a transient portion of length $T_t$ before the steady-state solution of period $T_s$ is achieved. The numerical solution contains $N_s$ time steps of size $dt$ within one period of the steady-state solution, but only $N_{\omega}$ points with $\Delta t$ spacing are required to decompose the $N_{\omega}$ frequencies of interest without aliasing.}
\label{fig:wave}
\end{figure}

This section describes an important contribution from \citet{Martinietal21} that allows us to compute the action of the resolvent operator for a set of desired frequencies while time-stepping the equations only once. Integrating \eqref{eqn:direct} typically generates a transient response $T_t$ before obtaining the desired steady-state solution, as shown in figure \ref{fig:wave}. The length of $T_t$ affects the length of time stepping and the accuracy of the output, as discussed in $\S$\ref{sec:trans_err}. The discrete nature of time stepping encourages the usage of discrete Fourier transform (DFT) where $\hat{\bm q}_s(\omega)$ can be obtained for a base frequency, $\omega_{min}$, and its harmonics, $n\omega_{min},$ where $n \in \mathbb{Z}$. The DFT necessitates a specific time length of $T_s = 2\pi/\omega_{min}$ in order to accurately resolve the longest \edit{period} of interest. The number of snapshots within the steady-state period $T_s$ determines the lowest frequency that can be resolved.

\par In order to compute resolvent modes for all frequencies of interest 
\begin{equation}
\varOmega = \{0, \pm\omega_{min}, \pm2\omega_{min}, \pm3\omega_{min}, ..., \pm\omega_{max}\},
\label{eqn:varOmega}
\end{equation}
where $\omega_{max}$ represents the highest frequency of interest, the forcing term 
\begin{equation}
\bm f = \sum_{\omega_j \in \varOmega} \hat{\bm f}_je^{\text{i}\omega_j t}
\label{eqn:forcing_harmonic}
\end{equation}
must include all frequencies in $\varOmega$. The minimum number of snapshots within the $T_s$-period is $N_{\omega} = 2 \lceil \frac{\omega_{max}}{\omega_{min}} \rceil$ according to Nyquist's theorem \citep{Nyquist28}. Performing time integration of \eqref{eqn:direct} results in computing $N_s$ steady-state snapshots within the $T_s$-period, where typically $N_s \ge N_{\omega}$, as the time step $(dt)$ is chosen to ensure sufficient integration accuracy. Ultimately, by choosing $N_{\omega}$ steady-state snapshots, we can determine the Fourier coefficients by taking a DFT.

\par To elaborate on the previous point, assume a set of snapshots $\bm Q_{N_s} = \{\bm q_1, \bm q_2, \bm q_3, ..., \bm q_{N_s}\}$ (analogous to the pink dots in figure \ref{fig:wave}), where $\bm q_j$ represents the $j^{th}$ steady-state snapshot in the time domain. The fast Fourier transform (FFT) can efficiently compute $\hat{\bm Q}_{N_s} = \{\hat{\bm q}_1, \hat{\bm q}_2, \hat{\bm q}_3, ..., \hat{\bm q}_{N_s}\}$. However, the maximum resolved frequency within $\hat{\bm Q}_{N_s}$ surpasses $\omega_{max}$ since typically $N_{\omega} \sim O(10^2)$, and $N_s \sim O(10^3-10^5)$. Therefore, an optimal size to resolve all $\omega \in \varOmega$ without aliasing is to consider $N_{\omega}$ equally spaced snapshots in $\bm Q_{N_{\omega}} = \{\bm q_1, \bm q_2, \bm q_3, ..., \bm q_{N_{\omega}}\}$ (analogous to the cyan dots in figure \ref{fig:wave}). Taking the FFT of $\bm Q_{N_{\omega}}$ yields $\hat{\bm Q}_{N_{\omega}} = \{\hat{\bm q}_1, \hat{\bm q}_2, \hat{\bm q}_3, ..., \hat{\bm q}_{N_{\omega}}\}$, where each member $\hat{\bm q}_j$ represents the solution to $(\text{i}\omega_j \bm I - \bm A)\hat{\bm q}_j = \hat{\bm f}_j$, with $\omega_j \in \varOmega$.

\begin{figure}
\centering
\includegraphics[width=\textwidth]{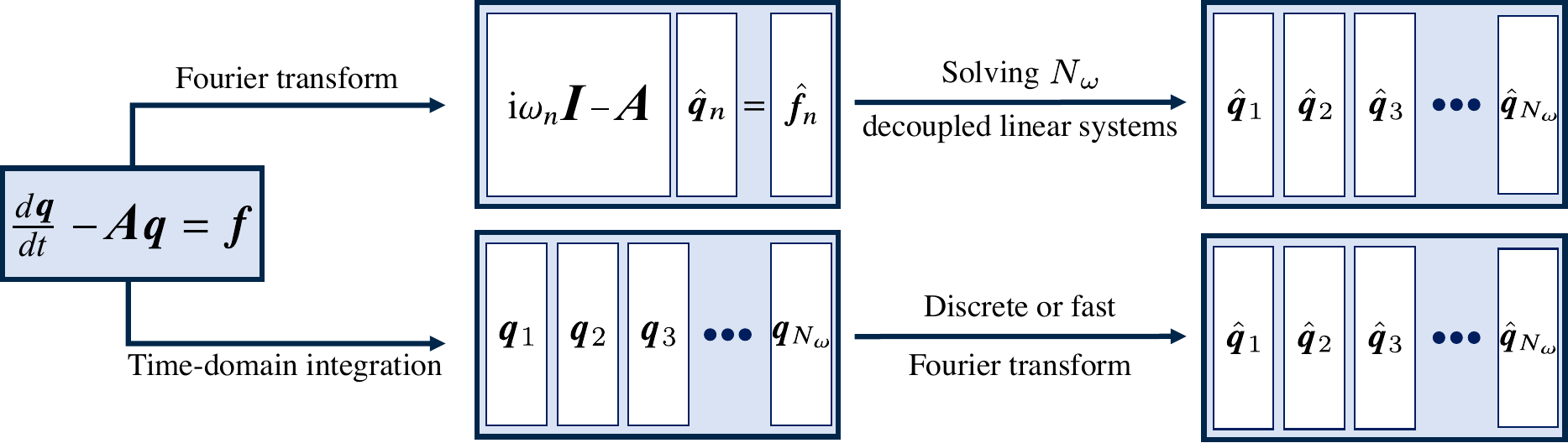}
\caption{Flowchart depicting the action of $\bm R$ on $N_{\omega}$ inputs for the RSVD-LU (upper route) and the RSVD-$\Delta t$ (bottom route) algorithms. Both routes produce the same result, but the bottom route is computationally advantageous for large systems.}
\label{fig:flowchart}
\end{figure}

To avoid leakage, the equidistant snapshots within $\bm Q_{N_{\omega}}$ need to span the entire $T_s$ period, \ie
\begin{equation}
T_s = dt \times N_s = \Delta t \times N_{\omega}.
\end{equation}
For a given pair ($\omega_{min}, \omega_{max}$), 
\begin{equation}
\Delta t = \frac{T_s}{N_{\omega}} = \frac{2\pi/\omega_{min}}{2 \lceil \frac{\omega_{max}}{\omega_{min}} \rceil}
\end{equation}
is predetermined, so $dt$ must be selected such that $\frac{N_s}{N_{\omega}} \in \mathbb{N}$.

Figure \ref{fig:flowchart} demonstrates the equivalence between computing the action of $\bm R$ for a range of frequencies in both the RSVD-LU and RSVD-$\Delta t$ algorithms. Starting from the LNS equations, the upper route involves applying a Fourier transform before solving $N_{\omega}$ decoupled linear systems to compute the action of the resolvent operator on $N_{\omega}$ forcing inputs. The bottom route involves integrating the LNS equations in the time domain, followed by a Fourier transform to generate the same output as the upper route. All frequencies of interest, $\omega \in \varOmega$, are included in the forcing so that the time stepping is performed only once, and the response at each frequency is obtained using a DFT or FFT.

\section{RSVD-\texorpdfstring{$\Delta t$}{deltat}: RSVD with time stepping} \label{sec:RSVDt}

Our algorithm, which we refer to as RSVD-$\Delta t$, uses time stepping to eliminate the computational bottleneck within the RSVD algorithm for large systems. Specifically, solving the direct and adjoint LNS equations to apply the action of $\bm R$ and $\bm R^*$ circumvents the need for LU decomposition, improving the scaling of the algorithm (see $\S$\ref{sec:comp_theory}), enabling resolvent analysis for the large systems typical of three-dimensional flows. RSVD-$\Delta t$ is outlined in Algorithm \ref{alg:alg_RSVDt} and described in what follows. 


\begin{algorithm} 
\caption{RSVD-$\Delta t$}
\label{alg:alg_RSVDt}
\begin{algorithmic}[1]

\State \textbf{Inputs:} $\bm A, k, q, \varOmega,$ TSS,\,$ dt, T_t$
    
    \State $\hat{\boldsymbol{\varTheta}} \leftarrow \;$randn$(N,k,N_{\omega})$ 
    \label{algt:stp1}
    \Comment{Create random test matrices}
    
    \State $\hat{\bm Y}  \leftarrow \;$\texttt{DirectAction}$(\bm A,\hat{\boldsymbol{\varTheta}}, $TSS,\,$dt, T_t)$ \label{algt:stp2}
    \Comment{Sample the range of $\bm R$}
    
    \If{$q > 0$}  \Comment{Optional power iteration} \label{algt:stp3_1}
    \State $\hat{\bm Y} \leftarrow \;  $\texttt{PI}$(\bm A, \hat{\bm Y}, q, $TSS,\,$dt, T_t)$
    \Comment{Algorithm \ref{alg:PI} with time stepping}
    \label{algt:stp3_2}
    
    \EndIf
    
    \State $\hat{\bm Q} \leftarrow \;$qr$_{\varOmega}$$(\hat{\bm Y})$ 
    \label{algt:stp4}
    \Comment{Build the orthonormal subspace $\hat{\bm Q}$}
    
    \State $\hat{\bm S}  \leftarrow \;$ \texttt{AdjointAction}$(\bm A^*,\hat{\bm Q}, $TSS,\,$dt, T_t)$  \label{algt:stp5}
    \Comment{Sample the image of $\bm R$}
    
    \State $(\tilde{\bm U} , \boldsymbol{\varSigma} , \bm V ) \leftarrow \;$svd$_{\varOmega}$$(\hat{\bm S})$  \label{algt:stp6}
    \Comment{Obtain $\boldsymbol{\varSigma} , \bm V$}
    
    \State $\bm U  \leftarrow \; (\hat{\bm Q} \tilde{\bm U})_{\varOmega} $
    \Comment{Recover $\bm U$} \label{algt:stp7}

\State \textbf{Outputs:} $\bm U , \boldsymbol{\varSigma} , \bm V $ for all $\omega \in \varOmega$

\NoNumber{\footnotesize{Algorithm $3$: Inputs include: linearized operator $\bm A$, number of modes $k$, number of power iterations $q$, frequency range $\varOmega$, TSS abbreviation for time-stepping scheme (\eg backward Euler), time step $dt$, and the transient length $T_t$. Outputs include: $k$ response modes $\bm U$, $k$ forcing modes $\bm V$ and the corresponding gains $\boldsymbol{\varSigma}$. Here, $k, q, \varOmega$ are common parameters with RSVD-LU. $(\cdot)_{\varOmega}$ means the function is separately applied to each $\omega \in \varOmega$. \texttt{DirectAction} and \texttt{AdjointAction} are functions that solve the direct and adjoint LNS equations, respectively, with a predefined forcing. \texttt{PI} is a function that performs the power iteration.}}

\end{algorithmic}
\end{algorithm}


\par As in the standard RSVD algorithm \edit{(Algorithm \ref{alg:alg_RSVD})}, the first step is to create random forcing matrices to sketch $\bm R$. Since our time-stepping approach computes all frequencies of interest at once, a separate test matrix $\hat{\varTheta} \in \mathbb{C}^{N \times k}$ is generated for each frequency $\omega \in \varOmega$ (line \ref{algt:stp1}). Next (line \ref{algt:stp2}), the $\mathtt{DirectAction}$ function solves the LNS equations forced by the set of test matrices in the time domain to obtain the sketch $\hat{\bm Y}$ of the resolvent operator $\bm R$ for all $\omega \in \varOmega$. 
Line \ref{algt:stp3_1} checks whether or not power iteration is desired, and if so (\ie $q > 0$), line \ref{algt:stp3_2} jumps to algorithm \ref{alg:PI} to increase the accuracy of resolvent modes. All instances of applying the action of the resolvent operator or its adjoint in Algorithm \ref{alg:PI} are performed via time stepping. In line \ref{algt:stp4}, an orthonormal subspace $\hat{\bm Q}$ is constructed for the sketch at each frequency via QR decomposition. Note that the $\varOmega$ subscript indicates that the operation is performed separately for each frequency $\omega \in \varOmega$. Next, in line \ref{algt:stp5}, the $\mathtt{AdjointAction}$ function solves the adjoint LNS equations forced by the set of $\hat{\bm Q}$ matrices in the time domain to sample the image of the resolvent operator $\bm R$ for all $\omega \in \varOmega$. Finally, the estimates of the k leading right singular vector $\bm V$ and gains $\boldsymbol{\varSigma}$ are obtained via an economy SVD of the $N \times k$ matrix $\hat{\bm S}$ (line \ref{algt:stp6}), and left singular vectors $\bm U$ are recovered in line \ref{algt:stp7}.

\edit{In the context of resolvent analysis, the RSVD algorithm can be slightly modified to enhance the accuracy of the response modes. One approach, as described by \citet{Ribeiroetal20}, involves performing an additional direct action after the final SVD in the standard RSVD algorithm to compute more accurate response modes. The cost of this additional step is equivalent to performing a direct action.

On the other hand, the costliest steps in executing the RSVD-$\Delta t$ algorithm are typically the direct and adjoint actions, while the costs of the QR and SVD steps are almost negligible (see table \ref{tab:tab2}). This suggests that we can perform an SVD instead of QR in line \ref{algt:stp4} and use the left singular vectors as the response modes instead of recovering them later in line \ref{algt:stp7}. The modes we obtain this way are more accurate than the recovered versions. Therefore, if one does not wish to perform an additional action, we recommend performing SVD in line \ref{algt:stp4} and ignoring line \ref{algt:stp7}.}

\section{Computational complexity} \label{sec:comp_theory}

The primary advantage of the RSVD-$\Delta t$ algorithm is its reduced computational cost. In this section, we discuss the CPU and memory cost scaling of applying the action of the resolvent operator via time stepping and compare it to LU-based approaches, as summarized in table \ref{tab:tab0}. We assume that the LNS equations are discretized using a sparse scheme such as finite differences, finite volume, or finite elements. Once the linearized operator $\bm A$ is constructed, the goal is to solve the linear system given by
\begin{equation}
(\text{i}\omega \bm I - \bm A)\bm x = \bm b
\label{eqn:solve}
\end{equation}
to compute the action of $\bm R$ on $\bm b$.

\subsection{CPU cost}

\par Direct solvers find the solution of \eqref{eqn:solve} to machine precision. A common approach is to find the LU decomposition of $(\text{i}\omega \bm I - \bm A)$ and solve the decomposed system via back substitution. The process of computing lower and upper triangular matrices with full or partial pivoting can be extremely expensive for large systems \citep{Duffetal17} and is often the dominant cost of solving a linear system \citep{MarquetLarsson15}. Once the LU decomposition is obtained, solving the LU-decomposed system is typically comparatively inexpensive. The theoretical cost scaling of LU decomposition of the sparse matrices that arise from collocation-based discretization methods (like finite differences) is $O(N^{1.5})$ and $O(N^2)$ for two-dimensional and three-dimensional systems, respectively \citep{Amestoyetal19}. The larger scaling exponent and number of grid points present in a three-dimensional problem make the LU decomposition of the corresponding linear operator costly. Optimized algorithms for computing LU decomposition are available in open-source software packages such as LAPACK \citep{Andersonetal99}, MUMPS \citep{Amestoyetal01}, PARDISO \citep{Schenketal01}, and Hypre \citep{FalgoutYang02}, which are designed to leverage massive parallelization. The LU decomposition becomes increasingly dominant (compared to solving the LU-decomposed system or other algorithmic steps) as the size of the system increases for both the standard Arnoldi-based method and the RSVD-LU algorithm, reducing the computational advantage of the latter.

\begin{table}
\begin{center}
\begin{tabular}{cccccc}
Problem size & \hspace{1mm} Action of $\bm R$ & \hspace{1mm} CPU time & \hspace{1mm} Memory \\ \hline \\
\multirow{2}{*}{Two-dimensional} & \hspace{1mm} {LU decomposition} & \hspace{1mm} $O(N^{1.5})$ & \hspace{1mm} $O(N^{1.2})$ \\
 & \hspace{1mm} {time stepping} & \hspace{1mm} $O(N)$ & \hspace{1mm} $O(N)$ \vspace{2mm} \\ 
\multirow{2}{*}{Three-dimensional} & \hspace{1mm} {LU decomposition} & \hspace{1mm} $O(N^2)$ & \hspace{1mm} $O(N^{1.6})$ \\
 & \hspace{1mm} {time stepping} & \hspace{1mm} $O(N)$ & \hspace{1mm} $O(N)$ \\
\end{tabular}
\end{center}
\caption{The scaling of CPU time and memory requirements with respect to $N$ for computing the action of $\bm R$ (or $\bm R^*$) using time stepping and LU decomposition.}
\label{tab:tab0}
\end{table}

\par Iterative solvers contain convergence criteria that can be adjusted to reduce computational cost at the expense of a less accurate solution. The performance of iterative solvers strongly depends on the condition number $\kappa$, the ratio between the largest and smallest eigenvalues of a matrix. Matrices with condition numbers of great than $\sim 10^4$ are considered to be ill-conditioned \citep{Saad03}, which can cause slow convergence and numerical stability issues for iterative solvers \citep{Skeel79}. The LNS operator $\bm A$ is typically a sparse but ill-conditioned matrix. When $\omega$ is small, $(\text{i}\omega \bm I - \bm A)$ inherits the ill-conditioning of $\bm A$, making the use of an iterative solver challenging. The conditioning improves as $\omega$ increases, so the lowest frequencies control the overall cost of using an iterative method to compute resolvent modes. In addition to the condition number, other properties such as the size, sparsity pattern, and density (or sparsity ratio) of a matrix can also ease or aggravate the situation \citep{TrefethenBau97}. 

In principle, iterative solvers are attractive when solving \eqref{eqn:solve} up to machine precision is unnecessary, as is the case when using the RSVD algorithm, which is already an approximation. The main challenge remains the typically high condition number of $(\text{i}\omega \bm I - \bm A)$, as explained above. One potential solution is the common practice of using a preconditioner \citep{Saad03_2}. Preconditioners are matrices that are multiplied on the left, right, or both sides of the target matrix to decrease its condition number and thus increase the convergence of iterative solvers. The methods of computing preconditioners and numerous related theories and practices are neatly summarized in a few surveys \citep{Axelsson85, Benzi02, PearsonPestana20}. Despite numerous advancements in this field, not all matrices have effective preconditioners. Some recent studies \citep{Houtmanetal23} are exploring the use of iterative methods to solve \eqref{eqn:solve} within the context of resolvent analysis, but direct methods, especially LU decomposition, have long been the dominant choice \citep{Moarrefetal13, Jeunetal16, Schmidtetal18, Ribeiroetal20}.

\par The cost of time-stepping methods rely on integrating the LNS equations in the time domain. Time-stepping of ODEs (such as the one in \eqref{eqn:direct}) has a long history and is a mature field \citep{Haireretal93, WannerHairer96, TrefethenBau97}. Herein, two classes -- implicit and explicit integration schemes -- are available and widely used in the scientific computing community.

\par Implicit integrators possess better stability properties but require a system of the form \begin{equation}
\boldsymbol{\mathcal A}\bm x = \bm b
\label{eqn:ODE}
\end{equation}
to be solved at every iteration. Here, $\bm b \in \mathbb{C}^{N \times k}$ is a function of the solution at previous time and the exogenous forcing (if present), and $\boldsymbol{\mathcal A} \in \mathbb{C}^{N \times N}$ is the temporal discretized operator, which is a function of the linear operator $\bm A$. For example, $\boldsymbol{\mathcal A}$ can be written as a first-order polynomial of the form $\boldsymbol{\mathcal A} = c_1 \bm I + c_2\bm A$ for multi-step methods, where constants are determined based on integration scheme and time step, \eg $\boldsymbol{\mathcal A} = \bm I - dt\bm A$ for backward Euler. A superficial comparison between \eqref{eqn:ODE} and \eqref{eqn:solve} indicates that implicit time steppers suffer from the same issues elaborated above. However, the key difference is that $\bm A$ is multiplied by the (small) time step $dt$, so the ill-conditioning of $\bm A$ is largely overwhelmed by the ideal conditioning of the identity matrix $\bm I$. This improved conditioning makes possible the application of iterative solvers. Implicit integrators require at least one LU decomposition of $\boldsymbol{\mathcal A}$ for direct solvers or a preconditioner for indirect solvers, which are not $O(N)$ operations. However, this one-time cost is often small enough that it is overwhelmed by other operations such that the observed computational complexity remains $O(N)$.

\par For explicit integrators, the solution at each time step is an explicit function of the solution (and exogenous forcing) at previous time steps. Accordingly, a solution of a linear system is not required, and each step contains only inexpensive sparse matrix-vector products for a linear ODE such as \eqref{eqn:solve}, making each step rapidly computable. The downside of explicit methods is that they are less numerically stable and often require many small steps to ensure stability for stiff systems \citep{SuliMayers03}. Nevertheless, the drastically smaller cost of each step for explicit integrators often outweighs the disadvantage of requiring many small steps, and many computational fluid dynamics codes are equipped with explicit integrators such as Runge--Kutta schemes. 

\par Explicit integrators involve repeatedly multiplying the sparse matrix ${\bm A}$ with vectors during the time-stepping process, which scales like $O(N)$. The number of times these operations must be performed over a fixed time interval is determined by the time step $dt$. If the time step is set by a CFL-like stability condition, then it scales like $O(N^{-1/2})$ and $O(N^{-1/3})$ for two- and three-dimensional problems, respectively, and the total number of time steps $N_t$ in a fixed time interval scales as the inverse of these values. The overall CPU cost scales like $O(N N_t)$, leading to $O(N^{3/2})$ and $O(N^{4/3})$ scaling for two- and three-dimensional problems, respectively. In practice, however, the time step is chosen to control the error associated with the highest frequency of interest rather than being determined by a CFL condition, as discussed in $\S$\ref{sec:truncation_err}, and is thus independent of $N$. Accordingly, we observe linear scaling when using explicit integrators in practice.

\subsection{Memory requirements} 

\par Supercomputers and parallel solvers can keep the hope of computing the LU decomposition of massive and poorly conditioned systems alive; however, massive calculations require massive storage, and memory becomes the top issue \citep{Davisetal16}. Generally, direct solvers are more robust than iterative solvers but can consume significant memory due to the fill-in process of factorization \citep{MarquetLarsson15}. The memory requirement associated with LU decomposition for resolvent analysis has been empirically observed to scale like $O(N^{1.2})$ and $O(N^{1.6})$ for two-dimensional and three-dimensional systems, respectively \citep{Towneetal22}. The exponents are not guaranteed and can become better or worse depending on the system of interest.

\par Explicit integration schemes have certain advantages over implicit integration schemes. Explicit schemes typically do not require much space for sparse matrix-vector products. The required memory is mainly used to store the forcing and response modes in Fourier space which scales like $O(N)$, as will be discussed in $\S$\ref{sec:FourierSum}. On the other hand, implicit integration schemes, in addition to the Fourier space matrices, require memory for solving \eqref{eqn:ODE}, which depends heavily on the sparsity of the LU-decomposed matrices or the iterative methods employed. For some systems, these methods may scale worse than $O(N)$, resulting in increased memory requirements.

\subsection{Matrix-free implementation} 

\par So far, we have assumed that the LNS matrix $\bm A$ is explicitly formed. In contrast to the standard frequency-domain approaches including the RSVD-LU algorithm, our time-stepping approach can be applied in a matrix-free manner using any code with linear direct and adjoint capabilities without explicitly forming $\bm A$ \citep{dePandoetal12, Martinietal21}. In this case, the cost scaling of our algorithm will follow that of the underlying Navier-Stokes code, which is again typically linear with the problem dimension. 

\section{Sources of error in the RSVD-\texorpdfstring{$\Delta t$}{deltat} algorithm} \label{sec:ErrSource}

Next, we identify sources of error within the RSVD-$\Delta t$ algorithm, which stem from the RSVD approximation and the time-stepping approach used to compute the action of $\bm R$. By effectively addressing these sources of error, the RSVD-$\Delta t$ method can be optimized for improved efficiency.

\subsection{RSVD approximation}

RSVD offers estimates of the resolvent modes rather than exact ground truth. The accuracy of these estimates is extensively discussed in \citet{Halkoetal11}, and it naturally depends on the gain separation, \edit{defined as the ratio $\sigma_{i}/\sigma_{i+1}$, where $\sigma_i$ is the $i^{th}$ singular value.} As mentioned earlier, incorporating power iteration and employing a few extra test vectors beyond the desired number of modes can improve the accuracy of the resolvent modes. In many cases, the approximation error of RSVD is the primary source of error in RSVD-$\Delta t$, such that it accurately reproduces the results of the RSVD-LU algorithm.

\subsection{Time stepping sources of error} \label{sec:TS_Err}

When computing the action of $\bm R$ and $\bm R^*$ using time stepping, two types of errors are introduced in addition to the RSVD approximation\edit{: truncation and transient errors.}

\subsubsection{Truncation error} \label{sec:truncation_err}

The first source of time-stepping error is the truncation error of the numerical integration schemes used to solve the time-domain equations. Common approaches include classical numerical integration schemes such as Runge--Kutta, implicit/explicit Euler, Adams-Moulton family, and others \citep{Haireretal93, WannerHairer96}. These methods introduce truncation errors resulting from the approximation of Taylor series expansions. Hence, a chosen time step introduces an expected truncation error, with higher-order schemes providing greater precision.

\par Local truncation error (LTE) is derived for ODEs as
\begin{equation}
LTE = C\frac{d^pf(t)}{dt^p}O(dt^p),
\label{eqn:truncation}
\end{equation}
where $C$ is a constant, and $p$ is the order of the time-stepping scheme. In this study, our focus is on ODEs with harmonic forcing $f(t) = \hat{f}e^{\text{i}\omega t}$. Substituting the forcing term into \eqref{eqn:truncation}, we observe that 
\begin{equation}
LTE \propto O((\omega dt)^p).
\label{eqn:truncation2}
\end{equation}
This equation indicates that for a fixed time step $dt$, the error in the computed resolvent modes will be frequency dependent and vary as $\omega^p$. Therefore, in addition to satisfying any stability constraints, the time step $dt$ must be selected such that $\omega_{max} dt$ is sufficiently small to obtain accurate resolvent modes up to the maximum desired frequency $\omega_{max}$. 

\subsubsection{Transient error} \label{sec:trans_err}

The second source of time-stepping error arises from the unwanted transient response. The solution of \eqref{eqn:direct} can be written as a sum of its transient and steady-state components, 
\begin{equation}
\bm q(t) = \bm q_t(t) + \bm q_s(t),
\label{eqn:gen_sol}
\end{equation}
where the transient part $\bm q_{t}$ decays to zero as $t \to \infty$ and the steady-state part $\bm q_{s}$ is $T$-periodic, \ie $\bm q_{s}(t+T) = \bm q_{s}(t)$. Taking the Fourier transform of each part leads to 
\begin{equation}
\hat{\bm q}(\omega) = \hat{\bm q}_t(\omega) + \hat{\bm q}_s(\omega).
\label{eqn:gen_sol_Fourier}
\end{equation}
Only the steady-state solution is desired, so any non-zero transient part constitutes an error in our representation of the action of the resolvent operator (or its adjoint) on the prescribed forcing. The transient response can be understood as the response of the system to an initial condition that is not synced with the forcing applied to the system. It may initially grow for non-normal systems like the LNS equations \citep{Schmid07} but eventually decays at the rate of the least-damped eigenvalue of $\bm A$.  

\par We define the transient error as the ratio between the norms of the transient and steady-state responses,
\begin{equation} 
\epsilon = \frac{||\bm q_t||}{||\bm q_s||},
\label{eqn:et}
\end{equation} 
where the $l^2$-norms can be replaced with $||\cdot||_q$ for non-identity weight matrices. In cases where we solve \eqref{eqn:direct} with a zero initial condition (which is often the case), \ie $\bm q(0) = \bm q_t(0) + \bm q_s(0) = 0$, the transient error is initially one,
\begin{equation}
\epsilon(0) = \frac{||\bm q_t(0)||}{||\bm q_s(0)||} = 1.
\end{equation}
In the long term, the transient error approaches zero,
\begin{equation}
\lim_{t\to\infty} \epsilon(t) = \lim_{t\to\infty} \frac{||\bm q_t(t)||}{||\bm q_s(t)||} = 0,
\end{equation}
since $||\bm q_s||$ remains bounded.

\par The eigenspectrum of the linearized system $\bm A$ provides insights into the long-term response of the homogeneous system. Any initial perturbation will eventually follow the least-damped mode. However, in practice, computing the eigenspectrum of $\bm A$ is challenging, especially for large systems. Even obtaining a small number of eigenvalues using the Krylov-Schur method can be cumbersome. Therefore, a practical approach to understanding the long-term behavior of a system is to simulate the homogeneous ODE
\begin{equation}
\frac{d\bm q_h}{dt} - \bm A\bm q_h = 0,
\label{eqn:homogeneous}
\end{equation}
initialized with a random state \citep{ErikssonRizzi85, Edwardsetal94}. A random perturbation represents a worst-case scenario, as it excites all the slow modes of $\bm A$. By monitoring the norm of $\bm q_h$ over time, we can estimate the slowest decay rate, which corresponds to the real part of the least-damped eigenvalue of $\bm A$. This also gives us an indication of the expected magnitude of the transient error. Performing a DFT on one cycle of the transient response allows us to determine the anticipated level of transient error within the desired frequency range. 

\par While it is possible to simply wait for the transient error to naturally decay over time, this approach comes with increased CPU cost, as it requires longer simulation durations. In $\S$\ref{sec:EffTrans}, we will present an efficient method to achieve a smaller transient error within a shorter time frame.

\section{Optimizing the RSVD-\texorpdfstring{$\Delta t$}{deltat} algorithm} \label{sec:Optimizing}

In this section, we present several approaches aimed at reducing the CPU cost and memory requirements of the RSVD-$\Delta t$ algorithm. These approaches, combined with the improved cost scaling of RSVD-$\Delta t$ compared to the RSVD-LU algorithm as discussed in $\S$\ref{sec:comp_theory}, are crucial in facilitating affordable resolvent analysis of complex three-dimensional flows.

\subsection{Minimizing memory requirements} \label{sec:MinMem}

First, we describe several strategies to minimize the memory required to compute resolvent modes for a given problem\edit{: streaming Fourier sums and shortcuts for real-valued matrices.}

\subsubsection{Streaming Fourier sums} \label{sec:FourierSum}

\begin{figure}
\centering
\includegraphics[width=\textwidth]{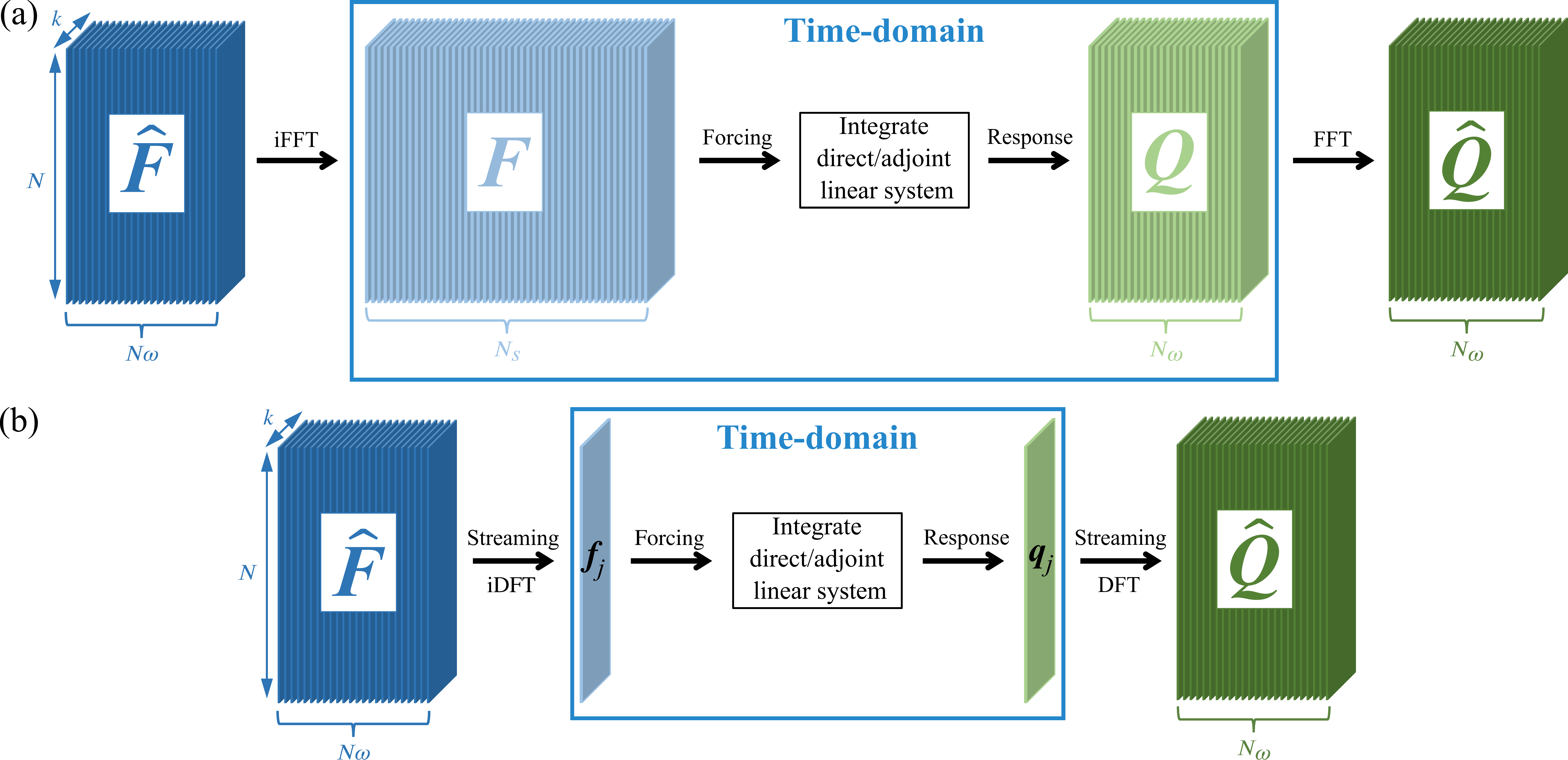}
\caption{Schematic of the action of $\bm R$ with (a) FFT/iFFT and (b) streaming DFT/iDFT methods to transform between the Fourier and time domains.}
\label{fig:FA_scheme}
\end{figure}

A straightforward implementation of computing the action of $\bm R$ (or $\bm R^*$) via time stepping entails $(i)$ transferring the forcing from Fourier space to the time domain, $\hat{\bm F} \xrightarrow{\text{iFFT}} \bm F$, $(ii)$ performing integration to obtain the steady-state solutions saved with a specific time interval, as explained in $\S$\ref{sec:rangeFreq}, and $(iii)$ transferring the response back to frequency space, $\bm Q \xrightarrow{\text{FFT}} \hat{\bm Q}$. A schematic of these steps is displayed in figure \ref{fig:FA_scheme}(a).

\par The first step requires zero-padding $\hat{\bm F} \in \mathbb{C}^{N \times k \times N_{\omega}}$ since $\bm F \in \mathbb{C}^{N \times k \times N_s}$ is required at all $N_s \gg N_{\omega}$ points in the period associated with the time step $dt \ll \Delta t$ required for accurate time stepping. The iFFT is computationally efficient but storing its output requires a minimum memory allocation of $O(NkN_s)$, excluding space for the iFFT calculations themselves. $\hat{\bm F}$ is automatically discarded before proceeding to the second step. In step $(ii)$, $\bm f_j \in \bm F$ is used to force the linear system at each time step until the transient ends, and the steady-state responses are stored in $\bm Q$. After integration, $\bm F$ is no longer needed
and is removed. Lastly, obtaining $\hat{\bm Q}$ from $\bm Q$ using an FFT requires an $O(NkN_{\omega})$ space to store the output. Overall, a minimum memory allocation of $O(NkN_s) + O(NkN_{\omega})$ is necessary to store both $\bm F$ and $\bm Q$ simultaneously.

\par The memory requirements of this process can be significantly reduced by leveraging streaming Fourier sums, as in the streaming SPOD algorithm proposed by \citet{SchmidtTowne19}.  This procedure is shown schematically in figure \ref{fig:FA_scheme}(b). In the streaming approach, a new forcing snapshot is created before each time step and promptly removed afterward. Also, the contribution to the Fourier modes of the response is computed only at specific time steps, after which the snapshot of the solution can be discarded. This eliminates the need to permanently store any data in the time domain, reducing the memory requirement to $2\times O(NkN_{\omega})$ for storing $\hat{\bm F}$ and $\hat{\bm Q}$. The streaming implementation utilizes the DFT formulation to create forcing inputs and compute the effect of steady-state response data on the ensemble of Fourier coefficients, as demonstrated in the following.

\begin{table}
\begin{center}
\begin{tabular}{ccc}
\vspace{2mm}
$\mathcal F/\mathcal F^{-1}$ & \hspace{1mm} CPU time & \hspace{1mm} Memory  \\ 
\hline
iFFT & \hspace{1mm} $Nk \times O(N_slog(N_s))$  & \hspace{1mm} $O(NkN_s)$ \\             
FFT  & \hspace{1mm} $Nk \times O(N_{\omega}log(N_{\omega}))$ & \hspace{1mm} $O(NkN_{\omega})$ \vspace{2mm} \\ 
Streaming iDFT & \hspace{1mm} $Nk\times O(N_{\text{total}}N_{\omega})$  & \hspace{1mm} $O(NkN_{\omega})$ \\             
Streaming DFT  & \hspace{1mm} $Nk\times O(N_{\omega}^2)$ & \hspace{1mm} $O(NkN_{\omega})$ \\
\end{tabular}
\end{center}
\caption{Comparison of CPU time and memory requirements using FFT/iFFT and streaming DFT/iDFT methods transfer back and forth between Fourier space and time domain. $N_{\text{total}} = N_t + N_s$ is the total number of time steps including transient and steady-state parts.}
\label{tab:tab1}
\end{table}

\par At each time step, the instantaneous forcing is created from its Fourier mode using the definition of the inverse Fourier transform, 
\begin{equation}
\bm f_p = \sum_{s = 1}^{N_{\omega}} \bm Z'_{ps}\hat{\bm f}_s,
\label{eqn:strm_IDFT}
\end{equation}
where $\bm Z'_{ps} = exp(-2\pi $i$/N_s)^{(p-1)(s-1)}$. The integer $p$ $(1 \leq p \leq N_s)$ specifies the phase of the periodic forcing at the current time step. Here, $\hat{\bm f}_s \in \mathbb{C}^{N\times k \times N_{\omega}}$ denotes Fourier modes that are accessible from memory. The sum is taken over every $\omega \in \varOmega$, and it outputs the $p^{th}$ time domain snapshot $\bm f_p \in \mathbb{C}^{N\times k}$. This process continues in a loop of size $N_s$ until the transient is passed and steady-state data is computed.

\par The response Fourier modes can be computed from the time-domain steady-state solutions in a similar streaming fashion. Following the definition of the DFT, each temporal snapshot $\bm q_l$ within the steady-state response contributes to each each Fourier mode according to the partial sum 
\begin{equation}
\left[\hat{\bm q}_s\right]_{r} = \left[\hat{\bm q}_s\right]_{r-1} + \bm Z_{ls}\bm q_r = \sum_{l = 1}^r \bm Z_{ls}\bm q_l,
\label{eqn:strm_DFT}
\end{equation}
where $\bm Z_{ls} = exp(-2\pi \text{i}/N_{\omega})^{(l-1)(s-1)}, 1 \le (l,s) \le N_{\omega}$. Here, $\left[\hat{\bm q}_s\right]_{r}$ represents the sum of contributions up to $\bm q_r$, which is the $r^{th}$ steady-state response and should be removed after adding its contribution to the sum. The partial sum is complete once $r = N_{\omega}$, \ie the effect of all $N_{\omega}$ steady-state data is included. 

\par A subtle but important difference between the iDFT matrix $\bm Z' \in \mathbb{C}^{N_{\omega} \times N_s}$ and the DFT matrix $\bm Z \in \mathbb{C}^{N_{\omega} \times N_{\omega}}$ is their sizes: $\bm Z$ is used to generate $N_s$ temporal snapshots of the forcing from $N_{\omega}$ Fourier modes, while $\bm Z'$ is used to convert $N_{\omega}$ temporal snapshots of the steady-state solution into $N_{\omega}$ Fourier modes. The steaming process of the adjoint equations is identical, except the equations are integrated backward in time and indices within the Fourier sums are adjusted accordingly. 

\par The CPU time and memory requirement of the FFT/iFFT and streaming DFT/iDFT approaches are summarized in table \ref{tab:tab1}. Although the streaming method incurs slightly higher CPU cost due to the efficiency of the FFT algorithm, this CPU overhead is negligible compared to the cost of taking a time step. Moreover, the memory savings of the streaming method can be substantial; the ratio of the memory required by the iFFT and streaming iDFT methods used to create the forcing snapshots scales like $O(N_s/N_{\omega})$, where $N_{\omega} \sim O(10^2)$, and $N_s \sim O(10^3-10^5)$ are typical values. Overall, the substantial memory benefit of the streaming method outweighs the small CPU penalty, especially for large systems.

\subsubsection{Optimal cost for real-valued matrices}

\edit{The linear operator $\bm A$ is often real-valued. Indeed, this is usually case for LNS operator, except when considering a non-zero wavenumber in a Fourier-transformed homogeneous direction or when using complex-valued non-reflecting boundary conditions \citep{Colonius04, Hu08}. When $\bm A$ is real-valued, memory requirements can be significantly reduced. Having} $\bm R_{\edit{\omega}} = (\text{i}\omega \bm I - \bm A)^{-1} = \bm U \boldsymbol{\varSigma} \bm V^*$, the resolvent operator corresponding to $-\omega$ can be written as
\begin{equation}
\bm R_{-\omega} =  (-\text{i}\omega \bm I - \bm A)^{-1} = (\overline{\text{i}\omega \bm I} - \overline{\bm A})^{-1} = \overline{(\text{i}\omega \bm I - \bm A)^{-1}}  = \overline{\bm R}_{\omega} = \overline{\bm U} \boldsymbol{\varSigma} \overline{\bm V}^*,
\label{eqn:realA}
\end{equation}
where $\bar{(\cdot)}$ denotes the complex conjugate and $\bm A = \overline{\bm A}$ when $\bm A$ is real-valued. Equation \eqref{eqn:realA} proves that the gains of positive and negative frequencies are symmetric and the resolvent modes are complex conjugates of one another. Therefore, computing the resolvent modes for positive $\omega \in \varOmega$ naturally provides results for negative frequencies. This symmetry halves the CPU cost for the RSVD-LU algorithm but does not reduce the memory requirement. On the other hand, in the case of RSVD-$\Delta t$, the memory requirements are halved, but there is no significant reduction in the CPU, as further elaborated.

\par Since the frequencies of interest become $\varOmega_+ = \{0, +\omega_{min}, +2\omega_{min}, ..., +\omega_{max}\}$, the total number of frequencies becomes $\lfloor \frac{N_{\omega}}{2} \rfloor + 1$. In this scenario, only Fourier coefficients corresponding to $\omega \in \varOmega_+$ are saved and the memory storage required for both input and output matrices ($\hat{\bm F}$ and $\hat{\bm Q}$ discussed in $\S$\ref{sec:FourierSum}) is halved. In terms of CPU, generating the forcing and computing the response is twice as fast but the speed-up is not significant as the time stepping remains identical to the complex-valued case.

\subsubsection{An additional option for reducing memory} \label{sec:group}

If additional memory savings are required, the memory requirements of RSVD-$\Delta t$ can be sharply reduced by dividing the frequencies of interest into multiple sets at the expense of additional CPU cost. For instance, when the frequencies are divided into $d$ equal groups, the memory requirement is reduced by a factor of $d$. The penalty of doing so is that the CPU time scales proportionally with $d$, since the entire algorithm needs to be repeated for each group of frequencies. The RSVD-LU algorithm offers no such opportunity to reduce memory requirements, \eg to make a particular calculation possible on a given computer, at the expense of higher CPU cost.

\subsection{Minimizing the CPU cost: efficient transient removal} \label{sec:EffTrans}

Within the time-stepping process, the removal of the transient responses is crucial and is naturally accomplished through the long-time integration of \eqref{eqn:direct}, as discussed in $\S$\ref{sec:trans_err}. Nonetheless, certain LNS operators exhibit a painfully slow decay rate, resulting in lengthy transient durations and costly time stepping. Therefore, we present an efficient transient removal strategy to minimize the CPU cost.

\par Our strategy uses the differing evolution of the steady state and transient parts of the solution to directly compute and remove the transient from the solution. Considering two solutions of \eqref{eqn:direct}, $\bm q_1 = \bm q(t_1)$ and $\bm q_2 = \bm q(t_1 + \Delta t)$, we can express them in terms of their steady-state and transient parts, as in \eqref{eqn:gen_sol}, as 
\begin{equation}
\begin{aligned}
\bm q_1 &= \bm q_{s, 1} + \bm q_{t, 1},
\\
\bm q_2 &= \bm q_{s, 2} + \bm q_{t, 2},
\label{eqn:1}
\end{aligned}
\end{equation}
where $\bm q_{s, 1}, \bm q_{s, 2}, \bm q_{t, 1},$ and $\bm q_{t, 2}$ are four unknowns. Applying a prescribed forcing in \eqref{eqn:direct} at a single frequency $\omega$ yields 
\begin{equation}
\bm q_{s, 2} = \bm q_{s, 1}e^{\text{i}\omega \Delta t}.
\label{eqn:2}
\end{equation}
Also, the transient response follows the form of a homogenous response, resulting in
\begin{equation}
\bm q_{t, 2} = e^{\bm A \Delta t}\bm q_{t, 1}.
\label{eqn:3}
\end{equation}
Simplifying \eqref{eqn:1}, \eqref{eqn:2}, and \eqref{eqn:3} for $\bm q_{t, 1}$, we obtain
\begin{equation}
(\bm I - e^{-\text{i}\omega \Delta t}e^{\bm A \Delta t}) \bm q_{t, 1} = \bm b,
\label{eqn:4}
\end{equation}
where $\bm b = \bm q_1 - \bm q_2e^{-\text{i}\omega \Delta t}$ is known from the time-stepping solution. Equation \eqref{eqn:4} holds for any two points in time with arbitrary separation $\Delta t$. The exact steady-state solution with no transient error is obtained by solving \eqref{eqn:4} for $\bm q_{t,1}$ and using \eqref{eqn:1} to obtain $\bm q_{s,1} = \bm q_1 - \bm q_{t,1}$. 

\par The prescribed forcing in RSVD-$\Delta t$ consists of a range of frequencies, hence, it requires a pre-processing step to enable the transient removal strategy. We utilize $\bm Q = \{\bm q_1, \bm q_2, \bm q_3, ..., \bm q_{N_{\omega}}\}$ to construct $\hat{\bm Q} \in \mathbb{C}^{N\times N_{\omega}}$, where the snapshots are equidistant with a time interval of $\Delta t$. Additionally, we define $\bm Q^{\Delta t} = \{\bm q_2, \bm q_3, \bm q_4, ..., \bm q_{N_{\omega}+1}\}$ as a shifted matrix, resulting in $\hat{\bm Q}^{\Delta t} \in \mathbb{C}^{N\times N_{\omega}}$. Here, $\hat{\bm q}_j \in \hat{\bm Q}$ represents $\bm q_1$ in the above equations, while $\hat{\bm q}_j^{\Delta t} \in \hat{\bm Q}^{\Delta t}$ represents $\bm q_2$, both oscillating at the same frequency. Therefore, a single time stepping is sufficient to obtain \eqref{eqn:4} for all $\omega \in \varOmega$.

\par We emphasize two crucial aspects of our strategy. Firstly, it functions as a post-processing step that comes into play after acquiring simulation snapshots, which encompass both transient and steady-state components. Its primary aim is to selectively remove the transient portion. Secondly, our strategy does not introduce any modifications to the LNS operator. Instead, it is tailored to solving equations that leave the steady-state response unaffected. 

\par Solving \eqref{eqn:4} can be computationally expensive, particularly for large systems, even if we assume that computing $e^{\bm A \Delta t}$ is feasible. To address this issue, one possible approach is to choose a small $\Delta t$ and expand the exponential term as $e^{\bm A \Delta t} = \sum_j{\frac{(\bm A \Delta t)^j}{j!}}$. However, this leads to solving a similar linear system to \eqref{eqn:solve}, which we wish to avoid. Another approach is to leverage iterative methods (\eg GMRES) when $\Delta t$ is sufficiently large. Although the solution may converge within a reasonable time frame, solving similar systems needs to be repeated for all test vectors and frequencies. To overcome these challenges, we propose employing Petrov-Galerkin (or Galerkin) projection to obtain an affordable, approximate solution of \eqref{eqn:4}.

\par Consider a low-dimensional representation of the transient response as 
\begin{equation}
\bm q_{t, 1} = \boldsymbol{\phi} \boldsymbol{\beta}_1,
\label{eqn:5}
\end{equation}
where $\boldsymbol{\phi} \in \mathbb{C}^{N \times r}$, with $r \ll N$, is an orthonormal trial basis spanning the transient response and $\boldsymbol{\beta}_1 \in \mathbb{C}^{r}$ represents the coefficients describing the transient in this basis. By substituting \eqref{eqn:5} into \eqref{eqn:4}, the linear system 
\begin{equation}
(\bm I - e^{-\text{i}\omega \Delta t}e^{\bm A \Delta t}) \boldsymbol{\phi} \boldsymbol{\beta}_1 = \bm b
\label{eqn:6}
\end{equation}
is overdetermined. Petrov-Galerkin projection with test basis $\boldsymbol{\psi} \in \mathbb{C}^{N \times r}$ is employed to close \eqref{eqn:6}, giving
\begin{equation}
\boldsymbol{\psi}^*(\bm I - e^{-\text{i}\omega \Delta t}e^{\bm A \Delta t}) \boldsymbol{\phi} \boldsymbol{\beta}_1 = \boldsymbol{\psi}^* \bm b.
\label{eqn:7}
\end{equation}
Solving \eqref{eqn:7} for $\boldsymbol{\beta}_1$ and inserting the solution into \eqref{eqn:5} yields 
\begin{equation}
\bm q_{t, 1} = \boldsymbol{\phi} (\boldsymbol{\psi}^*\boldsymbol{\phi} - e^{-\text{i}\omega \Delta t} \tilde{\bm M})^{-1} \boldsymbol{\psi}^*\bm b,
\label{eqn:8}
\end{equation}
where 
\begin{equation}
\tilde{\bm M} = \boldsymbol{\psi}^* e^{\bm A \Delta t} \boldsymbol{\phi} \in \mathbb{C}^{r \times r}
\label{eqn:9}
\end{equation}
is a reduced matrix that maps the coefficients. The advantage of this strategy is that it allows for the computation of the inverse of $(\boldsymbol{\psi}^*\boldsymbol{\phi} - e^{-\text{i}\omega \Delta t}\tilde{\bm M})$ due to its reduced dimension. Obtaining $\tilde{\bm M}$ is also an efficient process, involving two steps: $(i)$ integrating the columns of $\boldsymbol{\phi}$ over $\Delta t$, and $(ii)$ projecting $e^{\bm A \Delta t} \boldsymbol{\phi}$ onto the columns of $\boldsymbol{\psi}$. The construction cost of $\tilde{\bm M}$ for each $\omega \in \varOmega$ is primarily determined by the first step. Specifically, when the number of columns in $\boldsymbol{\phi}$ is $r = N_{\omega}$ and $\Delta t = T_s/N_{\omega}$, the total cost of constructing $\tilde{\bm M}$ for all $\omega \in \varOmega$ is equivalent to integrating the LNS equations for an additional $T_s$ duration. \edit{While it is possible to use a Taylor expansion of $e^{\bm A \Delta t}$ to compute $e^{\bm A \Delta t} \boldsymbol{\phi}$, the number of terms required for the expansion can become excessive when $\Delta t > dt$. Therefore, we opt for time integration as a more practical alternative.}

\par Galerkin projection is a special case of the above procedure in which the test and trial functions are the same, \ie $\boldsymbol{\phi}$ is also the test function. Using this strategy with either Galerkin or Petrov-Galerkin projections, the accuracy of the solution relies on the ability of the column space of $\boldsymbol{\phi}$ to adequately span the transient response. Thus, the challenge lies in constructing an appropriate basis to accurately capture the transient behavior. Before the introduction of appropriate trial bases, we note that one can construct a new $\boldsymbol{\phi}$ for each $\omega \in \varOmega$, however, the bases that we define later are universal for all frequencies. Hence, the reduced matrix $\tilde{\bm M}$ is constructed once for all frequencies. Subsequently, \eqref{eqn:8} obtains transient responses at each frequency and updates the steady-state responses. 

\par Given the rapid decay of most terms in the transient response, it is advantageous to utilize the least-damped eigenvectors of $\bm A$ as the chosen trial basis. By excluding the least-damped eigenvectors, we effectively increase the decay rate of the transient response. Let $\lambda_1$ denote the least-damped eigenvalue of $\bm A$, with $\bm V_1$ representing the corresponding eigenvector. We define $\boldsymbol{\phi} = {\bm V_1}$, thereby removing the transient component projected onto $\bm V_1$. As a result, the norm of the updated transient, obtained by subtracting this projection, follows the decay rate associated with the second least-damped eigenvalue of $\bm A$. Similarly, the trial basis $\boldsymbol{\phi}$ can encompass the first $r-1$ least-damped eigenvectors, $\boldsymbol{\phi} = \mathrm{orth} \{\bm V_1, \bm V_2, ..., \bm V_{r-1} \}$, leading to a decay rate governed by the $r^{th}$ least-damped eigenvalue of $\bm A$. For this particular trial basis, Petrov-Galerkin projection can be utilized, where $\boldsymbol{\psi}$ incorporates the adjoint eigenvectors. This approach ensures the complete elimination of transient projection onto the least-damped modes. To be clear, this procedure does not eliminate the impact of these modes on the steady-state response, but only on the transient response.  

\par The main challenge associated with this trial basis is the computational cost of computing the least-damped eigenvectors (and adjoint eigenvectors in the case of Petrov-Galekin projection), especially for large systems, even when using algorithms designed for this purpose, \eg Krylov-based methods \citep{ErikssonRizzi85, Edwardsetal94}. Overall, the least-damped modes of $\bm A$ are most helpful for systems that suffer from only a few slowly decaying modes.

\par Another powerful trial basis is formed by stacking the snapshots into a matrix during the integration of the LNS equations, resulting in $\boldsymbol{\phi} = \mathrm{orth} \{\bm q_1, \bm q_2, \bm q_3, ..., \bm q_r\}$ (an orthogonalization of the matrix of snapshots). Specifically, $\boldsymbol{\phi}$ can be constructed as the union of $\hat{\bm Q}$ and $\hat{\bm Q}^{\Delta t}$ as a reliable trial basis. Performing QR decomposition on this matrix is essential to ensure orthogonality. As the LNS equations are allowed to run for a longer duration, $\boldsymbol{\phi}$ becomes an increasingly effective trial basis, providing improved estimates of the transient responses across all frequencies $\omega \in \varOmega$. We have observed that this basis is \edit{more} accurate for higher frequencies compared to lower ones. 

\par A feature of our transient-removal approach is its flexibility in incorporating multiple trial bases. For instance, by considering the matrix of least-damped eigenvectors of $\bm A$ in $\boldsymbol{\phi}_1$ and the on-the-fly snapshots in $\boldsymbol{\phi}_2$, a combined trial basis $\boldsymbol{\phi} = \boldsymbol{\phi}_1 \cup \boldsymbol{\phi}_2$ can be constructed and orthogonalized. The combination of trial bases, with $\boldsymbol{\phi}_2$ being highly effective at higher frequencies, offers benefits at lower frequencies. 

\par The expected transient error remaining before and after applying our transient removal approach can be estimated using a preprocessing step. We begin by integrating the homogeneous system \eqref{eqn:homogeneous} using a random initial condition with unit norm. By employing \eqref{eqn:1}, \eqref{eqn:2}, and \eqref{eqn:3}, and assuming $\bm q_s = 0$, we can apply either Petrov-Galerkin or Galerkin projection to calculate the updated transient norms. This approach is feasible when $\boldsymbol{\phi}$ does not depend on real-time simulation, such as when it represents the matrix of least-damped eigenvectors. However, if $\boldsymbol{\phi}$ consists of snapshots, we must generate synthetic snapshots. To accomplish this, we set $\bm q_{s, 0} = -\bm q_{t, 0}$ to ensure the initial snapshot $\bm q_0 = \bm q_{s, 0} + \bm q_{t, 0}$ equals zero. Subsequent snapshots are obtained by superimposing the transient responses (from the homogeneous simulation) onto steady-state responses generated as $\bm q_{s, j} = e^{\text{i}\omega j\Delta t}\bm q_{s, 1}$, where $\Delta t$ is the time-distance between snapshots. Using this technique, we can construct $\boldsymbol{\phi}$ for varying periods and assess the efficacy of the transient removal strategy. The updated transient error, similar to \eqref{eqn:et}, is computed as the ratio of norms between the updated transient and steady-state responses, which monotonically decreases after the transient growth phase. This iterative process is performed for all $\omega \in \varOmega$, necessitating the generation of fresh snapshots for the steady-state responses while keeping the transient response fixed. The computational expense associated with obtaining this \emph{a priori} error estimate is primarily determined by the integration of the homogeneous system and typically constitutes less than 5\% of the overall cost of executing the complete algorithm for computing the resolvent modes. We illustrate the application of this strategy using various trial bases in $\S$\ref{sec:test_cases}.

\section{Test cases} \label{sec:test_cases}

In this section, the RSVD-$\Delta t$ algorithm is tested using two problems. First, the accuracy of the algorithm and the effectiveness of the transient removal strategy are verified using the complex Ginzburg-Landau equation. Second, the computational efficiency and scalability of the algorithm are demonstrated and compared to that of the RSVD-LU algorithm using a three-dimensional discretization of a round jet.

\subsection{Complex Ginzburg-Landau equation}

The complex Ginzburg-Landau equation was initially derived for analytical studies of Poiseuille flow \citep{StewartsonStuart71} and has subsequently been used more generally as a convenient model of a flow susceptible to non-modal amplification \citep{HuntCrighton91, Bagherietal09, ChenRowley11, Cavalierietal19}. Here, we use it as an inexpensive test case to validate our algorithm. The complex Ginzburg-Landau system follows the form of \eqref{eqn:linsys_fil} with
\begin{equation}
\begin{gathered}
\bm A = -\nu \frac{\partial}{\partial x} + \gamma \frac{\partial^2}{\partial x^2}+ \mu(x), 
\\
\mu(x) = (\mu_0 - c_{\mu}^2) + \frac{\mu_2}{2} x^2,
\\
\bm B = \bm C = \bm I.
\label{eqn:GL_obj}
\end{gathered}
\end{equation} 
Following \citet{Bagherietal09}, we set  $\gamma = 1 - \text{i}, \nu = 2 + 0.2\text{i}, \mu_0 = 0.38, c_{\mu} = 0.2,$ and $\mu_2 = -0.01$. These parameters ensure global stability and provide a large gain separation between the leading mode and the rest of the modes at the peak frequency \citep{Bagherietal09}. To explicitly build the $\bm A$ operator, a central finite difference method is used to discretize $x \in [-100, 100]$ using $N = 500$ grid points. \edit{The domain is sufficiently extended in both $\pm x$ directions such that it resembles an unbounded domain without a need for explicit boundary conditions \citep{Bagherietal09}. The weight matrix $\bm{W}$ is set to the identity on account of the uniform grid.}

\subsubsection{RSVD-$\Delta t$ validation: assessing the transient and truncation errors}

\begin{figure}
\centering
\includegraphics[width=\textwidth]{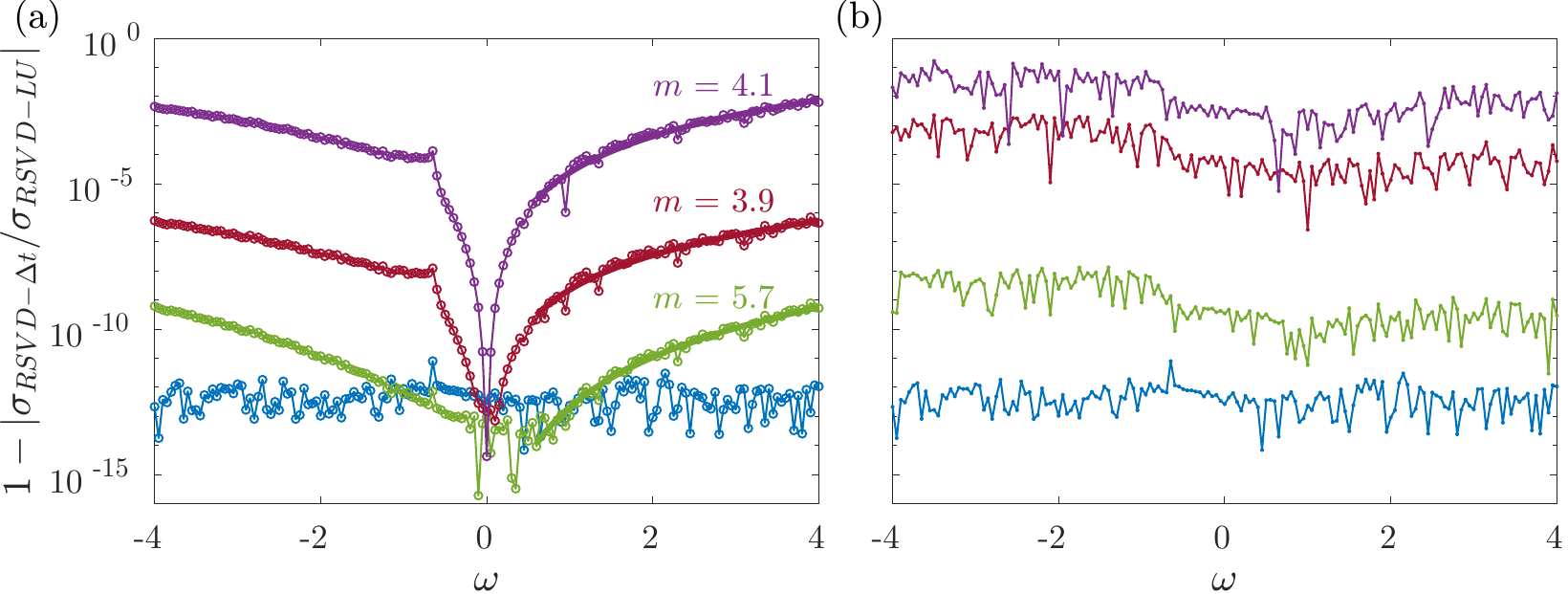}
\caption{Relative error between gains computed using the RSVD-LU and RSVD-$\Delta t$ algorithms for the Ginzburg-Landau problem: (a) $T_t$ = 5000 and \{TSS, $dt$\} = \{BDF4, 0.1\} (purple), \{BDF4, 0.01\} (red), (BDF6, 0.01) (green), and \{BDF6, 0.001\} (blue) varies; (b) \{BDF6, 0.001\} is fixed and $T_t$ varies as 500 (purple), 1000 (red), 2500 (green), and 5000 (blue). In (a), the exponents m are shown for the best-fit exponential within $\omega \in [0.6, 4]$.}
\label{fig:validation}
\end{figure}

\par The RSVD-$\Delta t$ outcome must replicate the RSVD-LU outcome up to machine precision when cutting both sources of errors described in $\S$\ref{sec:TS_Err}. Truncation error depends on the integration scheme and the time step, while the transient error depends on the length of the simulation. Therefore, using a tiny time step with a high-order integration scheme and a lengthy transient duration should eliminate the errors due to time integration. \edit{While the order of magnitude of ``tiny'' time step and ``lengthy'' transient duration may vary depending on the problem setup, the key point is that time-stepping error can be reduced to machine precision accuracy if desired.}

\par \edit{Time-stepping errors are investigated by setting the number of test vectors to $k = 1$ and power iterations to $q = 0$. These minimal values are used since including additional test vectors or power iterations have no effect on the time-stepping error. This implies that whether we are computing the action of optimal or suboptimal modes, each mode will exhibit a similar order of error due to time stepping.} The desired set of frequencies is $\varOmega \in [-4, 4]$ with $\Delta \omega = 0.05$. The gains of the Ginzburg-Landau system are computed using RSVD and RSVD-$\Delta t$ and the relative errors for various cases are shown in figure \ref{fig:validation}. The minimum error is near machine precision when BDF6, $dt = 10^{-3}$, and $T_t = 5000$ is used, validating the RSVD-$\Delta t$ algorithm.

\par By decreasing the order of the integration scheme or increasing the time step, the truncation error becomes larger, and hence, the error in the computed gains becomes larger. In figure \ref{fig:validation}(a), the transient length is held fixed at $T_t = 5000$ and the gains are obtained using \{BDF6, $dt = 10^{-2}$\}, \{BDF4, $dt = 10^{-2}$\}, and \{BDF4, $dt = 10^{-1}$\}. For all four cases, the relative error is around $O(10^{-13})$ at $\omega = 0$, confirming that the transient effect is negligible. Moving away from zero frequency, the errors increase like $O(\omega^{\sim 4})$ and $O(\omega^{\sim 6})$ for the BDF4 and BDF6 schemes, respectively, consistent with the theoretical asymptotic estimates \edit{given in \eqref{eqn:truncation2}}.

\par Figure \ref{fig:validation}(b) displays how the length of time that the transient is allowed to decay can affect the accuracy of the gains as a function of frequency. This time, the time-stepping scheme of \{BDF6, $dt = 10^{-3}$\} is held fixed, ensuring negligible truncation error, and the transient lengths are varied as $T_t = \{500, 1000, 2500, 5000\}$. Smaller values of $T_t$ leave more transient residual in the steady-state response. Longer transient lengths lead to smaller gain errors with a similar trend. The frequency distribution of the transient error depends on the eigenspectrum of the system. For example, a cluster of weakly damped modes around a specific frequency can lead to a peak transient error localized at the same frequency. In $\S$\ref{sec:jets}, the peak transient for the jet flows occurs near zero frequency.

\subsubsection{Efficient transient removal} 

\begin{figure}
\centering
\includegraphics[width=\textwidth]{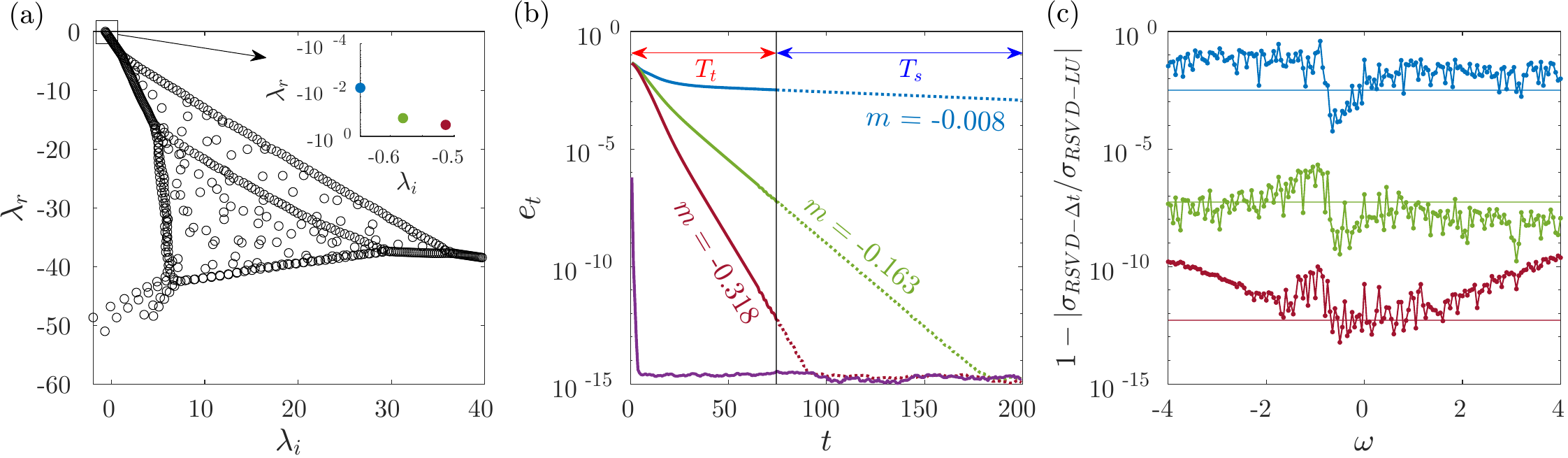}
\caption{Transient-removal for the Ginzburg-Landau test problem: (a) Spectrum of Ginzburg-Landau operator with a zoomed-in view of the three least-damped eigenvalues. (b) Transient error measurement: blue curve represents original decay, while green, red, and purple curves depict decay using Galerkin projection with $\boldsymbol{\phi}$ of $\bm V_1$, $\{\bm V_1, \bm V_2\}$, and a matrix of snapshots, respectively. (c) Relative error comparison between the RSVD-$\Delta t$ and RSVD-LU algorithms. Solid horizontal lines in (c) represent the expected transient error arising from the transient norm at the end of the $T_t$ (the black vertical line in (b)).}
\label{fig:trans_up}
\end{figure}

In this section, we demonstrate the transient removal strategy proposed in $\S$\ref{sec:EffTrans}. We apply this strategy to the same Ginzburg-Landau system for the same $\varOmega$ range described above and compare the results to the RSVD-LU results as a reference.

\par The eigenspectrum of the Ginzburg-Landau operator is shown in figure \ref{fig:trans_up}(a), and the three least-damped (and thus slowest decaying) modes have decays rates of $\lambda_{1, r} = -0.008, \lambda_{2, r} = -0.163$, and $\lambda_{3, r} = -0.318$, respectively, where the subscript $r$ indicates the real part of the eigenvalue $\lambda$. Figure \ref{fig:trans_up}(b) depicts the transient norm as a function of time, where $\epsilon$ is measured as follows: we initially obtain the \emph{true} steady-state solution by integrating \eqref{eqn:direct} for \edit{a duration of 5000 time units} at $\omega = 0.5$ (similar results for other frequencies), ensuring that the natural decay has eliminated the transient response to machine precision and use the steady-state response to measure the transient errors.

\par The natural decay in this system occurs slowly, as illustrated in figure \ref{fig:trans_up}(b). By defining $\boldsymbol{\phi}_1$ as $\bm V_1$ and utilizing Galerkin projection, we remove the fraction of the transient decaying at the rate of $e^{\lambda_1 t}$, resulting in a noticeable change in the decay slope. Including the two least-damped modes with $\boldsymbol{\phi}_2 = \{\bm V_1, \bm V_2\}$ further steepens the decay rate, aligning closely with the corresponding least-damped eigenvalues shown in figure \ref{fig:trans_up}(a). However, it is the matrix of snapshots that proves to be the most effective, completely eliminating the transient within a short period of time.

\par We employ \{BDF6, $dt = 10^{-2}$\} to compute gains using RSVD-$\Delta t$, considering three cases of transient removal that are halted at $T_t = 75$: $(i)$ natural decay, $(ii)$ Galerkin projection with $\boldsymbol{\phi}_1$, and $(iii)$ Galerkin projection with $\boldsymbol{\phi}_2$. The error is measured as the relative difference in gain between the RSVD-LU and RSVD-$\Delta t$ algorithms, as depicted in figure \ref{fig:trans_up}(c). The plot clearly illustrates that smaller transient errors lead to reduced gain errors. In the first two cases, the transient error dominates, while in the third case, the transient error balances with the truncation error at lower frequencies, with truncation dominating at higher frequencies. Our findings indicate that the matrix of snapshots is an effective basis for representing and removing the transient.  

\subsubsection{Impact of power iteration}

\begin{figure}
\centering
\includegraphics[width=\textwidth]{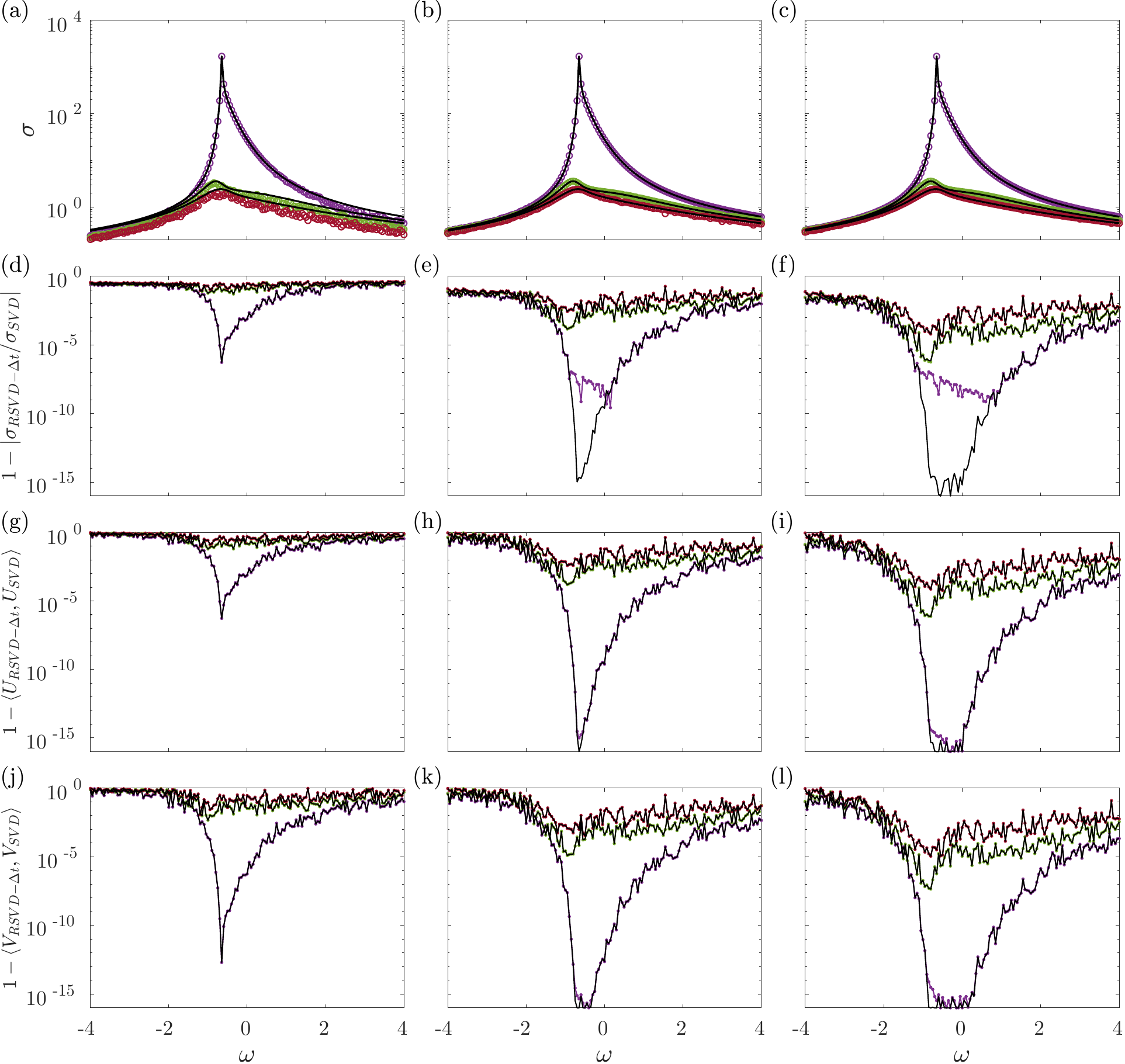}
\caption{Impact of power iteration on the Ginzburg-Landau gains: (a-c) the gains of the first three optimal modes using SVD (line) and RSVD-$\Delta t$ (circle), and (d-e) the relative error between them. \edit{The inner-product error between response (g-i) and focing (j-l) modes are plotted. (a,d,g,j), (b,e,h,k), and (c,f,i,l)} correspond to $q$ of 0, 1, and 2, respectively. Black lines in \edit{(d-l)} show the relative error between the RSVD-LU algorithm and SVD for reference. \edit{Purple, green, and red colors represent the first three leading modes, respectively.}}
\label{fig:power_it}
\end{figure}

Finally, we explore the impact of the number of power iterations $q$ on the accuracy of the solution.  For both the RSVD-LU and RSVD-$\Delta t$ algorithms, we set $k = 6$ and vary $q$ from 0 to 2. \edit{While the target number of modes in this analysis is three, we include three additional test vectors to enhance the overall accuracy of the modes.} Additionally, RSVD-$\Delta t$ uses a \edit{BDF6 integrator with $dt = 0.01$} and $T_t  = 100$. \edit{The transient length is two orders of magnitude shorter than the expected length based on the original decay rate. This reduction is a result of removing the slowest decaying component, which allows us to shorten the simulation duration while still achieving a time-stepping error of $O(10^{-8})$, as shown in figure \ref{fig:power_it}(c).} A standard Arnoldi-based approach \edit{(SVD-based with no approximation)} is used to provide a ground-truth reference for defining the error.    

\par The leading three singular values and corresponding relative errors are shown in figure \ref{fig:power_it}. One power iteration leads to a noticeable accuracy improvement. As expected, using one or more power iterations substantially improves the accuracy of both the RSVD-LU and RSVD-$\Delta t$ algorithms. The optimal singular value in particular improves dramatically for frequencies with a large gap between the optimal and suboptimal modes. The RSVD-LU errors approach machine precision near the peak frequency, while the RSVD-$\Delta t$ errors saturate at the floor set by the choice of integration parameters. For the rest of the modes and frequencies, the relative error between the RSVD-LU and RSVD-$\Delta t$ algorithms is smaller than the relative error between the RSVD-LU algorithm and the ground truth, so the relative errors are identical. \edit{Increasing the number of power iterations allows a broader range of frequencies to achieve machine precision accuracy, which aligns with the expectations for RSVD approximations \citep{Halkoetal11}. Increasing the number of power iterations from $q=1$ to 2 to 3 enables the leading modes to reach machine precision across the entire spectrum of interest, while also pushing suboptimal modes closer to machine precision. We have found one power iteration to be sufficient for most problems, and we recommend this as a default value for our algorithm. The expected error from power iteration is problem specific. For the Ginzburg-Landau problem, a single power iteration may reduce the error to below the 10\% threshold for the third mode across the frequency spectrum and achieve machine precision for the optimal mode at peak frequencies where a significant gap exists. Moreover, RSVD is generally more effective when a large gap exists between the first (or first few) singular values and the rest of the spectrum. In cases with closely spaced singular values—such as at the higher end of the spectrum in the Ginzburg-Landau problem—the modes converge more slowly.}

\edit{The accuracy of the modes, \ie the left and right singular vectors, is ensured when the gains closely match the ground truth \citep{Halkoetal11}. This relationship is illustrated in figure \ref{fig:power_it}, where we compare the inner-product relative error between modes computed via SVD and RSVD-$\Delta t$. The inner-product error between two unit-norm vectors $\bm{v_1}$ and $\bm{v_2}$ is formally defined as
\begin{equation}
e_{ip} = 1 - \langle \boldsymbol{\bm v_1}, \boldsymbol{v_2} \rangle.
\label{eqn:inner}
\end{equation}

By comparing the relative error of modes obtained from RSVD-LU and SVD (black lines in  figure \ref{fig:power_it}) across varying numbers of power iterations, we observe that when the gain relative error is below 10\%, the inner-product error for both the forcing and response modes is similarly small, mostly of the same order or better. The inner-product error between RSVD-$\Delta t$ and SVD aligns closely with that of RSVD-LU and SVD. However, the gain relative error between RSVD-$\Delta t$ and SVD remains bounded around $O(10^{-8})$, while the inner-product relative error can approach machine precision. This behavior is expected and can be understood through perturbation analysis.

As demonstrated by \citet{Stewart98, Stewart06}, when considering a perturbed matrix with a relative error of $O(\epsilon)$, a second-order convergence of $O(\epsilon^2)$ in the inner products is expected. Given that our algorithm's error is additive due to the time-stepping process, $O(\epsilon)$ error in the gains ensures $O(\epsilon^2)$ inner-product errors between the modes computed by RSVD-$\Delta t$ and RSVD-LU. However, fundamentally, our algorithm cannot exceed the accuracy achieved by RSVD-LU.}

\subsection{Round turbulent jet} \label{sec:jets}

A round jet is used to demonstrate the reduced cost and improved scaling of our algorithm. \edit{The mean flow is obtained from a large eddy simulation (LES) using the ``CharLES'' compressible flow solver developed by Cascade Technologies  \citep{Bresetal17, Bresetal18} and recently acquired by Cadence Design Systems, for a Mach number $M = \frac{U_j}{a} = 0.4$ and Reynolds number $Re = \frac{U_jD_j}{\nu_j} = 0.45\times10^6$.} Here, $U_j$ is the mean centerline velocity at the nozzle exit, $a$ is the ambient speed of sound, $\nu_j$ is the kinematic viscosity at the nozzle exit, and $D_j$ is the diameter of the nozzle. Validation of the LES simulation against experimental results and more details on the numerical setup are available in \citet{Bresetal18}.

\par
\begin{figure}
\centering
\includegraphics[width=\textwidth]{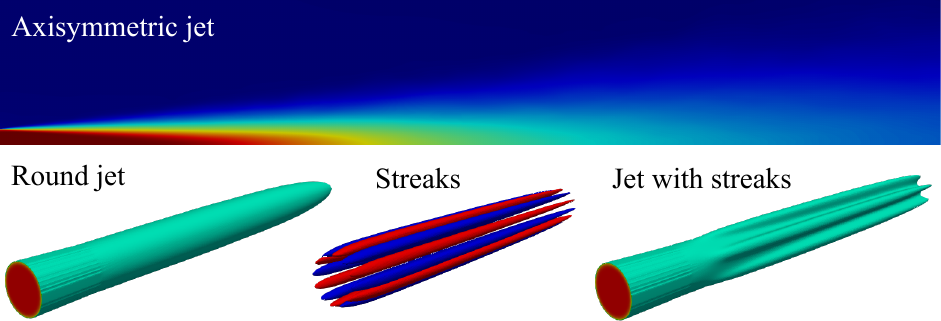}
\caption{The mean streamwise velocity of the axisymmetric jet, three-dimensional round jet, and jet with streaks. The jet with streaks is obtained by adding the streaks with an azimuthal wavenumber of 6 to the mean flow of the round jet.}
\label{fig:jet_views}
\end{figure}

 \edit{The computation of the three-dimensional resolvent modes is performed within a region of interest defined by $x \in [0, 20]$ and $y \times z \in [-4, 4] \times [-4, 4]$. The spatial discretization of this region is accomplished using a grid with dimensions of $400 \times 140 \times 140$, respectively. The core region is more finely discretized than the downstream region, as we expect smaller structures within the area defined by $x \times y \times z \in [0, 5] \times [-1, 1] \times [-1, 1]$, which contains $200 \times 70 \times 70$ points. The weight matrix $\bm{W}$ accounts for the non-uniform grid but does not reweight the flow variables (which are already non-dimensionalized). For more physically meaningful energy norms, one might consider a kinetic energy norm of Chu's compressible energy norm \citep{Chuetal65}, both of which have been utilized in the literature \citep{Towneetal18, Schmidtetal18, Schlander24}. The precise choice of norm is expected to have little impact on the results since the flow is subsonic \citep{Karbanetal20, VogelCoder22}.} The mean flow is obtained by revolving the axisymmetric mean flow around the streamwise axis, as depicted in figure \ref{fig:jet_views}. The domain is large enough to accommodate sizable low-frequency structures, and the mesh is resolved to capture structures that emerge in the response modes up to Strouhal ($St$) number of 1, where $St = \frac{\omega D_j}{2\pi U_j}$ is the non-dimensional form of frequency. The range $St \in [0, 1]$ is wide enough to include the \edit{range of frequencies that are of interest in this jet} \citep{Schmidtetal18}. \edit{Eddy viscosity can enhance the effectiveness of resolvent-based models by partially accounting for unmodeled Reynolds stresses \citep{ReynoldsHussain72, Pickeringetal21}. One simple way to approximate this effect is by reducing the Reynolds number. \citet{Pickeringetal21} found that $Re = 1000$ was effective for the same jet configuration considered here and we adopt this value in our paper. The effect of the Reynolds number on the resolvent gains and modes of a jet is thoroughly documented in \citet{Schmidtetal18}.}

\par The LNS equations are expressed in terms of specific volume, the three velocity components, and pressure, which can be compactly represented as \edit{$\bm q = [\xi, u_x, u_r, u_{\theta}, p]^T (x, r, \theta, t)$}. The three-dimensional state in the frequency domain is 
\begin{equation}
\edit{\bm q'(x, y, z, t) = \sum_{\omega} \hat{\bm q}_{\omega}(x, y, z) e^{\text{i}\omega t},}
\label{eqn:var_fourier_3D}
\end{equation}
and each mode is characterized by its frequency $\omega$.

\begin{figure}
\centering
\includegraphics[width=\textwidth]{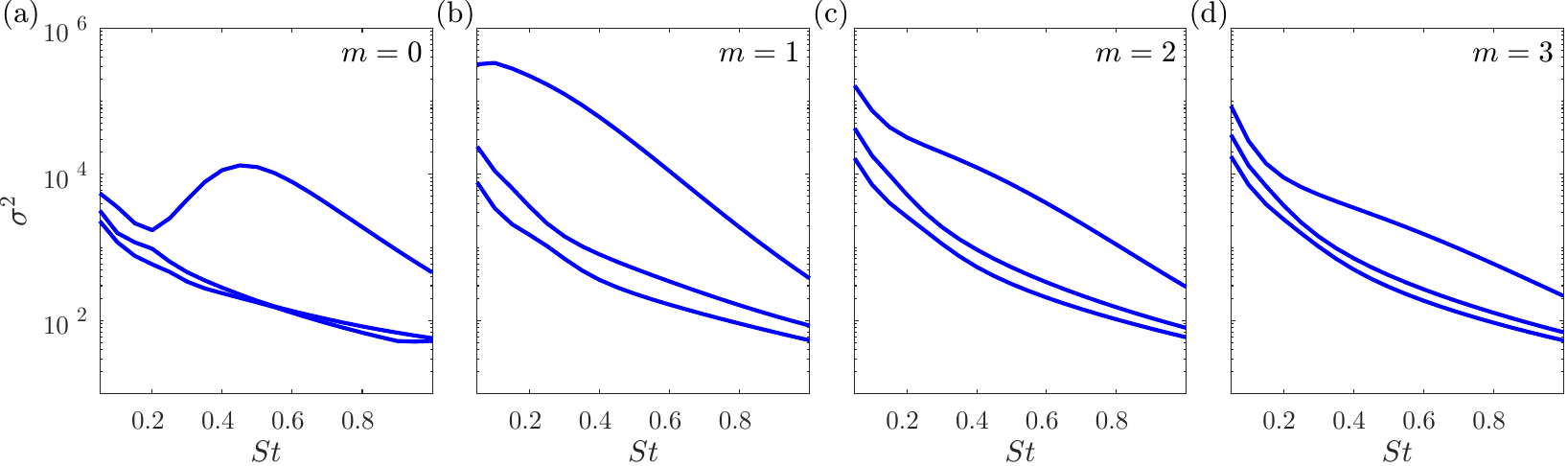}
\caption{Three leading gains of the axisymmetric jet for four azimuthal wavenumbers.}
\label{fig:gains_2D}
\end{figure}

\par To validate our three-dimensional results, we also perform a axisymmetric resolvent analysis of the same jet for a set of azimuthal wavenumbers in which the symmetry in the azimuthal direction is exploited. The mean flow is obtained on the symmetry plane with cylindrical coordinates $(x, r)$. The axisymmetric state 
\begin{equation}
\edit{\bm q'(x, r, \theta, t) = \sum_{m,\omega} \hat{\bm q}_{m,\omega}(x, r) e^{\text{i}m\theta} e^{\text{i}\omega t}}
\label{eqn:var_fourier}
\end{equation}
is characterized by the pair $(m, \omega)$, where $m$ denotes azimuthal wavenumber. \edit{The domain of interest for resolvent analysis is $x \times r \in [0, 20] \times [0, 4]$, which captures the core flow region while being surrounded by a sponge layer to minimize boundary reflections. The boundary conditions in the sponge region are designed to absorb outgoing waves effectively, reducing reflections and numerical artifacts \citep{Mani12, Schmidtetal17}. The domain is discretized using fourth-order summation-by-parts finite differences \citep{MattssonNordstrom04} with a $400 \times 100$ grid in the streamwise $x$ and radial $r$ directions, respectively. To improve accuracy, the grid resolution is higher in the core region where the primary flow features are concentrated. Note that we employed SBP differentiation operators but not an SAT implementation of the boundary conditions. However, this choice has minimal impact on our results due to the sponge region and Dirichlet boundary conditions, which mitigates the influence of boundary conditions \citep{Abgralletal20}. Similar to the three-dimensional case, a non-uniform grid setup is employed with a higher resolution in the core region. A grid-convergence study verifies the relative error between gains with this mesh and twice the number of grid points is less than 1\% for $0 \leq St \leq 0.7$ and less than 10\% for $0.7 < St \leq 1$. The larger errors at higher frequencies is due to small structures and could be eliminated by refining the mesh. However, we wish to keep the total size of the system as small as possible to keep the cost of the RSVD-LU algorithm manageable for comparison with our algorithm.} The remaining parameters are kept the same as in the three-dimensional discretization of the jet.

Figure \ref{fig:gains_2D} shows the gains (squared singular values) for $m = 0, 1, 2, 3$. The dominant mechanisms for each wavenumber are analyzed in detail in \citet{Schmidtetal18} and \citet{Pickeringetal21}. The optimal mode when $m = 0, St \ge 0.2$ corresponds to Kelvin-Helmholtz (KH) instability. At $m = 0$, the KH modes are overtaken by Orr-type modes for $St < 0.2$. At $m > 0$, streaks become the dominant response and continue to prevail as the primary instability at low frequencies $St \rightarrow 0$. \edit{Orr-type modes are characterized by perturbation structures that are initially oriented at an angle against the mean shear, leading to energy amplification through algebraic growth. These structures are subsequently reoriented along the direction of the mean shear as they evolve downstream. The resulting perturbation shape is distinctly tilted, with upstream portions leaning against the shear and downstream portions aligned with it, reflecting the interplay between shear-induced tilting and growth, as shown in the work of \citet{Pickeringetal20}.} The KH modes remain the most amplified response for the higher $St$-range when $m > 0$, causing the large separation between the leading mode and suboptimal modes. 

Similar gain trends are found in \citet{Schmidtetal18} and \citet{Pickeringetal20} for the same wavenumbers demonstrating the robustness of the outcome even though the computational domains, $Re$, state vector, sponge regions, and boundary conditions are slightly different. The gains and corresponding modes of the axisymmetric jet are used as a baseline for comparison to the three-dimensional jet.

\subsubsection{Resolvent modes for the jet} \label{sec:res3D}

\begin{figure}
\centering
\includegraphics[width=\textwidth]{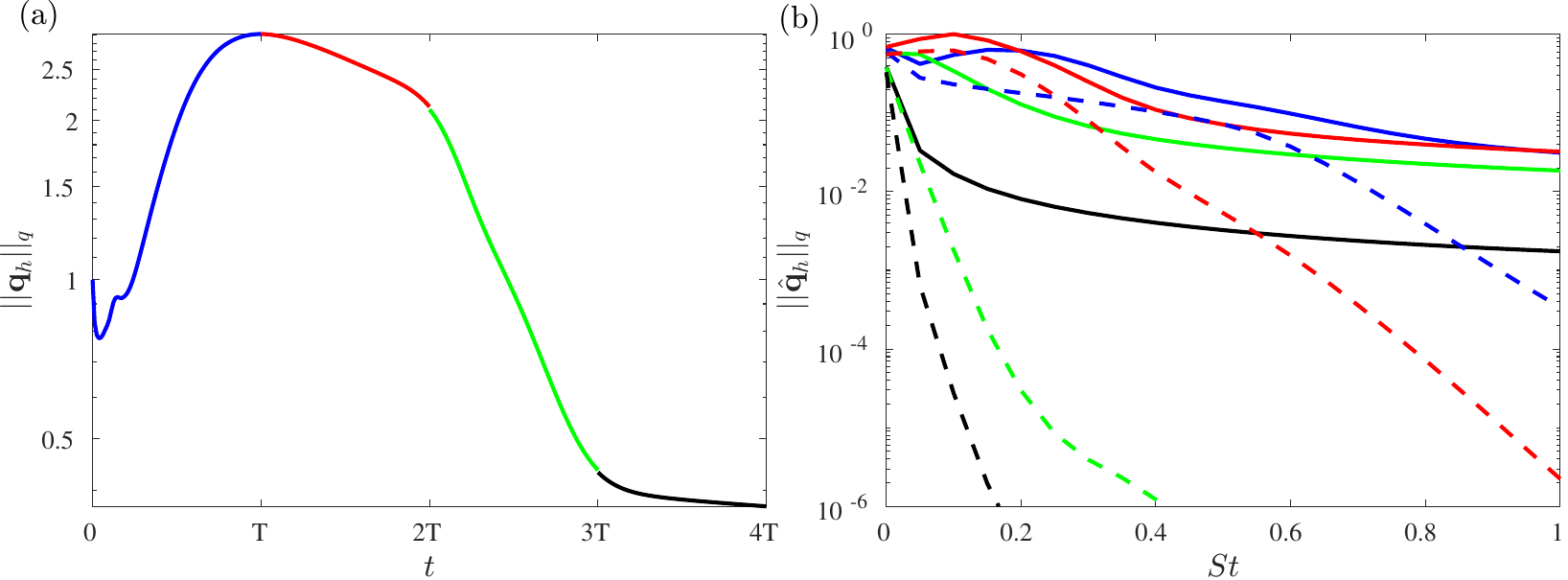}
\caption{Transient error estimates for the jet in (a) the time domain and (b) the frequency domain. Each colored period represents the duration utilized for obtaining norms in the frequency domain as shown in (b). Solid lines represent the natural decay, while dashed lines correspond to the transient removal strategy using Galerkin projection with the matrix of snapshots. \edit{The colors blue, red, green, and black correspond to the first, second, third, and fourth periods, respectively.}}
\label{fig:transient_3D}
\end{figure}

\par Resolvent modes for the three-dimensional round jet are computed for the same range of $St \in [0, 1]$ with $\Delta St = 0.05$. The six leading modes are of interest, so we set $k = 10$ and $q = 1$. \edit{The choice of $k$ is determined by the number of desired modes, with a few additional test vectors included to enhance accuracy. The number of power iterations $q$ may be increased if the modes at the frequencies of interest have not yet converged. In this case, an additional power iteration does not affect the results for $St \le 0.7$, making $q = 1$ an effective choice.} For the RSVD-$\Delta t$ algorithm, we use the classical $4^{th}$ order Runge--Kutta (RK4) integrator with $dt = 0.00625$. The steady-state interval is $T_s = 20$. Figure \ref{fig:transient_3D} shows the expected transient error in the time and frequency domains. The transient initially grows in time before slowly decaying in figure \ref{fig:transient_3D}(a). The resulting error in the frequency domain obtained from selecting each colored segment for computing resolvent modes is shown in figure \ref{fig:transient_3D}(b). Our transient removal strategy, using Galerkin projection with the matrix of snapshots, drastically reduces these errors for $St > 0$, as indicated by the dashed lines.  We select a transient duration of $T_t \approx 2T_s$ (green segment), for which the transient removal strategy brings the transient error below 1\% for $St > 0$.  

\par
\begin{figure}
\centering
\includegraphics[width=\textwidth]{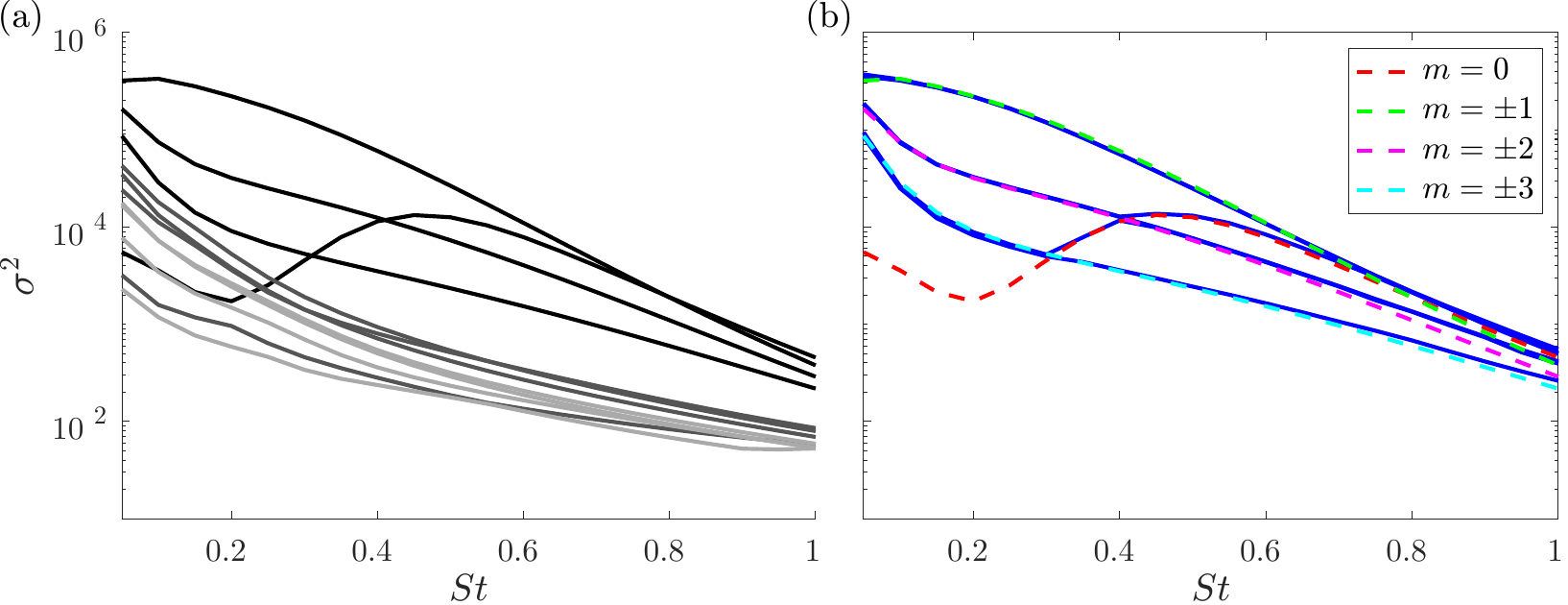}
\caption{Resolvent gains for the jet: (a) the union of the axisymmetric jet gains; (b) the optimal gains of the axisymmetric jet corresponding to various values of $m$ (dashed lines) overlaid on top of the six leading gains for the three-dimensional discretization (solid lines). \edit{In (a), optimal modes for each $m$ are colored black, gradually transitioning to grey for suboptimal modes.}}
\label{fig:gains_3D}
\end{figure}

 Figure \ref{fig:gains_3D} compares the gains of two-dimensional and three-dimensional discretizations of the jet. Due to the azimuthal symmetry of the problem, the gains of the three-dimensional problem are expected to be the union of the gains from the axisymmetric problem \citep{Sirovich87_2}. Since higher wavenumbers $(m > 3)$ have lower gains \citep{Pickeringetal21}, the union of the first four azimuthal wavenumbers is enough to match the leading modes of the three-dimensional system. The azimuthal symmetry makes modes corresponding to $m \neq 0$ appear in pairs for the three-dimensional problem. \edit{As a result, figure \ref{fig:gains_3D}(a) displays three distinct curves, even though six modes are represented; each curve corresponds to one of the two symmetrical modes that overlap. The optimal gain at $m = 0$ is not captured in the three-dimensional analysis, as it is suboptimal and has lower energy compared to the other modes when $St \le 0.3$ as shown in figure \ref{fig:gains_3D}(a).} The six computed modes appear in pairs for $St \le 0.3$, after which the gain of the $m=0$ mode becomes large enough to appear for the three-dimensional problem. Up to $St = 0.8$, the largest gains are associated with $m = \pm 1$. All of the modes that appear for the three-dimensional problem are KH modes; more \edit{($k > 10$)} resolvent modes would need to be computed to capture Orr modes that are buried beneath a slew of KH modes for each azimuthal wavenumber. The close match between the computed three-dimensional modes and the set of two-dimensional modes verifies that the three-dimensional calculations are properly capturing the known physics for this problem. The small mismatch at frequencies close to $St = 1$ is due to mild under-resolution of the grid for the compact structures that appear at these frequencies.  

\par
\begin{figure}
\centering
\includegraphics[width=\textwidth]{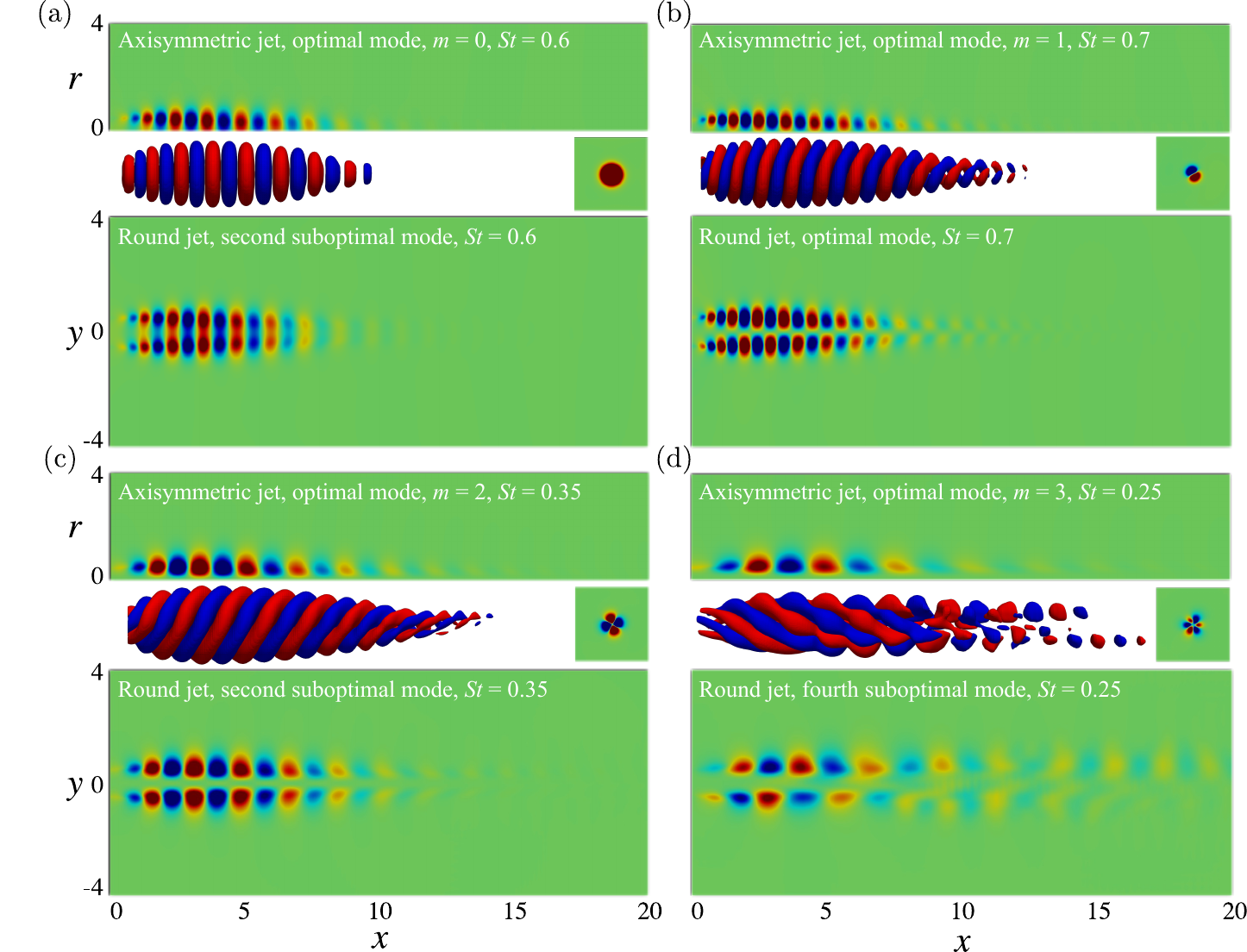}
\caption{Four groups of axisymmetric and three-dimensional pressure modes are shown, including axisymmetric views, three-dimensional iso-volume representations, and $x-y$ plane snapshots of the round jet. Cross-sections at $x = 5$ confirm the azimuthal wavenumber of the three-dimensional results. Color bar ranges are adjusted for visualization.}
\label{fig:modes23D}
\end{figure}

 Figure \ref{fig:modes23D} shows the pressure response modes at four $(St, m)$ pairs (other components such as velocity yield similar observations). Each panel shows, for one $(St, m)$ pair, contours of the two-dimensional mode computed leveraging symmetry, isocontours of the corresponding three-dimensional mode, and contours for cross sections of the three-dimensional mode in the $x-y$ and $y-z$ planes. These images show the wavepacket form of the modes, confirm the classification of each three-dimensional mode with a particular azimuthal wavenumber, and illustrate the match between the symmetric and three-dimensional results. As noted by \citet{Martinietal21}, symmetries such as the azimuthal homogeneity of the jet produce pairs of modes with equal gain that can be arbitrarily combined (under the constraint of orthogonality) to produce equally valid mode pairs. For visualization purposes, we have adjusted the phase and summed the mode pairs to best match those of the modes from the axisymmetric calculations. 

\subsubsection{Computational complexity comparison} \label{sec:jets_cost}

\begin{sloppypar} We showcase the superior computational efficiency and scalability of the RSVD-$\Delta t$ algorithm compared to the RSVD-LU algorithm using the three-dimensional jet by varying the discretized state dimension $N$. We set $k = 10$, $N_{\omega} = 21$, and $q = 0$ for both algorithms and $dt = 0.00625$, $T_t = 2T_s$, and $T_s = 20$ in the RSVD-$\Delta t$ algorithm as in $\S$\ref{sec:res3D}. The reported costs for the RSVD-LU algorithm includes only a single LU decomposition and the two solutions of the LU decomposed system (once for the direct system and once for the adjoint system) at each frequency of interest, highlighting the LU decomposition as the primary bottleneck in the RSVD-LU algorithm and similar methods utilizing LU decomposition to solve \eqref{eqn:solve}. The reported costs encompasses the entire RSVD-$\Delta t$ algorithm with a total integration length of $4T_s$ per action, including one extra period to account for the transient removal strategy, as explained in $\S$\ref{sec:EffTrans}. The RSVD-$\Delta t$ algorithm is implemented using PETSc \citep{Balayetal19}, while the LU decomposition in the RSVD-LU algorithm utilizes PETSc in conjunction with the MUMPS \citep{Amestoyetal01} external package. \edit{All calculations are performed on one Intel Xeon Gold 6154 processor on the University of Michigan's Great Lakes cluster, with wall time serving as a proxy for CPU time.} \end{sloppypar}

\par
\begin{figure}
\centering
\includegraphics[width=\textwidth]{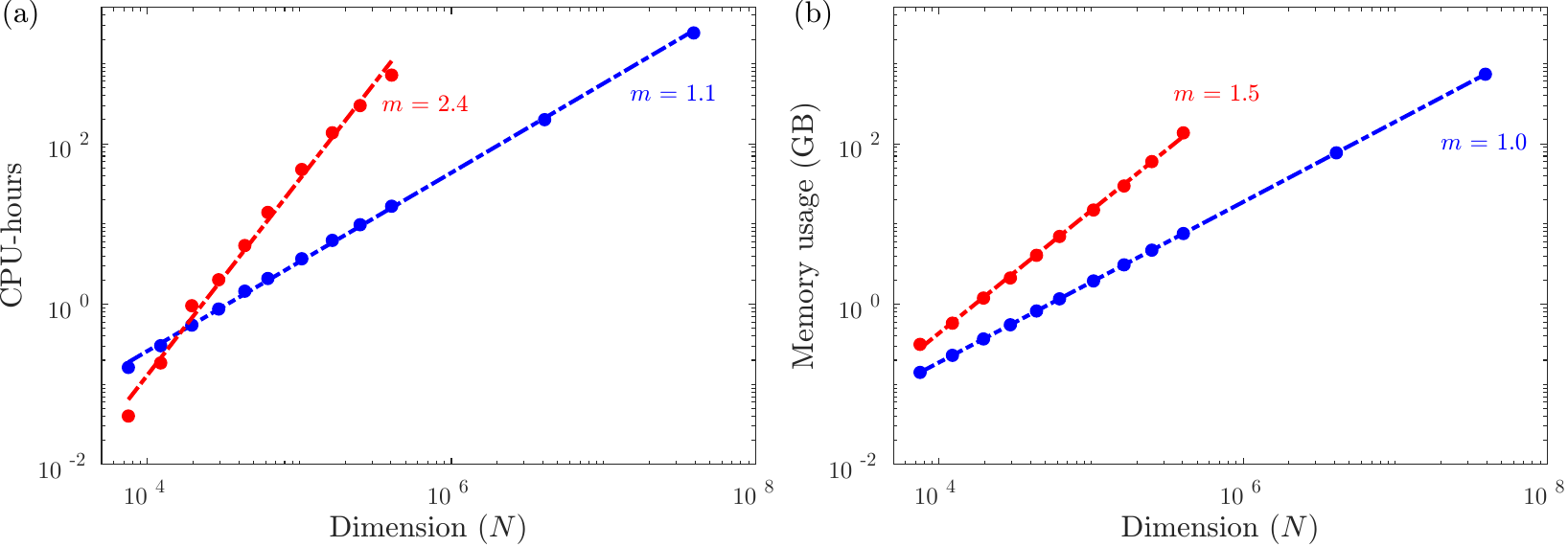}
\caption{Computational cost as a function of the state dimension $N$ for the three-dimensional jet: (a) CPU-hours and (b) memory usage  for the RSVD-LU (red) and RSVD-$\Delta t$ (blue) algorithms.}
\label{fig:scalings1}
\end{figure}

 The measured CPU time for both algorithms are shown in figure \ref{fig:scalings1}(a) as a function of the state dimension $N$. The RSVD-LU algorithm scales poorly, in fact exceeding the theoretical scaling of $O(N^2)$ for three-dimensional flows (refer to $\S$\ref{sec:comp_theory}) due to poor performance at low frequencies that has also been noted in other studies \citep{Pickeringetal20}. In contrast, the RSVD-$\Delta t$ algorithm achieves (near) linear scaling, $O(N^{1.1})$, confirming its scalability to large problems. The calculations could not be performed using RSVD-LU for the largest two grids dimensions exceeding 1 million require an excessive amount of memory beyond our cluster limits of \edit{3.5 TB} when employing RSVD-LU. Therefore, these two data points are exclusively reserved for RSVD-$\Delta t$ to validate the linear scaling retention during the transition to more realistic dimensions. The final point corresponds to the same grid utilized in our three-dimensional jet flow test case.

\par Similar observations can be made about the memory requirements of the two algorithms, shown in figure \ref{fig:scalings1}(b). The observed $O(N^{1.5})$ memory scaling for the RSVD-LU algorithm is better than the CPU counterpart, but it is still the main barrier to applying the RSVD-LU algorithm when the state dimension is of the order of 10 million or higher. The RAM peak usage is determined entirely by LU decomposition and drops after the decomposed matrices are obtained.  On the other hand, the memory scaling for the RSVD-$\Delta t$ algorithm is exactly linear with the state dimension $N$, consistent with the theoretic scaling determined in $\S$\ref{sec:comp_theory}.

\par Most of the grids considered in figure \ref{fig:scalings1} are under-resolved to make the RSVD-LU calculations tractable. To compare the two algorithms for a realistic grid, Table \ref{tab:tab2} compares the costs of RSVD-LU and RSVD-$\Delta t$ for $N \approx 39$ million (5 state variables $\times$ a $[400 \times 140^2]$ grid), which was used for the three-dimensional calculations in $\S$\ref{sec:jets_cost}, and $N_{\omega} = 21$, $k=10$, and $q=1$. The CPU and memory requirements of the RSVD-LU algorithm are intractable for this problem, so we estimate these costs by extrapolating the best-fit lines in figure \ref{fig:scalings1}.  Computing the action of the resolvent operator in the RSVD-LU algorithm involves both LU decomposition and solving the decomposed system, with both being extrapolated but the latter not depicted in figure \ref{fig:scalings1}. This implies that for $q = 1$, the CPU time includes a single LU decomposition and four times solving the LU-decomposed system. On the other hand, for RSVD-$\Delta t$, the CPU time and memory usage are directly taken from our simulation, which employed 300 \edit{Intel Xeon Gold 6154 processors on the University of Michigan's Great Lakes cluster.}

\par The RSVD-LU algorithm exhibits a CPU time that is more than three orders of magnitude higher than that of the RSVD-$\Delta t$ algorithm. Specifically, using 300 cores, the wall-time for RSVD-$\Delta t$ is approximately 61 hours ($<$ 3 days), while the RSVD-LU algorithm requires over 75 300 000 CPU-hours, which translates to around 251 000 hours ($\sim$ 28 years) wall-time, assuming perfect linear speed-up using 300 parallel cores. Of course, this wall-time can be brought down by increasing the number of cores, but it is clear that super-computing resources would be required to make the wall-time acceptable, which is antithetical to role of resolvent analysis as a reduced-order model. This disparity becomes even more pronounced as $N$ increases due to the linear CPU scaling of RSVD-$\Delta t$ and the quadratic scaling of the RSVD-LU algorithm for three-dimensional problems. Table \ref{tab:tab2} confirms that the time-stepping process accounts for nearly all of the CPU time in RSVD-$\Delta t$.

\par The memory improvements of the RSVD-$\Delta t$ algorithm are arguably even more important. The memory usage in the RSVD-LU algorithm exceeds that of RSVD-$\Delta t$ by more than two orders of magnitude. The minimum memory requirement for LU calculations surpasses 130 TB for the three-dimensional jet flow. This amount of memory is more than can be accessed even on most high-performance-computing clusters. In contrast, the memory usage in RSVD-$\Delta t$ is optimized to store only three matrices of size $N \times k \times N_{\omega}$, which can be accurately estimated based on the size of each float number in C/C++. For instance, with $N \approx 39$ million, $k = 10$, and $N_{\omega} = 21$, the RAM consumption for these matrices amounts to $\sim$ 0.75 TB (using double precision with 64-bit indices). Moreover, the RAM requirements of our algorithm can be further reduced at the expense of higher CPU cost if necessary as proposed in $\S$\ref{sec:group}, while no such trade-off exists for the RSVD-LU algorithm.

\section{Application: jet with streaks} \label{sec:application}

Finally, we apply the RSVD-$\Delta t$ algorithm to study the impact of streaks on other coherent structures within a turbulent jet. This is a fully three-dimensional problem for which results obtained using other algorithms are not available.

\par Streaks -- elongated regions of low-velocity fluid -- have historically been observed and studied in turbulent channel flows (see \citet{Mckeon17} and \citet{Jimenez18} and the references therein). More recently, in unbounded shear flows such as round jet flows, streaks have been shown to be generated via the evolution of optimal initial conditions that maximize the transient energy growth \citep{JimenezBrancher17}. \citet{Nogueiraetal19} and \citet{Pickeringetal20} showed that streaks emerge as the dominant structures in the SPOD and resolvent spectra of jets at very low frequencies when $m \ge 1$. Streaks are produced via a lift-up mechanism applied to the rolls or streamwise vortices that are usually excited near the nozzle exit. The presence of streaks within turbulence modifies the flow quite significantly. In particular, streaks are shown to stabilize the KH wavepackets in a parallel plane shear layer \citep{MarantCossu18} and Tollmien--Schlichting waves in the Blasius boundary layer \citep{CossuBrandt02}. Similar findings on a high-speed turbulent jet by \citet{Wangetal21} demonstrate the stabilizing effects of finite-amplitude streaks on KH wavepackets. In this study, we investigate the impact of streaks on the linear amplification and spatial structure of the Kelvin-Helmholtz wavepackets described by the leading resolvent modes via a secondary stability analysis.  

\begin{table}
\begin{center}
\begin{tabular}{cccccc}
Algorithm & \multicolumn{3}{c}{\hspace{2mm}{CPU time (hours)}} & \hspace{2mm} Memory (GB) \\
\cmidrule{1-5}
& \hspace{2mm} Total & Action of $\bm R$ and $\bm R^*$ & SVD/QR \\ \\
{RSVD-LU} & \hspace{2mm} $7.53\times10^7$ & $7.53\times10^7$ & 0.762 & \hspace{2mm}$1.33\times10^5$ \\ 
{RSVD-$\Delta t$} & \hspace{2mm} $1.83\times10^4$ & $1.83\times10^4$  & 0.762 & \hspace{2mm} $7.36\times10^2$ \\ 
\end{tabular}
\end{center}
\caption{Comparison of the RSVD-LU and RSVD-$\Delta t$ algorithms in terms of CPU time and memory usage for the three-dimensional jet with $N \approx 39M, N_{\omega} = 21, k = 10$, and $q = 1$. The action of $\bm R$ and $\bm R^*$ use time stepping for RSVD-$\Delta t$ and a direct solver for the RSVD-LU algorithm.}
\label{tab:tab2}
\end{table}

\par The streaks that will be added to the mean flow are obtained from an initial resolvent analysis of the mean flow; specifically, streaks are the optimal resolvent response at very low frequencies \citep{Pickeringetal20}. Due to the symmetry of the mean jet, streaks obtained from data via SPOD or computed using resolvent analysis are associated with a particular azimuthal wavenumber. Accordingly, we compute the streaks using our axisymmetric code, which produces the same results as the three-dimensional code but at a lower cost. We compute them for $(St, l) = (0, 6)$, where $l$ denotes the azimuthal periodicity of the streaks. This choice of $l = 6$ corresponds to one of the main cases studied in \citet{Wangetal21}.

\par The updated mean flow with the streaks added has 6-fold rotational symmetry and, following \citet{Sinhaetal16}, can be written as
\begin{equation}
\bar{\bm q}(x, r, \theta) = \sum_{j = -\infty}^{\infty} \hat{\bar{\bm q}}_{lj}(x, r) e^{ilj\theta}.
\label{eqn:6-fold}
\end{equation}
They proved that after plugging the Fourier ansatz of the resulting mean flow into the LNS equations, given an azimuthal wavenumber $m$, the associated axisymmetric mode $\hat{\bm q}_{m, \omega}$ can only couple with $\hat{\bm q}_{m - lj, \omega}$ for $j \in \mathbb{Z}$. In our problem, $l = 6$ and sorting the modes with the lowest azimuthal modes, we expect coupling of modes in sets of $\bm q_{\omega}^L = \{\hat{\bm q}_{L - lj, \omega}\}_{l = -\infty}^{l = \infty}$, where $L = \{-2, -1, 0, 1, 2, 3\}$ includes all possibilities. Indexing in this manner implies that the modes with $L = 0, 3$ are unpaired while $L = \pm 1, \pm 2$ will show up in pairs in the three-dimensional setup due to symmetry.

\begin{figure}
\centering
\includegraphics[width=\textwidth]{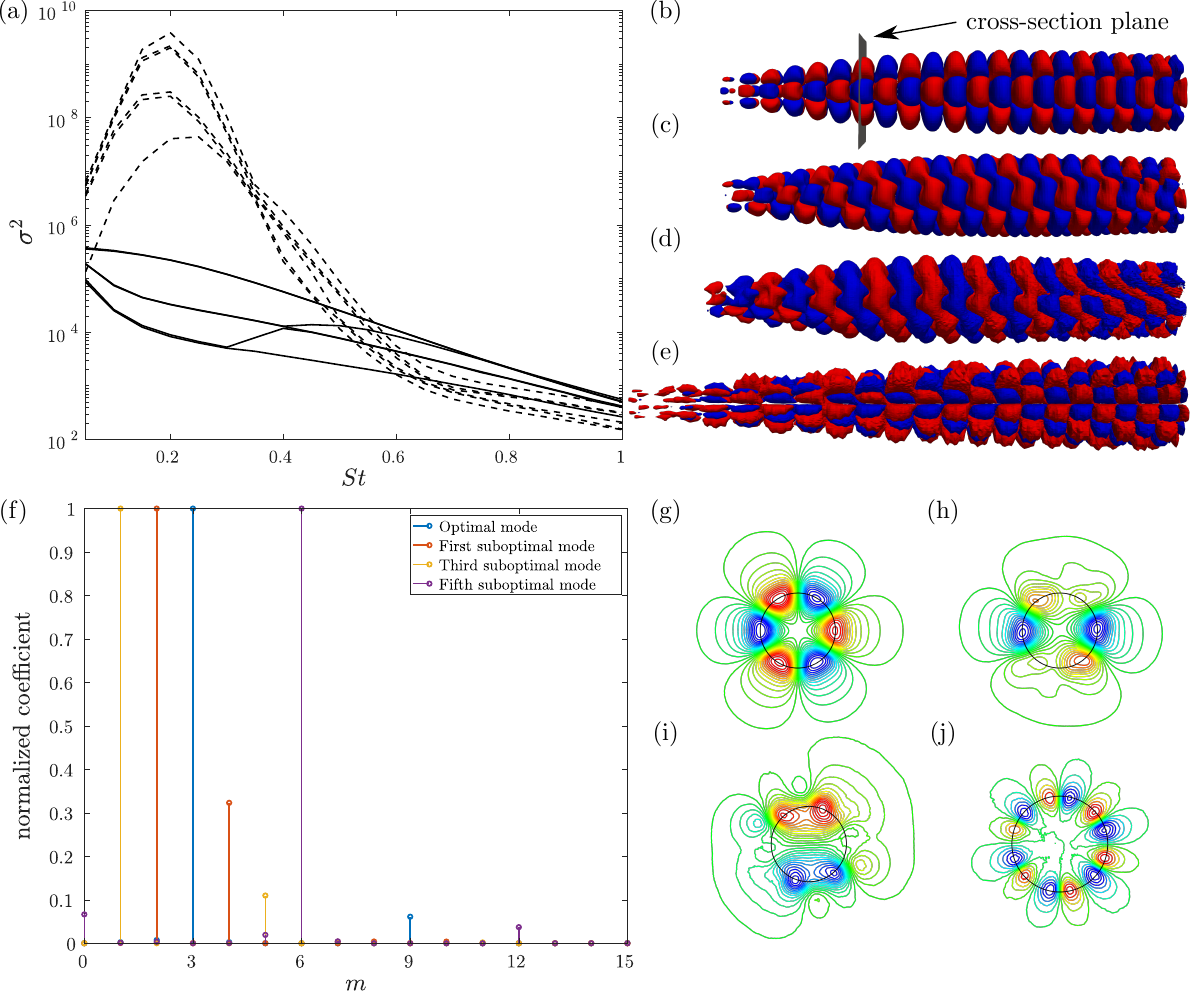}
\caption{Results for the jet with streaks: (a) resolvent gains for the round jet (solid line) and jet with streaks (dashed line); (b-e) the optimal, first, third, and fifth suboptimal pressure responses at $St = 0.2$; (g-j) contours of the pressure responses on cross-section at $x = 8.5$ corresponding to (b-e), respectively.  Fourier transforms are taken along the black circles shown to obtain the corresponding azimuthal wavenumber spectra for each mode shown in (f).}
\label{fig:strJet}
\end{figure}

\par \edit{The shape of the streaks is} sensitive to a few parameters including the viscosity (or equivalently turbulent Reynolds number or eddy-viscosity model if desired) and forcing region. In lieu of a more complex eddy-viscosity model, we use a constant turbulent Reynolds number of $Re =1000$. This value is close to the optimal frequency-dependent value determined by \citet{Pickeringetal21} for $St = 0$ as well as most of our frequency range of interest $St \in [0, 1]$ for the secondary stability problem. Additionally, the forcing region of the resolvent analysis used to compute the streaks must be limited to obtain streaks of finite streamwise length. If the domain is not limited, the forcing rolls that generate these streaks sustain them throughout the domain. After some trial and error, we limited the forcing region to $x, r \in [0, 1]\times[0, 1]$, which produced streaks with a location of peak amplitude $(x \in [5, 6])$ and overall shape consistent with the streak SPOD modes obtained by \citet{Nogueiraetal19}.

\par Once the axisymmetric streaks are computed, the three-dimensional streaks are obtained by revolving them around the $x-$axis with phase $e^{\text{i} l \theta}$ (see figure \ref{fig:jet_views}). \edit{The amplitude is defined as the ratio of the peak streamwise velocity of streaks over the maximum velocity at the nozzle exit. This serves as a free parameter that can be investigated across various values. According to \citet{Wangetal21}, the amplitude of these structures grows linearly over time. Therefore, no correct constant amplitude exists for our secondary analysis. The amplitude of streaks in our paper is set to 40\%, which is large enough to affect the modes compared to the round jet. The region of interest and grid points along with all the other parameters are the same as for the round jet.}

\par RSVD-$\Delta t$ is used to compute the resolvent modes for the modified mean flow. The number of test vectors is $k = 10$ and the gains are reported after $q = 2$ power iterations. \edit{For the same reasons mentioned for the round jet, we use 10 test vectors and are interested in computing the top six leading modes. Regarding the number of power iterations,} the first few leading modes converged after the first power iteration, but an extra power iteration is performed to ensure convergence since no ground truth results are available for comparison. The frequency range $St \in [0, 1]$ and discretization $\Delta St = 0.05$ are the same as used for the round jet in $\S$\ref{sec:jets}. The time-stepping scheme is RK4 with $dt = 0.00625$. Transient errors are held below $1\%$ for $St > 0$ via our transient removal strategy using Galerkin projection with the matrix of snapshots with a duration $T_t = 3 T_s$.

\par The gains for the round jet and jet with streaks are compared in figure \ref{fig:strJet}(a). The streaks have increased the gains by orders of magnitude for $St < 0.5$. Some of the gains appear in pairs, indicating mode pairs analogous to those described for the round jet, which arise due to the six-fold symmetry of the mean jet with streaks. The match occurs between the first and second suboptimal in addition to the third and fourth suboptimal modes. All modes almost coincide at $St = 0.35$ and continue decaying as $St$ increases.

\par The optimal, first, third, and fifth suboptimal pressure response modes at $St = 0.2$, where the leading gain is maximum, are shown in figure \ref{fig:strJet}. The second and fourth suboptimal modes are not shown since they are pairs with the first and third suboptimal modes, respectively. The three-dimensional iso-surfaces show KH wavepackets that are significantly altered by the streaks; characterizing the modes with the indexes defined earlier requires deeper investigation. To this end, cross-section contours at $x = 8.5$ are plotted. These plots are more complicated than the round jet due to the coupling between multiple azimuthal wavenumbers. We interpolate the pressure field on the circles shown on each contour plot to demonstrate the coupling azimuthal wavenumbers. Taking an FFT of the extracted data, the normalized coefficients are plotted against $m$ in \ref{fig:strJet}(f). This plot shows that the optimal mode is comprised of $L = 3$ with a larger weight and $L + l = 3 + 6 = 9$ with a smaller weight, which is consistent with our axisymmetric analysis. The first suboptimal mode includes $(L, L-l) = (2, -4)$, and its pair contains $(L, L+l) = (-2, 4)$, so both couplings and pairings are as expected. Similarly, the third mode is a coupling between $(L, L-l) = (1, -5)$, and the fourth mode is with $(L, L+l) = (-1, 5)$. Lastly, the fifth mode is unpaired and captures the $(L, L+l) = (0, 6)$ azimuthal wavenumbers with a small signature of $L + 2l = 12$.

\par From the perspective of computational cost, the jet with streaks is similar to the three-dimensional discretization of the round jet. Utilizing the RSVD-LU algorithm for the same grid with state dimension $N \approx 39$ million, the anticipated CPU time surpasses 75 million hours, as discussed in $\S$\ref{sec:jets_cost}. Nevertheless, leveraging RSVD-$\Delta t$ with $q = 2$ enabled us to complete the analysis within 37 thousand CPU-hours. Our computations used 300 cores, which results in a wall time of 28 years for the RSVD-LU algorithm and 123 hours for our algorithm. Additionally, memory requirements amount to more than 130 TB for the RSVD-LU algorithm and 0.75 TB for ours. \edit{As a point of comparison, the LES of a similar jet with $N \approx 64$ million consumed 464 thousand CPU hours \citep{Bresetal18}.} It is safe to say that this analysis would have been intractable using previous algorithms, demonstrating the promise of the RSVD-$\Delta t$ algorithm for extending the applicability of resolvent analysis to new problems in fluid mechanics. 

\section{Conclusions}  \label{sec:con}

This paper introduces RSVD-$\Delta t$, a novel algorithm designed for efficient computation of global resolvent modes in high-dimensional systems, particularly in the context of three-dimensional flows. By leveraging a time-stepping approach, RSVD-$\Delta t$ eliminates the reliance on LU decomposition that often hampers the scalability of current state-of-the-art algorithms. As a result, RSVD-$\Delta t$ not only enhances scalability but also extends the applicability of resolvent analysis to three-dimensional systems, overcoming previous computational limitations.

\par Scalability is of utmost importance for algorithms dealing with high-dimensional flows, and RSVD-$\Delta t$ excels in this regard. In contrast, the \edit{LU} decomposition of $(\text{i}\omega \bm I - \bm A)$ poses a significant computational challenge for the RSVD-LU algorithm, limiting its scalability with $O(N^2)$ scaling for 3D problems. The CPU demand of RSVD-$\Delta t$, on the other hand, exhibits linear proportionality to the state dimension.

\par In addition to CPU considerations, memory requirements play a crucial role in computing resolvent modes for large systems. The LU decomposition of $(\text{i}\omega \bm I - \bm A)$ is the primary contributor to peak memory usage in the RSVD-LU and other common algorithms. In contrast, the RSVD-$\Delta t$ algorithm primarily utilizes RAM to store input and output matrices in Fourier space, resulting in linear growth of memory consumption with dimension. To minimize the required memory, we utilize streaming calculations, which maintains low memory requirements with minimal computational impact. If memory limitations persist, the set of desired frequencies can be split into $d$ groups to further reduce the required memory by a factor of $d$.

\par The RSVD-$\Delta t$ algorithm contains three sources of error, each of which can be controlled by carefully selecting method parameters. The first arises from the RSVD approximation inherited from the RSVD algorithm. This error can be significantly reduced by employing power iteration and utilizing more test vectors than the desired number. The second source of error stems from the time integration method employed to compute the action of $\bm R$ and $\bm R^*$. Time-stepping errors encompass the transient response and truncation error. Truncation error arises from the numerical integration of the LNS equations and can be managed through careful selection of the time-stepping scheme and time step. The transient response emerges when the initial condition is not synchronized with the applied forcing, decaying over time but potentially requiring many periods to become sufficiently small. To expedite the removal of transients, a novel strategy is introduced involving the decomposition of snapshots into transient and steady-state components, with subsequent solving of equations for the transient. This computation is facilitated through Petrov-Galerkin and Galerkin projections. To ensure optimal performance, it is important to maintain a balance between truncation and transient errors. Focusing too much on reducing one source significantly while neglecting the other can lead to a waste of CPU time without an impact on the outcome. Also, keeping both errors smaller than the RSVD approximation error will not improve the accuracy of RSVD-$\Delta t$ with respect to SVD-based (true) results. By effectively eliminating both truncation and transient errors up to machine precision, RSVD-$\Delta t$ has been validated against the RSVD-LU algorithm using the complex Ginzburg-Landau equation.

\par The RSVD-$\Delta t$ algorithm is particularly valuable for analyzing three-dimensional flows, where other algorithms become impractical. The superior scalability of the RSVD-$\Delta t$ algorithm leads to an increasingly pronounced disparity in computational complexity compared to the RSVD-LU algorithm as the value of $N$ grows larger. As an example, we consider a moderately large state dimension of $N \approx 39$ million. Using the RSVD-LU algorithm for this problem would require an estimated 75 million CPU-hours and 130 TB of RAM. In contrast, the RSVD-$\Delta t$ algorithm required just 18,000 CPU-hours and 0.75 TB of RAM, a reduction of three and two orders of magnitude, respectively. In general, the benefits of the RSVD-$\Delta t$ algorithm are most pronounced for three dimensional flows and other large systems, while little advantage is gained for simple one- and two-dimensional flows.

\par Lastly, we leveraged the novel capabilities of the RSVD-$\Delta t$ algorithm to investigate the influence of streaks within the turbulent jet on the KH wavepackets. \edit{Using a procedure analogous to} a secondary stability analysis in which the steady streaks are added to the axisymmetric mean flow, we showed the significant impact of the streaks on the KH wavepackets. This included a substantial increase in gains within the range $St \in [0, 0.5]$, a change in the most amplified azimuthal wavenumber, and coupling of multiple azimuthal wavenumbers is some of the modes. Given the recently demonstrated presence of streaks in real jets, these finds warrant further investigation in the future. 

\par Our algorithm also has several implementation advantages. Our time-stepping approach enables matrix-free implementation, eliminating the explicit formation of the LNS matrix $\bm A$, instead directly utilizing built-in linear direct and adjoint capabilities available within many existing codes.  All operations within the RSVD-$\Delta t$ algorithm are amenable to efficient parallelization; we have optimized out implementation of the algorithm for parallel computing using the PETSc \citep{Balayetal19} and SLEPc \citep{Hernandezetal05} environments, facilitating full utilization of the computational power offered by modern high-performance clusters. Moreover, our code is designed to leverage GPUs, enabling the delegation of compute-intensive tasks to the GPU architecture for quicker and more efficient calculations. Finally, the efficiency and accuracy of the RSVD-$\Delta t$ algorithm could be further enhanced by incorporating strategies developed for the RSVD-LU algorithm. Notably, techniques proposed by \citet{Ribeiroetal20} and \citet{Houseetal22} can be integrated into our approach to use physical insight to select the initial test vectors instead of relying on entirely random ones. \edit{An open-source implementation of the RSVD-$\Delta t$ algorithm is available on GitHub (\url{https://github.com/AliFarghadan/RSVD-Delta-t}).}

\section*{Acknowledgements}
We would like to express our gratitude to Andr\'e Cavalieri for his invaluable feedback, insights, and contributions. We also acknowledge the University of Michigan's Great Lakes cluster for providing the essential computational resources that enabled us to conduct all computations for this research. A.F. and A.T. gratefully acknowledge funding for this work from the Michigan Institute for Computational Discovery and Engineering (MICDE) and AFOSR award number FA9550-20-1-0214.

\begin{appendices}

\section{RSVD-\texorpdfstring{$\Delta t$}{deltat} for the weighted resolvent operator}\label{appA}

For the sake of notational brevity, we have described resolvent analysis and the RSVD-$\Delta t$ algorithm in the absence of non-identity input, output, and weight matrices in the main text (see $\S$\ref{sec:resolvent}). In this appendix, we briefly explain the modifications required to include these additional matrices.  In this case, solving  the generalized Rayleigh quotient \eqref{eqn:optimization} is equivalent to computing the SVD of the weighted resolvent operator \citep{Towneetal18}
\begin{subequations}
\begin{align}
\tilde{\bm R} &= \bm W^{1/2}_q \bm C (\text{i}\omega \bm I - \bm A)^{-1} \bm B \bm W^{-1/2}_f, \label{eqn:subeq1}
\\
\tilde{\bm R} &= \tilde{\bm U} \boldsymbol{\varSigma} \tilde{\bm V}^*, \label{eqn:subeq2}
\end{align}
\end{subequations}

and further 
\begin{equation}
\begin{gathered}
\bm U  = \bm W^{-1/2}_q \tilde{\bm U},
\\
\bm V  = \bm W^{-1/2}_f \tilde{\bm V},
\end{gathered}
\label{eqn:UV_rec}
\end{equation}
where $\boldsymbol{\varSigma}$ contains the gains, and $\bm V$ and $\bm U$ are forcing and response modes, respectively. The resolvent operator is recovered as 
\begin{equation}
\bm R = \bm U \boldsymbol{\varSigma} \bm V^* \bm W_f.
\label{eqn:res_rec}
\end{equation}

\par
\begin{figure}
\centering
\includegraphics[width=\textwidth]{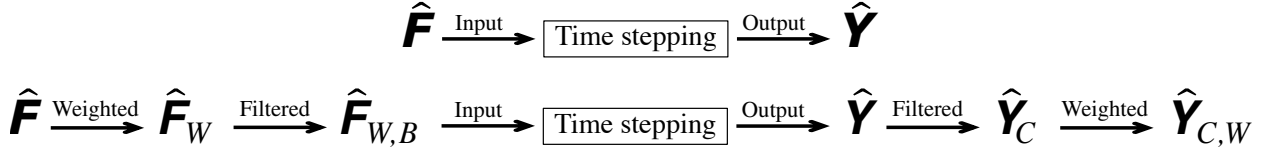}
\caption{The schematic of computing the action of $\bm R$ on top and the action of $\tilde{\bm R}$ on the bottom row.}
\label{fig:weighted_resolvent}
\end{figure}

 Time-stepping can effectively act as a surrogate for the action of the weighted resolvent operator $\tilde{\bm R}$ (or equivalently $\tilde{\bm R}^*$). In other words, our objective is to compute
\begin{equation}
\hat{\bm y} = \tilde{\bm R} \hat{\bm f} = \bm W^{1/2}_q \bm C (\text{i}\omega \bm I - \bm A)^{-1} \bm B \bm W^{-1/2}_f \hat{\bm f}
\label{eqn:action_Rmod}
\end{equation}
for all $\omega \in \varOmega$ using time stepping. The process begins by computing the product between $\hat{\bm f}_W = \bm W^{-1/2}_f \hat{\bm f}$ in Fourier space, followed by $\hat{\bm f}_{W,B} = \bm B \hat{\bm f}_W$. The products involving weight and input/output matrices are efficiently executed due to their sparsity. These operations are conducted for all $\omega \in \varOmega$ to obtain $\hat{\bm F}_{W,B}$. Subsequently, the action of $(\text{i}\omega \bm I - \bm A)^{-1}$ is computed on $\hat{\bm F}_{W,B}$ using time stepping to yield $\hat{\bm Y}$. The resulting output undergoes $\hat{\bm y}_C = \bm C \hat{\bm y}$ and $\hat{\bm y}_{C,W} = \bm W^{1/2}_q \hat{\bm y}_C$, which are repeated for all frequencies to obtain $\hat{\bm Y}_{C,W}$. Figure \ref{fig:weighted_resolvent} visually illustrates the order of calculations for $\bm R$ in the top row and $\tilde{\bm R}$ in the bottom row. An analogous process is utilized to compute the action of $\tilde{\bm R}^*$.

\section{Removing the least-damped modes using eigenvalues only}\label{appB}

The transient removal strategies described in $\S$\ref{sec:EffTrans} require a basis for the transient, either in the form of eigenvectors for the least-damped eigenvalues or data. In this appendix, we outline an alternative procedure to expedite the decay of transients that that uses knowledge of the least-damped eigenvalues themselves. Considering two solutions of \eqref{eqn:direct}, $\bm q_1 = \bm q(t_1)$ and $\bm q_2 = \bm q(t_1 + \Delta t)$, we can express them in terms of their steady-state and transient parts as 
\begin{equation}
\begin{aligned}
\bm q_1 &= \bm q_{s, 1} + \bm q_{t, 1},
\\
\bm q_2 &= \bm q_{s, 2} + \bm q_{t, 2},
\label{eqn:B1}
\end{aligned}
\end{equation}
where $\bm q_{s, 1}, \bm q_{s, 2}, \bm q_{t, 1},$ and $\bm q_{t, 2}$ are four unknowns. The transient parts can be written as 
\begin{equation}
\begin{aligned}
\bm q_{t, 1}  &= \bm q_{\lambda_1, 1} + \bm q_{rest, 1},
\\
\bm q_{t, 2}  &= \bm q_{\lambda_1, 2} + \bm q_{rest, 2},
\label{eqn:B2}
\end{aligned}
\end{equation}
where we assume the unknowns $\bm q_{\lambda_1, j}$ evolve as $\sim e^{\lambda_1 t}$, where $\lambda_1$ is the least-damped eigenvalue. Hence, 
\begin{equation}
\bm q_{\lambda_1, 2} = \bm q_{\lambda_1, 1}e^{\lambda_1 \Delta t},
\label{eqn:B3}
\end{equation}
where $\bm q_{\lambda_1, j}$ is essentially the projection of the transient response onto the least-damped eigenmode of $\bm A$ at $t = t_j$. The steady-state evolution at a prescribed forcing at a single frequency $\omega$ follows \eqref{eqn:2}. Therefore, in case of $||\bm q_{rest, j}|| = 0$, the system of equations is deterministic and $\bm q_{t, 1}$ can be found as 
\begin{equation}
\bm q_{t, 1} = \frac{\bm b}{c},
\label{eqn:B4}
\end{equation}
where $\bm b = \bm q_1 - \bm q_2e^{-\text{i}\omega \Delta t}$ is known from the time stepping and $c = {1 - e^{(\lambda_1 - \text{i}\omega) \Delta t}}$ is constant. Otherwise, \ie $||\bm q_{rest, j}|| \neq 0$, by simplifying terms, the transient part can be written as
\begin{equation}
\bm q_{t, 1} = \frac{\bm b}{c} - \frac{(1 - c)\bm q_{rest, 1} - \bm q_{rest, 2}e^{-\text{i}\omega \Delta t}}{c}.
\label{eqn:B5}
\end{equation}
Based on the fundamental assumption, the second term, which is unknown, decays faster than $e^{\lambda_{1, r} t}$. Therefore, by removing the first term $\frac{\bm b}{c}$, which is known, the residual eventually follows the second least-damped eigenvalue. If the forcing term encompasses a range of frequencies, the same relationships remain valid for each frequency after undergoing a DFT, and $\frac{\bm b}{c}$ can be separately eliminated for each $\omega \in \varOmega$. Note that the eigenvector associated with $\lambda_1$ was never used.

\par This procedure can be generalized to target the $d$ least-damped eigenmodes of $\bm A$. The solution at each time with arbitrary distances can be expanded as
\begin{equation}
\bm q_l = \bm q_{s, l} + \sum_{j = 1}^{d} \bm q_{\lambda_j, l} + \bm q_{rest, l},
\label{eqn:B6}
\end{equation}
for $1 \le l \le d+1$. Utilizing the same relationships, we can eliminate the slowest components, ensuring that the residual term decays faster than all $d$ modes. This procedure is developed to steepen the decay rate and shorten the transient length to meet the desired accuracy. The outcomes of this procedure closely resemble the output of the efficient transient strategy using Galerkin projection with the least-damped eigenmodes as the basis. The transient error can be estimated in a similar manner as described for the projection-based approach.

\end{appendices}

\bibliographystyle{jfm}
\bibliography{RSVDt}

\begin{thebibliography}{117}
\expandafter\ifx\csname natexlab\endcsname\relax\def\natexlab#1{#1}\fi
\def\au#1{#1} \def\ed#1{#1} \def\yr#1{#1}\def\at#1{#1}\def\jt#1{\textit{#1}}
  \def\bt#1{#1}\def\bvol#1{\textbf{#1}} \def\vol#1{#1} \def\pg#1{#1}
  \def\publ#1{#1}\def\arxiv#1{#1}\def\org#1{#1}\def\st#1{\textit{#1}}

\bibitem[Abgrall {\em et~al.\/}(2020)Abgrall, Nordstr{\"o}m, {\"O}ffner \&
  Tokareva]{Abgralletal20}
{\sc \au{Abgrall, R.}, \au{Nordstr{\"o}m, J.}, \au{{\"O}ffner, P.} \&
  \au{Tokareva, S.}} \yr{2020}  \at{Analysis of the sbp-sat stabilization for
  finite element methods part i: linear problems}.  \jt{Journal of Scientific
  Computing}  \bvol{85}~(2),  \pg{1--29}.

\bibitem[Adrian(2007)]{Adrian07}
{\sc \au{Adrian, R.~J.}} \yr{2007}  \at{Hairpin vortex organization in wall
  turbulence}.  \jt{Physics of Fluids}  \bvol{19}~(4).

\bibitem[Amestoy {\em et~al.\/}(2019)Amestoy, Buttari, l'Excellent \&
  Mary]{Amestoyetal19}
{\sc \au{Amestoy, P.~R.}, \au{Buttari, A.}, \au{l'Excellent, J.~Y.} \&
  \au{Mary, T.~A.}} \yr{2019}  \at{Bridging the gap between flat and
  hierarchical low-rank matrix formats: The multilevel block low-rank format}.
  \jt{SIAM Journal on Scientific Computing}  \bvol{41}~(3),  \pg{A1414--A1442}.

\bibitem[Amestoy {\em et~al.\/}(2001)Amestoy, Duff, L'Excellent \&
  Koster]{Amestoyetal01}
{\sc \au{Amestoy, P.~R}, \au{Duff, I.~S.}, \au{L'Excellent, J.~Y.} \&
  \au{Koster, J.}} \yr{2001}  \at{A fully asynchronous multifrontal solver
  using distributed dynamic scheduling}.  \jt{SIAM Journal on Matrix Analysis
  and Applications}  \bvol{23}~(1),  \pg{15--41}.

\bibitem[Anderson {\em et~al.\/}(1999)Anderson, Bai, Bischof, Blackford,
  Demmel, Dongarra, Du~Croz, Greenbaum, Hammarling, McKenney {\em
  et~al.\/}]{Andersonetal99}
{\sc \au{Anderson, E.}, \au{Bai, Z.}, \au{Bischof, C.}, \au{Blackford, L.~S.},
  \au{Demmel, J.}, \au{Dongarra, J.}, \au{Du~Croz, J.}, \au{Greenbaum, A.},
  \au{Hammarling, S.}, \au{McKenney, A.} \& \au{others}} \yr{1999} {\em LAPACK
  users' guide\/}.  \publ{SIAM}.

\bibitem[Arnoldi(1951)]{Arnoldi51}
{\sc \au{Arnoldi, W.~E.}} \yr{1951}  \at{The principle of minimized iterations
  in the solution of the matrix eigenvalue problem}.  \jt{Quarterly of Applied
  Mathematics}  \bvol{9}~(1),  \pg{17--29}.

\bibitem[Axelsson(1985)]{Axelsson85}
{\sc \au{Axelsson, O.}} \yr{1985}  \at{A survey of preconditioned iterative
  methods for linear systems of algebraic equations}.  \jt{BIT Numerical
  Mathematics}  \bvol{25}~(1),  \pg{165--187}.

\bibitem[Bagheri {\em et~al.\/}(2009)Bagheri, Henningson, Hoepffner \&
  Schmid]{Bagherietal09}
{\sc \au{Bagheri, S.}, \au{Henningson, D.~S.}, \au{Hoepffner, J.} \&
  \au{Schmid, P.~J.}} \yr{2009}  \at{Input-output analysis and control design
  applied to a linear model of spatially developing flows}.  \jt{Applied
  Mechanics Reviews}  \bvol{62}~(2).

\bibitem[Balay {\em et~al.\/}(2019)Balay, Abhyankar, Adams, Brown, Brune,
  Buschelman, Dalcin, Dener, Eijkhout, Gropp {\em et~al.\/}]{Balayetal19}
{\sc \au{Balay, S.}, \au{Abhyankar, S.}, \au{Adams, M.}, \au{Brown, J.},
  \au{Brune, P.}, \au{Buschelman, K.}, \au{Dalcin, L.}, \au{Dener, A.},
  \au{Eijkhout, V.}, \au{Gropp, W.} \& \au{others}} \yr{2019} {\em PETSc users
  manual\/}.  \publ{Argonne National Laboratory}.

\bibitem[Barthel {\em et~al.\/}(2022)Barthel, Gomez \& McKeon]{Bartheletal22}
{\sc \au{Barthel, B.}, \au{Gomez, S.} \& \au{McKeon, B.~J.}} \yr{2022}
  \at{Variational formulation of resolvent analysis}.  \jt{Physical Review
  Fluids}  \bvol{7}~(1),  \pg{013905}.

\bibitem[Benzi(2002)]{Benzi02}
{\sc \au{Benzi, M.}} \yr{2002}  \at{Preconditioning techniques for large linear
  systems: a survey}.  \jt{Journal of Computational Physics}  \bvol{182}~(2),
  \pg{418--477}.

\bibitem[Berg \& Nordstr{\"o}m(2012)]{BergNordstrom12}
{\sc \au{Berg, J.} \& \au{Nordstr{\"o}m, J.}} \yr{2012}  \at{Superconvergent
  functional output for time-dependent problems using finite differences on
  summation-by-parts form}.  \jt{Journal of Computational Physics}
  \bvol{231}~(20),  \pg{6846--6860}.

\bibitem[Br{\`e}s {\em et~al.\/}(2017)Br{\`e}s, Ham, Nichols \&
  Lele]{Bresetal17}
{\sc \au{Br{\`e}s, G.~A.}, \au{Ham, F.~E.}, \au{Nichols, J.~W.} \& \au{Lele,
  S.~K.}} \yr{2017}  \at{Unstructured large-eddy simulations of supersonic
  jets}.  \jt{AIAA journal}  \bvol{55}~(4),  \pg{1164--1184}.

\bibitem[Br{\`e}s {\em et~al.\/}(2018)Br{\`e}s, Jordan, Jaunet, Le~Rallic,
  Cavalieri, Towne, Lele, Colonius \& Schmidt]{Bresetal18}
{\sc \au{Br{\`e}s, G.~A.}, \au{Jordan, P.}, \au{Jaunet, V.}, \au{Le~Rallic,
  M.}, \au{Cavalieri, A. V.~G.}, \au{Towne, A.}, \au{Lele, S.~K.},
  \au{Colonius, T.} \& \au{Schmidt, O.~T.}} \yr{2018}  \at{Importance of the
  nozzle-exit boundary-layer state in subsonic turbulent jets}.  \jt{Journal of
  Fluid Mechanics}  \bvol{851},  \pg{83--124}.

\bibitem[Brynjell-Rahkola {\em et~al.\/}(2017)Brynjell-Rahkola, Tuckerman,
  Schlatter \& Henningson]{Brynjelletal17}
{\sc \au{Brynjell-Rahkola, M.}, \au{Tuckerman, L.~S.}, \au{Schlatter, P.} \&
  \au{Henningson, D.~S.}} \yr{2017}  \at{Computing optimal forcing using
  {L}aplace preconditioning}.  \jt{Communications in Computational Physics}
  \bvol{22}~(5),  \pg{1508--1532}.

\bibitem[Carpenter {\em et~al.\/}(1999)Carpenter, Nordstr{\"o}m \&
  Gottlieb]{Carpenteretal99}
{\sc \au{Carpenter, M.~H.}, \au{Nordstr{\"o}m, J.} \& \au{Gottlieb, D.}}
  \yr{1999}  \at{A stable and conservative interface treatment of arbitrary
  spatial accuracy}.  \jt{Journal of Computational Physics}  \bvol{148}~(2),
  \pg{341--365}.

\bibitem[Cavalieri {\em et~al.\/}(2019)Cavalieri, Jordan \&
  Lesshafft]{Cavalierietal19}
{\sc \au{Cavalieri, A. V.~G.}, \au{Jordan, P.} \& \au{Lesshafft, L.}} \yr{2019}
   \at{Wave-packet models for jet dynamics and sound radiation}.  \jt{Applied
  Mechanics Reviews}  \bvol{71}~(2).

\bibitem[Chavarin \& Luhar(2020)]{ChavarinLuhar20}
{\sc \au{Chavarin, A.} \& \au{Luhar, M.}} \yr{2020}  \at{Resolvent analysis for
  turbulent channel flow with riblets}.  \jt{AIAA Journal}  \bvol{58}~(2),
  \pg{589--599}.

\bibitem[Chen \& Rowley(2011)]{ChenRowley11}
{\sc \au{Chen, K.~K.} \& \au{Rowley, C.~W.}} \yr{2011}  \at{H2 optimal actuator
  and sensor placement in the linearised complex {G}inzburg--{L}andau system}.
  \jt{Journal of Fluid Mechanics}  \bvol{681},  \pg{241--260}.

\bibitem[Chu(1965)]{Chuetal65}
{\sc \au{Chu, B.-T.}} \yr{1965}  \at{On the energy transfer to small
  disturbances in fluid flow (part i)}.  \jt{Acta Mechanica}  \bvol{1}~(3),
  \pg{215--234}.

\bibitem[Colonius(2004)]{Colonius04}
{\sc \au{Colonius, T.}} \yr{2004}  \at{Modeling artificial boundary conditions
  for compressible flow}.  \jt{Annual Review of Fluid Mechanics}
  \bvol{36}~(1),  \pg{315--345}.

\bibitem[Cook \& Nichols(2023)]{CookNichols23}
{\sc \au{Cook, D.~A.} \& \au{Nichols, J.~W.}} \yr{2023}  \at{Three-dimensional
  receptivity of hypersonic sharp and blunt cones to free-stream planar waves
  using hierarchical input-output analysis}.  \jt{arXiv preprint
  arXiv:2306.03248} .

\bibitem[Cossu \& Brandt(2002)]{CossuBrandt02}
{\sc \au{Cossu, C.} \& \au{Brandt, L.}} \yr{2002}  \at{Stabilization of
  {T}ollmien--{S}chlichting waves by finite amplitude optimal streaks in the
  {B}lasius boundary layer}.  \jt{Physics of Fluids}  \bvol{14}~(8),
  \pg{L57--L60}.

\bibitem[Davis {\em et~al.\/}(2016)Davis, Rajamanickam \&
  Sid-Lakhdar]{Davisetal16}
{\sc \au{Davis, T.~A.}, \au{Rajamanickam, S.} \& \au{Sid-Lakhdar, W.~M.}}
  \yr{2016}  \at{A survey of direct methods for sparse linear systems}.
  \jt{Acta Numerica}  \bvol{25},  \pg{383--566}.

\bibitem[Dawson \& McKeon(2019)]{DawsonMcKeon19}
{\sc \au{Dawson, S. T.~M.} \& \au{McKeon, B.~J.}} \yr{2019}  \at{On the shape
  of resolvent modes in wall-bounded turbulence}.  \jt{Journal of Fluid
  Mechanics}  \bvol{877},  \pg{682--716}.

\bibitem[Duff {\em et~al.\/}(2017)Duff, Erisman \& Reid]{Duffetal17}
{\sc \au{Duff, I.~S.}, \au{Erisman, A.~M.} \& \au{Reid, J.~K.}} \yr{2017} {\em
  Direct methods for sparse matrices\/}.  \publ{Oxford University Press}.

\bibitem[Dunford \& Schwartz(1958)]{DunfordSchwartz58}
{\sc \au{Dunford, N.} \& \au{Schwartz, J.~T.}} \yr{1958} {\em Linear operators
  part I: general theory\/}.  \publ{John Wiley \& Sons}.

\bibitem[Edwards {\em et~al.\/}(1994)Edwards, Tuckerman, Friesner \&
  Sorensen]{Edwardsetal94}
{\sc \au{Edwards, W.~S.}, \au{Tuckerman, L.~S.}, \au{Friesner, R.~A.} \&
  \au{Sorensen, D.~C.}} \yr{1994}  \at{Krylov methods for the incompressible
  {N}avier-{S}tokes equations}.  \jt{Journal of computational physics}
  \bvol{110}~(1),  \pg{82--102}.

\bibitem[Eriksson \& Rizzi(1985)]{ErikssonRizzi85}
{\sc \au{Eriksson, L.~E.} \& \au{Rizzi, A.}} \yr{1985}  \at{Computer-aided
  analysis of the convergence to steady state of discrete approximations to the
  {E}uler equations}.  \jt{Journal of Computational Physics}  \bvol{57}~(1),
  \pg{90--128}.

\bibitem[Falgout \& Yang(2002)]{FalgoutYang02}
{\sc \au{Falgout, R.~D.} \& \au{Yang, U.~M.}} \yr{2002} hypre: A library of
  high performance preconditioners.  \bt{In {\em International Conference on
  Computational Science\/}},  \pg{pp. 632--641}.

\bibitem[G{\'o}mez {\em et~al.\/}(2016)G{\'o}mez, Sharma \&
  Blackburn]{Gomezetal16}
{\sc \au{G{\'o}mez, F.}, \au{Sharma, A.~S.} \& \au{Blackburn, H.~M.}} \yr{2016}
   \at{Estimation of unsteady aerodynamic forces using pointwise velocity
  data}.  \jt{Journal of Fluid Mechanics}  \bvol{804},  \pg{R4}.

\bibitem[Hairer {\em et~al.\/}(1993)Hairer, N{\o}rsett \& Wanner]{Haireretal93}
{\sc \au{Hairer, E.}, \au{N{\o}rsett, S.P.} \& \au{Wanner, G.}} \yr{1993} {\em
  Solving ordinary differential equations I: nonstiff problems\/}.
  \publ{Springer Berlin Heidelberg}.

\bibitem[Halko {\em et~al.\/}(2011)Halko, Martinsson \& Tropp]{Halkoetal11}
{\sc \au{Halko, N.}, \au{Martinsson, P.} \& \au{Tropp, J.~A.}} \yr{2011}
  \at{Finding structure with randomness: Probabilistic algorithms for
  constructing approximate matrix decompositions}.  \jt{SIAM Review}
  \bvol{53}~(2),  \pg{217--288}.

\bibitem[Hernandez {\em et~al.\/}(2005)Hernandez, Roman \&
  Vidal]{Hernandezetal05}
{\sc \au{Hernandez, V.}, \au{Roman, J.~E.} \& \au{Vidal, V.}} \yr{2005}
  \at{Slepc: A scalable and flexible toolkit for the solution of eigenvalue
  problems}.  \jt{ACM Transactions on Mathematical Software (TOMS)}
  \bvol{31}~(3),  \pg{351--362}.

\bibitem[Herrmann {\em et~al.\/}(2021)Herrmann, Baddoo, Semaan, Brunton \&
  McKeon]{Herrmannetal21}
{\sc \au{Herrmann, B.}, \au{Baddoo, P.~J.}, \au{Semaan, R.}, \au{Brunton,
  S.~L.} \& \au{McKeon, B.~J.}} \yr{2021}  \at{Data-driven resolvent analysis}.
   \jt{Journal of Fluid Mechanics}  \bvol{918},  \pg{A10}.

\bibitem[House {\em et~al.\/}(2022)House, Skene, Ribeiro, Yeh \&
  Taira]{Houseetal22}
{\sc \au{House, D.}, \au{Skene, C.}, \au{Ribeiro, J. H.~M.}, \au{Yeh, C.-A.} \&
  \au{Taira, K.}} \yr{2022}  \at{Sketch-based resolvent analysis}.  \jt{AIAA
  Paper $\#$2022-3335} .

\bibitem[Houtman {\em et~al.\/}(2023)Houtman, Timme \& Sharma]{Houtmanetal23}
{\sc \au{Houtman, J.}, \au{Timme, S.} \& \au{Sharma, A.}} \yr{2023}
  \at{Resolvent analysis of a finite wing in transonic flow}.  \jt{Flow}
  \bvol{3},  \pg{E14}.

\bibitem[Hu(2008)]{Hu08}
{\sc \au{Hu, F.~Q.}} \yr{2008}  \at{Development of pml absorbing boundary
  conditions for computational aeroacoustics: A progress review}.
  \jt{Computers \& Fluids}  \bvol{37}~(4),  \pg{336--348}.

\bibitem[Hunt \& Crighton(1991)]{HuntCrighton91}
{\sc \au{Hunt, R.~E.} \& \au{Crighton, D.~G.}} \yr{1991}  \at{Instability of
  flows in spatially developing media}.  \jt{Proceedings of the Royal Society
  of London. Series A: Mathematical and Physical Sciences}  \bvol{435}~(1893),
  \pg{109--128}.

\bibitem[Hutchins \& Marusic(2007)]{HutchinsMarusic07}
{\sc \au{Hutchins, N.} \& \au{Marusic, I.}} \yr{2007}  \at{Large-scale
  influences in near-wall turbulence}.  \jt{Philosophical Transactions of the
  Royal Society A: Mathematical, Physical and Engineering Sciences}
  \bvol{365}~(1852),  \pg{647--664}.

\bibitem[Jeun {\em et~al.\/}(2016)Jeun, Nichols \& Jovanovi{\'c}]{Jeunetal16}
{\sc \au{Jeun, J.}, \au{Nichols, J.~W.} \& \au{Jovanovi{\'c}, M.~R.}} \yr{2016}
   \at{Input-output analysis of high-speed axisymmetric isothermal jet noise}.
  \jt{Physics of Fluids}  \bvol{28}~(4),  \pg{047101}.

\bibitem[Jim{\'e}nez(2018)]{Jimenez18}
{\sc \au{Jim{\'e}nez, J.}} \yr{2018}  \at{Coherent structures in wall-bounded
  turbulence}.  \jt{Journal of Fluid Mechanics}  \bvol{842},  \pg{P1}.

\bibitem[Jimenez-Gonzalez \& Brancher(2017)]{JimenezBrancher17}
{\sc \au{Jimenez-Gonzalez, J.~I.} \& \au{Brancher, P.}} \yr{2017}
  \at{Transient energy growth of optimal streaks in parallel round jets}.
  \jt{Physics of Fluids}  \bvol{29}~(11),  \pg{114101}.

\bibitem[Jordan \& Colonius(2013)]{JordanColonius13}
{\sc \au{Jordan, P.} \& \au{Colonius, T.}} \yr{2013}  \at{Wave packets and
  turbulent jet noise}.  \jt{Annual Review of Fluid Mechanics}  \bvol{45}~(1),
  \pg{173--195}.

\bibitem[Jovanovic(2004)]{Jovanovic04}
{\sc \au{Jovanovic, M.~R.}} \yr{2004} {\em Modeling, analysis, and control of
  spatially distributed systems\/}.  \publ{University of California, Santa
  Barbara}.

\bibitem[Jovanovi{\'c}(2021)]{Jovanovic21}
{\sc \au{Jovanovi{\'c}, M.~R.}} \yr{2021}  \at{From bypass transition to flow
  control and data-driven turbulence modeling: An input--output viewpoint}.
  \jt{Annual Review of Fluid Mechanics}  \bvol{53}~(1),  \pg{010719--060244}.

\bibitem[Kamal {\em et~al.\/}(2023)Kamal, Lakebrink \& Colonius]{Kamaletal23}
{\sc \au{Kamal, O.}, \au{Lakebrink, M.~T.} \& \au{Colonius, T.}} \yr{2023}
  \at{Global receptivity analysis: physically realizable input--output
  analysis}.  \jt{Journal of Fluid Mechanics}  \bvol{956},  \pg{R5}.

\bibitem[Karban {\em et~al.\/}(2020)Karban, Bugeat, Martini, Towne, Cavalieri,
  Lesshafft, Agarwal, Jordan \& Colonius]{Karbanetal20}
{\sc \au{Karban, U.}, \au{Bugeat, B.}, \au{Martini, E.}, \au{Towne, A.},
  \au{Cavalieri, A. V.~G.}, \au{Lesshafft, L.}, \au{Agarwal, A.}, \au{Jordan,
  P.} \& \au{Colonius, T.}} \yr{2020}  \at{Ambiguity in mean-flow-based linear
  analysis}.  \jt{Journal of Fluid Mechanics}  \bvol{900},  \pg{R5}.

\bibitem[Kato(2013)]{Kato13}
{\sc \au{Kato, T.}} \yr{2013} {\em Perturbation theory for linear operators\/}.
   \publ{Springer Science \& Business Media}.

\bibitem[Lesshafft {\em et~al.\/}(2019)Lesshafft, Semeraro, Jaunet, Cavalieri
  \& Jordan]{Lesshafftetal19}
{\sc \au{Lesshafft, L.}, \au{Semeraro, O.}, \au{Jaunet, V.}, \au{Cavalieri, A.
  V.~G.} \& \au{Jordan, P.}} \yr{2019}  \at{Resolvent-based modeling of
  coherent wave packets in a turbulent jet}.  \jt{Physical Review Fluids}
  \bvol{4}~(6),  \pg{063901}.

\bibitem[Li \& Malik(1996)]{LiMalik96}
{\sc \au{Li, F.} \& \au{Malik, M.~R.}} \yr{1996}  \at{On the nature of {PSE}
  approximation}.  \jt{Theoretical and Computational Fluid Dynamics}
  \bvol{8}~(4),  \pg{253--273}.

\bibitem[Lumley(1967)]{Lumley67}
{\sc \au{Lumley, J.~L.}} \yr{1967}  \at{The structure of inhomogeneous
  turbulent flows}.  \jt{Atmospheric Turbulence and Radio Wave Propagation}
  \pg{pp. 166--178}.

\bibitem[Mani(2012)]{Mani12}
{\sc \au{Mani, A.}} \yr{2012}  \at{Analysis and optimization of numerical
  sponge layers as a nonreflective boundary treatment}.  \jt{Journal of
  Computational Physics}  \bvol{231}~(2),  \pg{704--716}.

\bibitem[Marant \& Cossu(2018)]{MarantCossu18}
{\sc \au{Marant, M.} \& \au{Cossu, C.}} \yr{2018}  \at{Influence of optimally
  amplified streamwise streaks on the {K}elvin--{H}elmholtz instability}.
  \jt{Journal of Fluid Mechanics}  \bvol{838},  \pg{478--500}.

\bibitem[Marquet \& Larsson(2015)]{MarquetLarsson15}
{\sc \au{Marquet, O.} \& \au{Larsson, M.}} \yr{2015}  \at{Global wake
  instabilities of low aspect-ratio flat-plates}.  \jt{European Journal of
  Mechanics-B/Fluids}  \bvol{49},  \pg{400--412}.

\bibitem[Martini {\em et~al.\/}(2020)Martini, Cavalieri, Jordan, Towne \&
  Lesshafft]{Martinietal20}
{\sc \au{Martini, E.}, \au{Cavalieri, A. V.~G.}, \au{Jordan, P.}, \au{Towne,
  A.} \& \au{Lesshafft, L.}} \yr{2020}  \at{Resolvent-based optimal estimation
  of transitional and turbulent flows}.  \jt{Journal of Fluid Mechanics}
  \bvol{900},  \pg{A2}.

\bibitem[Martini {\em et~al.\/}(2022)Martini, Jung, Cavalieri, Jordan \&
  Towne]{Martinietal22}
{\sc \au{Martini, E.}, \au{Jung, J.}, \au{Cavalieri, A. V.~G.}, \au{Jordan, P.}
  \& \au{Towne, A.}} \yr{2022}  \at{Resolvent-based tools for optimal
  estimation and control via the {W}iener--{H}opf formalism}.  \jt{Journal of
  Fluid Mechanics}  \bvol{938},  \pg{E2}.

\bibitem[Martini {\em et~al.\/}(2021)Martini, Rodr{\'\i}guez, Towne \&
  Cavalieri]{Martinietal21}
{\sc \au{Martini, E.}, \au{Rodr{\'\i}guez, D.}, \au{Towne, A.} \&
  \au{Cavalieri, A. V.~G.}} \yr{2021}  \at{Efficient computation of global
  resolvent modes}.  \jt{Journal of Fluid Mechanics}  \bvol{919},  \pg{A3}.

\bibitem[Mattsson \& Nordstr{\"o}m(2004)]{MattssonNordstrom04}
{\sc \au{Mattsson, K.} \& \au{Nordstr{\"o}m, J.}} \yr{2004}  \at{Summation by
  parts operators for finite difference approximations of second derivatives}.
  \jt{Journal of Computational Physics}  \bvol{199}~(2),  \pg{503--540}.

\bibitem[McKeon(2017)]{Mckeon17}
{\sc \au{McKeon, B.~J.}} \yr{2017}  \at{The engine behind (wall) turbulence:
  perspectives on scale interactions}.  \jt{Journal of Fluid Mechanics}
  \bvol{817},  \pg{P1}.

\bibitem[McKeon \& Sharma(2010)]{McKeonSharma10}
{\sc \au{McKeon, B.~J.} \& \au{Sharma, A.~S.}} \yr{2010}  \at{A critical-layer
  framework for turbulent pipe flow}.  \jt{Journal of Fluid Mechanics}
  \bvol{658},  \pg{336--382}.

\bibitem[Moarref {\em et~al.\/}(2013)Moarref, Sharma, Tropp \&
  McKeon]{Moarrefetal13}
{\sc \au{Moarref, R.}, \au{Sharma, A.~S.}, \au{Tropp, J.~A.} \& \au{McKeon,
  B.~J.}} \yr{2013}  \at{Model-based scaling of the streamwise energy density
  in high-{R}eynolds-number turbulent channels}.  \jt{Journal of Fluid
  Mechanics}  \bvol{734},  \pg{275--316}.

\bibitem[Monokrousos {\em et~al.\/}(2010)Monokrousos, {\AA}kervik, Brandt \&
  Henningson]{Monokrousosetal10}
{\sc \au{Monokrousos, A.}, \au{{\AA}kervik, E.}, \au{Brandt, L.} \&
  \au{Henningson, D.~S.}} \yr{2010}  \at{Global three-dimensional optimal
  disturbances in the {B}lasius boundary-layer flow using time-steppers}.
  \jt{Journal of Fluid Mechanics}  \bvol{650},  \pg{181--214}.

\bibitem[Morra {\em et~al.\/}(2019)Morra, Semeraro, Henningson \&
  Cossu]{Morraetal19}
{\sc \au{Morra, P.}, \au{Semeraro, O.}, \au{Henningson, D.~S.} \& \au{Cossu,
  C.}} \yr{2019}  \at{On the relevance of {R}eynolds stresses in resolvent
  analyses of turbulent wall-bounded flows}.  \jt{Journal of Fluid Mechanics}
  \bvol{867},  \pg{969--984}.

\bibitem[Nogueira {\em et~al.\/}(2019)Nogueira, Cavalieri, Jordan \&
  Jaunet]{Nogueiraetal19}
{\sc \au{Nogueira, P. A.~S.}, \au{Cavalieri, A. V.~G.}, \au{Jordan, P.} \&
  \au{Jaunet, V.}} \yr{2019}  \at{Large-scale streaky structures in turbulent
  jets}.  \jt{Journal of Fluid Mechanics}  \bvol{873},  \pg{211--237}.

\bibitem[Nyquist(1928)]{Nyquist28}
{\sc \au{Nyquist, H.}} \yr{1928}  \at{Certain topics in telegraph transmission
  theory}.  \jt{Transactions of the American Institute of Electrical Engineers}
   \bvol{47}~(2),  \pg{617--644}.

\bibitem[de~Pando {\em et~al.\/}(2012)de~Pando, Sipp \& Schmid]{dePandoetal12}
{\sc \au{de~Pando, M.~F.}, \au{Sipp, D.} \& \au{Schmid, P.~J.}} \yr{2012}
  \at{Efficient evaluation of the direct and adjoint linearized dynamics from
  compressible flow solvers}.  \jt{Journal of Computational Physics}
  \bvol{231}~(23),  \pg{7739--7755}.

\bibitem[Pearson \& Pestana(2020)]{PearsonPestana20}
{\sc \au{Pearson, J.~W.} \& \au{Pestana, J.}} \yr{2020}  \at{Preconditioners
  for krylov subspace methods: An overview}.  \jt{GAMM-Mitteilungen}
  \bvol{43}~(4),  \pg{e202000015}.

\bibitem[Pickering {\em et~al.\/}(2020)Pickering, Rigas, Nogueira, Cavalieri,
  Schmidt \& Colonius]{Pickeringetal20}
{\sc \au{Pickering, E.}, \au{Rigas, G.}, \au{Nogueira, P. A.~S.},
  \au{Cavalieri, A. V.~G.}, \au{Schmidt, O.~T.} \& \au{Colonius, T.}} \yr{2020}
   \at{Lift-up, {K}elvin--{H}elmholtz and {O}rr mechanisms in turbulent jets}.
  \jt{Journal of Fluid Mechanics}  \bvol{896},  \pg{A2}.

\bibitem[Pickering {\em et~al.\/}(2021)Pickering, Rigas, Schmidt, Sipp \&
  Colonius]{Pickeringetal21}
{\sc \au{Pickering, E.}, \au{Rigas, G.}, \au{Schmidt, O.~T.}, \au{Sipp, D.} \&
  \au{Colonius, T.}} \yr{2021}  \at{Optimal eddy viscosity for resolvent-based
  models of coherent structures in turbulent jets}.  \jt{Journal of Fluid
  Mechanics}  \bvol{917},  \pg{A29}.

\bibitem[Ran {\em et~al.\/}(2021)Ran, Zare \& Jovanovi{\'c}]{Ranetal21}
{\sc \au{Ran, W.}, \au{Zare, A.} \& \au{Jovanovi{\'c}, M.~R.}} \yr{2021}
  \at{Model-based design of riblets for turbulent drag reduction}.  \jt{Journal
  of Fluid Mechanics}  \bvol{906},  \pg{A7}.

\bibitem[Reynolds \& Hussain(1972)]{ReynoldsHussain72}
{\sc \au{Reynolds, W.~C.} \& \au{Hussain, A. K. M.~F.}} \yr{1972}  \at{The
  mechanics of an organized wave in turbulent shear flow. part 3. theoretical
  models and comparisons with experiments}.  \jt{Journal of Fluid Mechanics}
  \bvol{54}~(2),  \pg{263--288}.

\bibitem[Ribeiro {\em et~al.\/}(2020)Ribeiro, Yeh \& Taira]{Ribeiroetal20}
{\sc \au{Ribeiro, J. H.~M.}, \au{Yeh, C.-A.} \& \au{Taira, K.}} \yr{2020}
  \at{Randomized resolvent analysis}.  \jt{Physical Review Fluids}
  \bvol{5}~(3),  \pg{033902}.

\bibitem[Rolandi {\em et~al.\/}(2024)Rolandi, Ribeiro, Yeh \&
  Taira]{Rolandietal24}
{\sc \au{Rolandi, L.~V.}, \au{Ribeiro, J. H.~M.}, \au{Yeh, C.-A.} \& \au{Taira,
  K.}} \yr{2024}  \at{An invitation to resolvent analysis}.  \jt{Theoretical
  and Computational Fluid Dynamics}  \pg{pp. 1--37}.

\bibitem[Saad(2003{\natexlab{{\em a\/}}})]{Saad03_2}
{\sc \au{Saad, Y.}} \yr{2003{\natexlab{{\em a\/}}}}  \at{Finding exact and
  approximate block structures for ilu preconditioning}.  \jt{SIAM Journal on
  Scientific Computing}  \bvol{24}~(4),  \pg{1107--1123}.

\bibitem[Saad(2003{\natexlab{{\em b\/}}})]{Saad03}
{\sc \au{Saad, Y.}} \yr{2003{\natexlab{{\em b\/}}}} {\em Iterative methods for
  sparse linear systems\/}.  \publ{SIAM}.

\bibitem[Sasaki {\em et~al.\/}(2022)Sasaki, Cavalieri, Hanifi \&
  Henningson]{Sasakietal22}
{\sc \au{Sasaki, K.}, \au{Cavalieri, A. V.~G.}, \au{Hanifi, A.} \&
  \au{Henningson, D.~S.}} \yr{2022}  \at{Parabolic resolvent modes for streaky
  structures in transitional and turbulent boundary layers}.  \jt{Physical
  Review Fluids}  \bvol{7}~(10),  \pg{104611}.

\bibitem[Schenk {\em et~al.\/}(2001)Schenk, G{\"a}rtner, Fichtner \&
  Stricker]{Schenketal01}
{\sc \au{Schenk, O.}, \au{G{\"a}rtner, K.}, \au{Fichtner, W.} \& \au{Stricker,
  A.}} \yr{2001}  \at{Pardiso: a high-performance serial and parallel sparse
  linear solver in semiconductor device simulation}.  \jt{Future Generation
  Computer Systems}  \bvol{18}~(1),  \pg{69--78}.

\bibitem[Schlander {\em et~al.\/}(2024)Schlander, Rigopoulos \&
  Papadakis]{Schlander24}
{\sc \au{Schlander, R.~K.}, \au{Rigopoulos, S.} \& \au{Papadakis, G.}}
  \yr{2024}  \at{Resolvent analysis of turbulent flow laden with low-inertia
  particles}.  \jt{Journal of Fluid Mechanics}  \bvol{985},  \pg{A27}.

\bibitem[Schmid(2007)]{Schmid07}
{\sc \au{Schmid, P.~J.}} \yr{2007}  \at{Nonmodal stability theory}.  \jt{Annual
  Review of Fluid Mechanics}  \bvol{39}~(1),  \pg{129--162}.

\bibitem[Schmid(2010)]{Schmid10}
{\sc \au{Schmid, P.~J.}} \yr{2010}  \at{Dynamic mode decomposition of numerical
  and experimental data}.  \jt{Journal of Fluid Mechanics}  \bvol{656},
  \pg{5--28}.

\bibitem[Schmid(2022)]{Schmid22}
{\sc \au{Schmid, P.~J.}} \yr{2022}  \at{Dynamic mode decomposition and its
  variants}.  \jt{Annual Review of Fluid Mechanics}  \bvol{54},  \pg{225--254}.

\bibitem[Schmid \& Henningson(2001)]{SchmidHenningson01}
{\sc \au{Schmid, P.~J.} \& \au{Henningson, D.~S.}} \yr{2001} {\em Stability and
  transition in shear flows\/}.  \publ{Springer, New York}.

\bibitem[Schmidt \& Towne(2019)]{SchmidtTowne19}
{\sc \au{Schmidt, O.~T.} \& \au{Towne, A.}} \yr{2019}  \at{An efficient
  streaming algorithm for spectral proper orthogonal decomposition}.
  \jt{Computer Physics Communications}  \bvol{237},  \pg{98--109}.

\bibitem[Schmidt {\em et~al.\/}(2017)Schmidt, Towne, Colonius, Cavalieri,
  Jordan \& Br{\`e}s]{Schmidtetal17}
{\sc \au{Schmidt, O.~T.}, \au{Towne, A.}, \au{Colonius, T.}, \au{Cavalieri, A.
  V.~G.}, \au{Jordan, P.} \& \au{Br{\`e}s, G.~A.}} \yr{2017}  \at{Wavepackets
  and trapped acoustic modes in a turbulent jet: coherent structure eduction
  and global stability}.  \jt{Journal of Fluid Mechanics}  \bvol{825},
  \pg{1153--1181}.

\bibitem[Schmidt {\em et~al.\/}(2018)Schmidt, Towne, Rigas, Colonius \&
  Br{\`e}s]{Schmidtetal18}
{\sc \au{Schmidt, O.~T.}, \au{Towne, A.}, \au{Rigas, G.}, \au{Colonius, T.} \&
  \au{Br{\`e}s, G.~A.}} \yr{2018}  \at{Spectral analysis of jet turbulence}.
  \jt{Journal of Fluid Mechanics}  \bvol{855},  \pg{953--982}.

\bibitem[Sinha {\em et~al.\/}(2016)Sinha, Gudmundsson, Xia \&
  Colonius]{Sinhaetal16}
{\sc \au{Sinha, A.}, \au{Gudmundsson, K.}, \au{Xia, H.} \& \au{Colonius, T.}}
  \yr{2016}  \at{Parabolized stability analysis of jets from serrated nozzles}.
   \jt{Journal of Fluid Mechanics}  \bvol{789},  \pg{36--63}.

\bibitem[Sipp \& Marquet(2013)]{SippMarquet13}
{\sc \au{Sipp, D.} \& \au{Marquet, O.}} \yr{2013}  \at{Characterization of
  noise amplifiers with global singular modes: the case of the leading-edge
  flat-plate boundary layer}.  \jt{Theoretical and Computational Fluid
  Dynamics}  \bvol{27}~(5),  \pg{617--635}.

\bibitem[Sirovich(1987{\natexlab{{\em a\/}}})]{Sirovich87_1}
{\sc \au{Sirovich, L.}} \yr{1987{\natexlab{{\em a\/}}}}  \at{Turbulence and the
  dynamics of coherent structures. i. coherent structures}.  \jt{Quarterly of
  Applied Mathematics}  \bvol{45}~(3),  \pg{561--571}.

\bibitem[Sirovich(1987{\natexlab{{\em b\/}}})]{Sirovich87_2}
{\sc \au{Sirovich, L.}} \yr{1987{\natexlab{{\em b\/}}}}  \at{Turbulence and the
  dynamics of coherent structures. ii. symmetries and transformations}.
  \jt{Quarterly of Applied Mathematics}  \bvol{45}~(3),  \pg{573--582}.

\bibitem[Skeel(1979)]{Skeel79}
{\sc \au{Skeel, R.~D.}} \yr{1979}  \at{Scaling for numerical stability in
  {G}aussian elimination}.  \jt{Journal of the ACM (JACM)}  \bvol{26}~(3),
  \pg{494--526}.

\bibitem[Stewart(1993)]{Stewart93}
{\sc \au{Stewart, G.~W.}} \yr{1993}  \at{On the early history of the singular
  value decomposition}.  \jt{SIAM review}  \bvol{35}~(4),  \pg{551--566}.

\bibitem[Stewart(1998)]{Stewart98}
{\sc \au{Stewart, G.~W.}} \yr{1998} {\em Perturbation theory for the singular
  value decomposition\/}.  \publ{Citeseer}.

\bibitem[Stewart(2006)]{Stewart06}
{\sc \au{Stewart, M.}} \yr{2006}  \at{Perturbation of the svd in the presence
  of small singular values}.  \jt{Linear Algebra and its Applications}
  \bvol{419}~(1),  \pg{53--77}.

\bibitem[Stewartson \& Stuart(1971)]{StewartsonStuart71}
{\sc \au{Stewartson, K.} \& \au{Stuart, J.~T.}} \yr{1971}  \at{A non-linear
  instability theory for a wave system in plane {P}oiseuille flow}.
  \jt{Journal of Fluid Mechanics}  \bvol{48}~(3),  \pg{529--545}.

\bibitem[S{\"u}li \& Mayers(2003)]{SuliMayers03}
{\sc \au{S{\"u}li, E.} \& \au{Mayers, D.~F.}} \yr{2003} {\em An introduction to
  numerical analysis\/}.  \publ{Cambridge university press}.

\bibitem[Symon {\em et~al.\/}(2019)Symon, Sipp \& McKeon]{Symonetal19}
{\sc \au{Symon, S.}, \au{Sipp, D.} \& \au{McKeon, B.~J.}} \yr{2019}  \at{A tale
  of two airfoils: resolvent-based modelling of an oscillator versus an
  amplifier from an experimental mean}.  \jt{Journal of Fluid Mechanics}
  \bvol{881},  \pg{51--83}.

\bibitem[Taira {\em et~al.\/}(2017)Taira, Brunton, Dawson, Rowley, Colonius,
  McKeon, Schmidt, Gordeyev, Theofilis \& Ukeiley]{Tairaetal17}
{\sc \au{Taira, K.}, \au{Brunton, S.~L.}, \au{Dawson, S. T.~M.}, \au{Rowley,
  C.~W.}, \au{Colonius, T.}, \au{McKeon, B.~J.}, \au{Schmidt, O.~T.},
  \au{Gordeyev, S.}, \au{Theofilis, V.} \& \au{Ukeiley, L.~S.}} \yr{2017}
  \at{Modal analysis of fluid flows: An overview}.  \jt{AIAA Journal}
  \bvol{55}~(12),  \pg{4013--4041}.

\bibitem[Theofilis(2011)]{Theofilis11}
{\sc \au{Theofilis, V.}} \yr{2011}  \at{Global linear instability}.  \jt{Annual
  Review of Fluid Mechanics}  \bvol{43},  \pg{319--352}.

\bibitem[Thomareis \& Papadakis(2018)]{ThomareisPapadakis18}
{\sc \au{Thomareis, N.} \& \au{Papadakis, G.}} \yr{2018}  \at{Resolvent
  analysis of separated and attached flows around an airfoil at transitional
  {R}eynolds number}.  \jt{Physical Review Fluids}  \bvol{3}~(7),  \pg{073901}.

\bibitem[Towne(2016)]{Towne16}
{\sc \au{Towne, A.}} \yr{2016}  \at{Advancements in jet turbulence and noise
  modeling: accurate one-way solutions and empirical evaluation of the
  nonlinear forcing of wavepackets}. PhD thesis, California Institute of
  Technology.

\bibitem[Towne \& Colonius(2015)]{TowneColonius15}
{\sc \au{Towne, A.} \& \au{Colonius, T.}} \yr{2015}  \at{One-way spatial
  integration of hyperbolic equations}.  \jt{Journal of Computational Physics}
  \bvol{300},  \pg{844--861}.

\bibitem[Towne {\em et~al.\/}(2015)Towne, Colonius, Jordan, Cavalieri \&
  Bres]{Towneetal15}
{\sc \au{Towne, A.}, \au{Colonius, T.}, \au{Jordan, P.}, \au{Cavalieri, A.~V.}
  \& \au{Bres, G.~A.}} \yr{2015}  \at{Stochastic and nonlinear forcing of
  wavepackets in a {M}ach 0.9 jet}.  \jt{AIAA Paper $\#$2015-2217} .

\bibitem[Towne {\em et~al.\/}(2020)Towne, Lozano-Dur{\'a}n \&
  Yang]{Towneetal20}
{\sc \au{Towne, A.}, \au{Lozano-Dur{\'a}n, A.} \& \au{Yang, X.}} \yr{2020}
  \at{Resolvent-based estimation of space--time flow statistics}.  \jt{Journal
  of Fluid Mechanics}  \bvol{883},  \pg{A17}.

\bibitem[Towne {\em et~al.\/}(2019)Towne, Rigas \& Colonius]{Towneetal19}
{\sc \au{Towne, A.}, \au{Rigas, G.} \& \au{Colonius, T.}} \yr{2019}  \at{A
  critical assessment of the parabolized stability equations}.  \jt{Theoretical
  and Computational Fluid Dynamics}  \bvol{33},  \pg{359--382}.

\bibitem[Towne {\em et~al.\/}(2022)Towne, Rigas, Kamal, Pickering \&
  Colonius]{Towneetal22}
{\sc \au{Towne, A.}, \au{Rigas, G.}, \au{Kamal, O.}, \au{Pickering, E.} \&
  \au{Colonius, T.}} \yr{2022}  \at{Efficient global resolvent analysis via the
  one-way {N}avier-{S}tokes equations}.  \jt{Journal of Fluid Mechanics}
  \bvol{948},  \pg{A9}.

\bibitem[Towne {\em et~al.\/}(2018)Towne, Schmidt \& Colonius]{Towneetal18}
{\sc \au{Towne, A.}, \au{Schmidt, O.~T.} \& \au{Colonius, T.}} \yr{2018}
  \at{Spectral proper orthogonal decomposition and its relationship to dynamic
  mode decomposition and resolvent analysis}.  \jt{Journal of Fluid Mechanics}
  \bvol{847},  \pg{821--867}.

\bibitem[Trefethen \& Bau~III(1997)]{TrefethenBau97}
{\sc \au{Trefethen, L.~N.} \& \au{Bau~III, D.}} \yr{1997} {\em Numerical linear
  algebra\/}.  \publ{Siam}.

\bibitem[Vershynin(2018)]{Vershynin18}
{\sc \au{Vershynin, R.}} \yr{2018} {\em High-dimensional probability: An
  introduction with applications in data science\/}.  \publ{Cambridge
  University Press}.

\bibitem[Vishnampet {\em et~al.\/}(2015)Vishnampet, Bodony \&
  Freund]{Vishnampetetal15}
{\sc \au{Vishnampet, R.}, \au{Bodony, D.~J.} \& \au{Freund, J.~B.}} \yr{2015}
  \at{A practical discrete-adjoint method for high-fidelity compressible
  turbulence simulations}.  \jt{Journal of Computational Physics}  \bvol{285},
  \pg{173--192}.

\bibitem[Vogel \& Coder(2022)]{VogelCoder22}
{\sc \au{Vogel, E.~A.} \& \au{Coder, J.~G.}} \yr{2022}  \at{A novel entropy
  normalization scheme for characterization of highly compressible flows}.
  \jt{Theoretical and Computational Fluid Dynamics}  \bvol{36}~(4),
  \pg{641--670}.

\bibitem[Wang {\em et~al.\/}(2021)Wang, Lesshafft, Cavalieri \&
  Jordan]{Wangetal21}
{\sc \au{Wang, C.}, \au{Lesshafft, L.}, \au{Cavalieri, A. V.~G.} \& \au{Jordan,
  P.}} \yr{2021}  \at{The effect of streaks on the instability of jets}.
  \jt{Journal of Fluid Mechanics}  \bvol{910},  \pg{A14}.

\bibitem[Wanner \& Hairer(1996)]{WannerHairer96}
{\sc \au{Wanner, G.} \& \au{Hairer, E.}} \yr{1996} {\em Solving ordinary
  differential equations II: stiff and differential-algebraic problems\/}.
  \publ{Springer Berlin Heidelberg}.

\bibitem[Yeh {\em et~al.\/}(2020)Yeh, Benton, Taira \& Garmann]{Yehetal20}
{\sc \au{Yeh, C.-A.}, \au{Benton, S.~I.}, \au{Taira, K.} \& \au{Garmann,
  D.~J.}} \yr{2020}  \at{Resolvent analysis of an airfoil laminar separation
  bubble at {Re} = 500 000}.  \jt{Physical Review Fluids}  \bvol{5}~(8),
  \pg{083906}.

\bibitem[Yeh \& Taira(2019)]{YehTaira19}
{\sc \au{Yeh, C.-A.} \& \au{Taira, K.}} \yr{2019}  \at{Resolvent-analysis-based
  design of airfoil separation control}.  \jt{Journal of Fluid Mechanics}
  \bvol{867},  \pg{572--610}.

\bibitem[Zhou {\em et~al.\/}(2022)Zhou, Xu \& Jim{\'e}nez]{Zhouetal22}
{\sc \au{Zhou, Z.}, \au{Xu, C.} \& \au{Jim{\'e}nez, J.}} \yr{2022}
  \at{Interaction between near-wall streaks and large-scale motions in
  turbulent channel flows}.  \jt{Journal of Fluid Mechanics}  \bvol{940},
  \pg{A23}.

\bibitem[Zhu \& Towne(2023)]{ZhuTowne23}
{\sc \au{Zhu, M.} \& \au{Towne, A.}} \yr{2023}  \at{Recursive one-way
  {N}avier-{S}tokes equations with {PSE}-like cost}.  \jt{Journal of
  Computational Physics}  \bvol{473},  \pg{111744}.

\end{thebibliography}

\end{document}